\documentclass[fleqn,usenatbib,onecolumn]{mnras}

\usepackage{newtxtext,newtxmath}

\usepackage[T1]{fontenc}

\DeclareRobustCommand{\VAN}[3]{#2}
\let\VANthebibliography\thebibliography
\def\thebibliography{\DeclareRobustCommand{\VAN}[3]{##3}\VANthebibliography}

\usepackage{graphicx}	

\newcommand{\curl}{\vec \nabla \times}
\newcommand{\vamb}{\vec v_\mathrm{amb}}
\newcommand{\vp}{\vec v_\mathrm{p}}

\newcommand{\ve}{\vec v_\mathrm{e}}
\newcommand{\nnp}{n_\mathrm{p}}
\newcommand{\nnn}{n_\mathrm{n}}
\newcommand{\nne}{n_\mathrm{e}}
\newcommand{\nnc}{n_\mathrm{c}}
\newcommand{\taupn}{\tau_\mathrm{pn}}
\newcommand{\Am}{\mathrm{Am}}

\title[Ambipolar diffusion in NS cores]{Three-dimensional numerical simulations of ambipolar diffusion in NS cores in the one-fluid approximation: instability of poloidal magnetic field}

\author[A.P. Igoshev \& R. Hollerbach]{
Andrei P. Igoshev\thanks{E-mail: a.igoshev@leeds.ac.uk, ignotur@gmail.com}
\& Rainer Hollerbach
\\
Department of Applied Mathematics, University of Leeds, LS2 9JT Leeds, UK
}

\date{Accepted XXX. Received YYY; in original form ZZZ}

\pubyear{2015}

\begin{document}
\label{firstpage}
\pagerange{\pageref{firstpage}--\pageref{lastpage}}
\maketitle

\begin{abstract}
We numerically model evolution of magnetic fields inside a neutron star under the influence of ambipolar diffusion in the weak-coupling mode in the one-fluid MHD approximation. Our simulations are three-dimensional and performed in spherical coordinates. Our model covers the neutron star core and includes crust where the magnetic field decay is due to Ohmic decay. We discover an instability of poloidal magnetic field under the influence of ambipolar diffusion. This instability develops in the neutron star core and grows on a timescale of 0.2 dimensionless times, reaching saturation by 2 dimensionless times. The instability leads to formation of azimuthal magnetic field with azimuthal wavenumber $m=14$ (at the moment of saturation) which keeps merging and reaches $m=4$ by 16 dimensionless times.
Over the course of our simulations (16 dimensionless times) the surface dipolar magnetic field decays, reaching 20~percent of its original value and keeps decaying. The decay timescale for the total magnetic energy is six dimensionless times. The ambipolar diffusion induces electric currents in the crust where these currents dissipate efficiently. Strong electric currents in the crust lead to heating, which could correspond to luminosities of $\approx 10^{29}$~erg~s$^{-1}$ during hundreds of Myrs for an initial magnetic field of $10^{14}$~G. Ambipolar diffusion leads to formation of small-scale magnetic fields at the neutron star surface.
\end{abstract}




\section{Introduction}
Neutron stars (NSs) are observed as vastly different astrophysical sources of transient and periodic nature which emit electromagnetic radiation ranging from radio to $\gamma$-rays. Isolated pulsars are best known for their periodic radio pulses \citep{Lorimer2012book}; Anomalous X-ray Pulsars and Soft Gamma Repeaters known as magnetars are mostly seen in X-rays and occasionally as transients in $\gamma$-rays and radio (for review see \citealt{KaspiBeloborodov2017}). The X-ray Dim Isolated Neutron stars (XDINs; for review see \citealt{Turolla2009ASSL}) and central compact objects (CCOs, for review see \citealt{Mayer2021A}) emit thermal X-ray radiation only. 

It was suggested by \cite{Harding2013FrPhy} that this observed diversity of NSs is explained by diversity of their magnetic field configurations and evolution. In this framework, magnetars have the strongest poloidal fields with comparable strength crust-confined toroidal fields, see e.g. \cite*{igoshev2021NatAs}. Central compact objects might have dipolar magnetic field suppressed by the fall-back \citep{Shabaltas2012ApJ,Vigano2012MNRAS,Igoshev2016MNRAS} or alternatively only small-scale magnetic fields generated as a result of a stochastic dynamo \citep{Gourgouliatos2020MNRAS,Igoshev2021ApJ}. Similarly an interpretation of observational properties for XDINs was suggested recently by \cite{DeGrandis2021ApJ} in the framework of magneto-thermal evolution of NSs \citep{PonsMiralles2009}. 
Thus, theoretical and computational studies of magnetic field evolution inside NSs is of great importance to decode and put into context NS observations. For a recent review of this problem, see \cite*{Igoshev2021Univ}.

NS magnetic field evolution is driven by Ohmic decay, Hall effect, ambipolar diffusion \citep{Goldreich1992}, and by effects related to superconductivity in the NS core \citep{Glampedakis2011MNRAS, Graber2015MNRAS}, see also recent work by \cite{Wood2022}. The magnetic field evolution is coupled to the thermal evolution \citep{PonsMiralles2009}. Some of these effects are studied reasonably well via analytical and numerical efforts. The magnetic field evolution driven by Ohmic decay and Hall effect in the NS crust was studied extensively by different groups in two and three dimensions \cite{Wareing2009PhPl,Wareing2010JPlPh,Hollerbach2002MNRAS,Hollerbach2004MNRAS,Gourgouliatos2016PNAS,Vigano2013MNRAS, Gourgouliatos2018ApJ,igoshev2021NatAs,Gourgouliatos2015MNRAS,Gourgouliatos2014MNRAS,Gourgouliatos2013MNRAS,Gourgouliatos2020MNRAS,Igoshev2021ApJ,Anzuini2022MNRAS}. Most of these efforts were nicely summarised in the recent review by \cite{Pons2019LRCA}.

Ambipolar diffusion and superconductivity are less studied.
Ambipolar diffusion could be one of the main drivers for evolution of magnetars with internal fields as strong as $5\times 10^{14}$~--~$10^{15}$~G. Under the influence of ambipolar diffusion their magnetic field will evolve on Myr timescales. Alternatively, radio pulsars become recycled in low-mass X-ray binaries due to accretion from the secondary star. These NSs become millisecond radio pulsars (MSPs). They are known for their significantly smaller magnetic fields, around $10^8$~--~$10^9$~G (for review see e.g. \citealt{Bhattacharya1991MSPreview}). These small magnetic fields might be related to the accretion process \citep{Alpar1982Natur} or produced as a result of ambipolar diffusion \citep{Cruces2019MNRAS}.

Ambipolar diffusion is a dynamic process in the NS core which requires the presence of both charged (electrons and protons) and neutral (neutrons) particles. Charged particle motion is mostly driven by electromagnetic fields while neutral particle motion is defined mostly by the NS gravitational potential. Simultaneously, two nuclear reactions take place: (1) neutrons decay into protons and electrons, and (2) protons and electrons merge into neutrons. Rates of these reactions differ for different depths inside the NS core.   

In this research we do not take into account effects of superfluidity and superconductivity. We plan to slowly increase the complexity of our simulations, distinguishing effects related to each individual process. As opposed to the superfluidity and superconductivity in the NS core, the ambipolar diffusion is theoretically understood reasonably well. Thus, meaningful three-dimensional numerical simulations are possible. The addition of superconductivity and superfluidity  lead to contradictory conclusions.
On the one hand, \cite{Elfritz2016MNRAS} used the formalism developed by \cite{Glampedakis2011MNRAS} and found that dissipation and expulsion in the NS core are weak, and thus strong magnetic fields in the core should survive for a long time. On the other hand, \cite{Dommes2017MNRAS} found that core magnetic field should be expelled on much shorter timescales due to buoyancy of proton vortices.

Studies of ambipolar diffusion were pioneered by \cite{HoyosReisenegger2008}, who established a general multifluid formalism and estimated relevant timescales in one dimension. They also performed first numerical simulations in the nonlinear regime. Later on, \cite{Castillo2017MNRAS} developed numerical simulations in spherical coordinates with axial symmetry, i.e. two-dimensional simulations. They adopted an approximation of motionless neutrons and considered the weak coupling mode. \cite{Castillo2017MNRAS} found that on short timescales the density of charged particles is perturbed by the ambipolar diffusion velocity in such a way that they create a pressure gradient which cancels the irrotational part of the magnetic force. Thus, the velocity field becomes solenoidal. This solenoidal velocity field drives longer evolution which converges to an equilibrium state. In this equilibrium state the toroidal magnetic field is confined to regions where the poloidal magnetic field lines are closed \citep{Castillo2017MNRAS}.

Later on, \cite{Passamonti2017MNRAS} performed numerical simulations of ambipolar diffusion in two dimensions in the one-fluid approximation. They studied short-term evolution when perturbations in number density of non-charged particles are formed as a response to the Lorentz force affecting the charged particles. They studied formation of these regular perturbations and resulting ambipolar velocity field as a function of NS core temperature. 
They found that in the weak coupling regime the chemical gradient partly cancels the Lorentz force. It means that the velocity field becomes solenoidal-dominated below $4\times 10^8$~K. \cite{Passamonti2017MNRAS} found that typical speeds of ambipolar diffusion are km/Myr in the normal matter case and much faster, $10^3$~km/Myr, in the superconducting case. In the case of superconductivity/superfluidity the suppression of the irrotational component occurs at higher temperatures below $9\times 10^8$~K. \cite{Passamonti2017MNRAS} did not model the long-term evolution of magnetic fields nor the influence of neutron star crust. 

Recently, \cite{Castillo2020MNRAS} performed two-dimensional simulations in the two-fluid approximation with inclusion of neutron motion. In these simulations they also found that magnetic field evolves toward the "Grad-Shafranov" equilibria. 
The behaviour of ambipolar diffusion in three dimensions is expected to differ from two dimensions because multiple instabilities are known for axisymmetric magnetic fields in three dimensions, see e.g. \cite{Tayler1973MNRAS,Markey1973MNRAS}.

The aim of our article is to model ambipolar diffusion in the weak coupling mode in three dimensions to study if it leads to formation of non-axisymmetric structures. Essentially we want to check if in three dimensions the ambipolar diffusion leads to an equilibrium state as it was found in axisymmetric simulations \cite{Castillo2020MNRAS}, or alternatively leads to complete decay of magnetic field.
We formulate a set of equations mostly following works by \cite{Goldreich1992} and \cite{Passamonti2017MNRAS}. In comparison to that work we assume a presence of weak magnetic field decay in the core caused by Ohmic losses, and we include the NS crust in our calculations. We solve the magnetic induction equation and equation for deviation from the $\beta$-equilibrium in three dimensions in spherical coordinates using novel spectral code \texttt{Dedalus}\footnote{https://dedalus-project.org}\citep{DedalusPaper,dedalusSphereI,dedalusSphereII}.

\section{Magnetic field evolution}
\label{s:magnetic_field_evolution}
\subsection{Key assumptions}
A few different approaches were suggested to simplify the system of equations describing the ambipolar diffusion. The recent notable cases include \cite{Castillo2020MNRAS} and \cite{Passamonti2017MNRAS}. We mostly follow the prescription by \cite{Passamonti2017MNRAS} with a few small changes. Here we summarise our key assumptions and show how they differ from the more recent equation set presented by \cite{Castillo2020MNRAS}.

Similarly to \cite{Castillo2020MNRAS} we aim at studying the evolution of sequential magneto-hydrostatic quasi-equilibrium states. In each of these states all forces applied to a fluid element are close to balancing each other. Each of these states is reached within a few Alfvén timescales i.e. within a few seconds of real time. The evolution of magnetic field proceeds on much longer timescales, $10^2$~--~$10^{10}$~years. We therefore do not follow the propagation of sound waves, gravity waves, or Alfvén waves.

Unlike \cite{Castillo2020MNRAS}, we neglect inertial terms in the continuity equation for particle densities. Moreover, we also neglect the advective term (baryon velocity), i.e. we assume that total $\nnn \vec v_\mathrm{n} = - \nnp \vec v_\mathrm{p}$. Overall, we work in the one-fluid MHD limit similarly to \cite{Passamonti2017MNRAS}. As noted by \cite{Castillo2020MNRAS}, this assumption might lead to underestimation of the timescale for ambipolar diffusion. In this work we are more interested in relaxing the axial symmetry assumption which was made in all previous simulations on this topic. In future work we plan to add equations describing the independent motion of the neutral component.

\subsection{Detailed derivation of equations}

Following the derivations by \cite{Goldreich1992,Passamonti2017MNRAS} we begin with the Maxwell–Faraday equation:
\begin{equation}
\frac{\partial \vec B}{\partial t} = - c \vec \nabla \times \vec E,    
\label{eq:mawell_farady}
\end{equation}
where $\vec B$ and $\vec E$ are the magnetic and electric fields, and $c$ is the speed of light.
In this work we assume that electric field evolves only under the influence of Ohmic decay and ambipolar diffusion (so keeping the first two terms in eq. 6 of \citealt{Passamonti2017MNRAS}):
\begin{equation}
\vec E = \frac{\vec j}{\sigma} - \frac{1}{c} \vp \times \vec B,
\label{eq:electric_field}
\end{equation}
where $\sigma$ is the electric conductivity, $\vec v_\mathrm{p}$ is the speed of protons, and $\vec j$ is the electric current density:
\begin{equation}
\vec j = e \nnc (\vp - \ve) = \frac{c}{4\pi} \vec \nabla \times \vec B,
\label{e:current}
\end{equation}
where $e$ is elementary charge, and $\nnc$ is the number density of charged particles. It is assumed that the number densities of protons and electrons are equal $\nnc = \nne \approx \nnp$ due to the electroneutrality. 

Combining eqs. (\ref{eq:mawell_farady}), (\ref{eq:electric_field}) and (\ref{e:current}) we derive:
\begin{equation}
\frac{\partial \vec B}{\partial t} = - \frac{c^2}{4\pi} \vec \nabla \times \left(\frac{1}{\sigma}\vec \nabla \times \vec B\right)
+ \vec \nabla \times (\vp \times \vec B).
\end{equation}

Now we replace $\vec B = \vec \nabla \times \vec A$ where $\vec A$ is the vector potential:
\begin{equation}
\curl \frac{\partial \vec A}{\partial t} = -\frac{c^2}{4\pi} \curl \left(\frac{1}{\sigma} \curl (\curl \vec A)\right)
+ \curl \left[ \vp \times (\curl \vec A) \right].
\end{equation}
Taking the `inverse curl' of this then yields:
\begin{equation}
\frac{\partial \vec A}{\partial t} = -\frac{c^2}{4\pi \sigma} \curl (\curl \vec A) + \vec v_\mathrm{amb} \times (\curl \vec A).
\label{eq:potent}
\end{equation}
Here, similarly to \cite{Passamonti2017MNRAS} we expand $\vec v_\mathrm{p} = \vec v_\mathrm{b} + x_n (\vec v_\mathrm{p} - \vec v_\mathrm{n})$. In this expression $x_n$ is the neutron fraction and $\vec v_\mathrm{b}$ is the speed of baryons. Further we assume that baryon speed is negligible and thus we replace $\vec v_\mathrm{p}\approx x_n (\vec v_\mathrm{p} - \vec v_\mathrm{n}) = \vamb$.

The velocity of ambipolar diffusion is written the same way as \cite{Passamonti2017MNRAS} (see Appendix~\ref{s:derivation_delta_mu} for more details):
\begin{equation}
\frac{1}{4\pi \nnc} (\curl \vec B) \times \vec B - \vec \nabla (\Delta \mu) = \frac{1}{x_n^2} \frac{m_p^* \vamb}{\taupn},
\label{eq:delta_mu_orig}
\end{equation}
where $\Delta \mu$ is the deviation from the $\beta$-equilibrium, $m_p^*$ is the effective proton mass, and $\taupn$ is the relaxation time for collision between protons and neutrons.
We rewrite this equation using the vector potential:
\begin{equation}
\vamb = \frac{x_n^2\taupn}{4\pi \nnc m_p^*} \left[ \left\{ \curl (\curl \vec A) \right\} \times (\curl \vec A) - 4\pi \nnc \vec \nabla (\Delta \mu) \right].
\label{eq:vamb_orig}
\end{equation}

Eq. (\ref{eq:potent}) is written as:
\begin{equation}
\frac{\partial \vec A}{\partial t} = -\frac{c^2}{4\pi \sigma} \curl (\curl \vec A) + \frac{x_n^2\taupn}{4\pi \nnc m_p^*} \left[ \curl (\curl \vec A) \times (\curl \vec A) - 4\pi \nnc \vec \nabla (\Delta \mu) \right] \times (\curl \vec A).
\label{eq:mf_evol_beginning}
\end{equation}
This is the first of two coupled equations. The second equation determines the deviation from the $\beta$-equilibrium. We derive this equation by taking the divergence of eq. (\ref{eq:delta_mu_orig}). We provide more details in Appendix~\ref{s:derivation_delta_mu}. This equation is written as:
\begin{equation}
\vec \nabla^2 (\Delta \mu)  = \frac{m_\mathrm{p}^* \lambda}{x_n^2 \nnc \taupn} \Delta \mu + \vec \nabla \cdot \left(\frac{1}{4\pi \nnc} (\curl \vec B) \times \vec B \right) - \frac{x_n \taupn \nnc}{m_\mathrm{p}^*} \left(-\vec \nabla (\Delta \mu) + \frac{1}{4\pi \nnc} (\curl \vec B) \times \vec B\right) \cdot \vec \nabla \left(\frac{m_\mathrm{p}^*}{x_n \nnc \taupn}\right).
\end{equation}
We simplify this equation by expanding brackets in the third term on the right side. We also assume that $m_p^*$ is constant over the core and $x_n \nnc \taupn$ only varies in the radial direction. In this case the equation is written as:
\begin{equation}
\vec \nabla^2 (\Delta \mu)  = \frac{m_\mathrm{p}^* \lambda}{x_n^2 \nnc \taupn} \Delta \mu + \vec \nabla \cdot \left(\frac{1}{4\pi \nnc} (\curl \vec B) \times \vec B \right) - \frac{x_n\taupn}{4\pi}\left(\left\{\curl \vec B \right\} \times \vec B\right) \cdot \frac{d}{dr} \left( \frac{1}{x_n \nnc \taupn} \right) + x_n \taupn \nnc \frac{\partial \Delta \mu}{\partial r} \frac{d}{dr} \left( \frac{1}{x_n \nnc \taupn} \right).
\end{equation}

\subsection{Neutron star physics}
To construct a NS model we solve numerically the Tolman–Oppenheimer–Volkoff equation \citep{OppenheimerVolkoff1939PhRv} using numerical fits to the equation of state developed by \cite{PearsonChamel2018MNRAS}. We use the BSk24 equation from a FORTRAN module\footnote{http://www.ioffe.ru/astro/NSG/BSk/index.html}. In our calculations we assume the central pressure of $10^{35}$~dyn~cm$^{-2}$, which corresponds to a NS with total mass of $1.46$~M$_\odot$, radius $R_\mathrm{NS} = 12.56$~km, and crust depth $h = 0.95$~km. Thus the core-crust transition occurs at radial distance $0.925$~R$_\mathrm{NS}$. 

The number density $\nnc$ of charged particles is computed as:
\begin{equation}
\nnc = Y_e x_n,
\end{equation}
where $Y_e$ is the fraction of electrons and $x_n$ is the baryon number density. 
The neutron fraction $x_n$ is computed as:
\begin{equation}
x_n = 1 - 2(Y_e + Y_\mu),
\end{equation}
where $Y_\mu$ is the fraction of muons. We subtract two times the electron fraction to remove the proton fraction.
We compute relaxation times $\taupn$ and $\tau_\mathrm{en}$ using equations provided by \cite{YakovlevShalybkov1990SvAL}:
\begin{equation}
\frac{1}{\tau_\mathrm{ep}} = 2.1\times 10^{16}\; T_9^2 \left(\frac{\rho_1}{\rho}\right)^{5/3} \;\mathrm{s}^{-1},
\end{equation}
where $\rho_1 = 2.8\times 10^{14}$~g cm$^{-3}$. A similar equation is used to compute the $\taupn$:
\begin{equation}
\frac{1}{\taupn} = 4.7\times 10^{18}\; T_9^2 \left(\frac{\rho_1}{\rho}\right)^{1/3} \;\mathrm{s}^{-1}.
\end{equation}
We compute the numerical value in Table~\ref{tab:coeff} as:
\begin{equation}
\taupn^0 = 2.13\times 10^{-19} T_9^{-2} \left( \frac{\rho_0}{\rho_1} \right)^{1/3} \; \mathrm{s}.
\label{e:taupn}
\end{equation}
The electrical conductivity of the core is computed as:
\begin{equation}
\sigma = \frac{e^2 \nnc \tau_{pe}}{m_e^*},
\end{equation}
where the effective electron mass is computed as $m_e^* = m_e \sqrt{1+x_e^2}$, where $x_e$ is the ratio between the Fermi momentum and the rest mass of the electron $m_e$. We obtain $x_e$ from the same FORTRAN code \texttt{bskfit18.f}. For the effective proton mass we assume $m_\mathrm{p}^* = 0.75 m_\mathrm{p}$ throughout the entire NS.

We use the following formalism to describe the change in reaction rate as a function of the deviation from the chemical equilibrium. We introduce $\lambda$ which is the change of reaction rate depending on deviation from chemical equilibrium $\lambda = (d\Gamma / d\Delta\mu)|_\mathrm{eq}$. We use the equation for the $\lambda$ factor by \cite{Sawyer1989PhRvD} similarly to \cite{Passamonti2017MNRAS}:
\begin{equation}
\lambda = 5\times 10^{27}\; T^6_8 \left(\frac{\rho}{\rho_1}\right)^{2/3} \; \; \mathrm{erg}^{-1}\; \mathrm{cm}^{-3}\; \mathrm{s}^{-1}  = \lambda_0 \left(\frac{\rho}{\rho_0}\right)^{2/3},
\end{equation}
where $\lambda_0$ is computed as:
\begin{equation}
\lambda_0 = 5\times 10^{33}\; T^6_9 \left(\frac{\rho_0}{\rho_1} \right)^{2/3}   \; \; \mathrm{erg}^{-1}\; \mathrm{cm}^{-3}\; \mathrm{s}^{-1}.
\label{e:lmb}
\end{equation}

\subsection{Timescale and dimensionless version of the equations}

\begin{figure*}
    \centering
    \begin{minipage}{0.49\linewidth}
    \includegraphics[width=0.99\columnwidth]{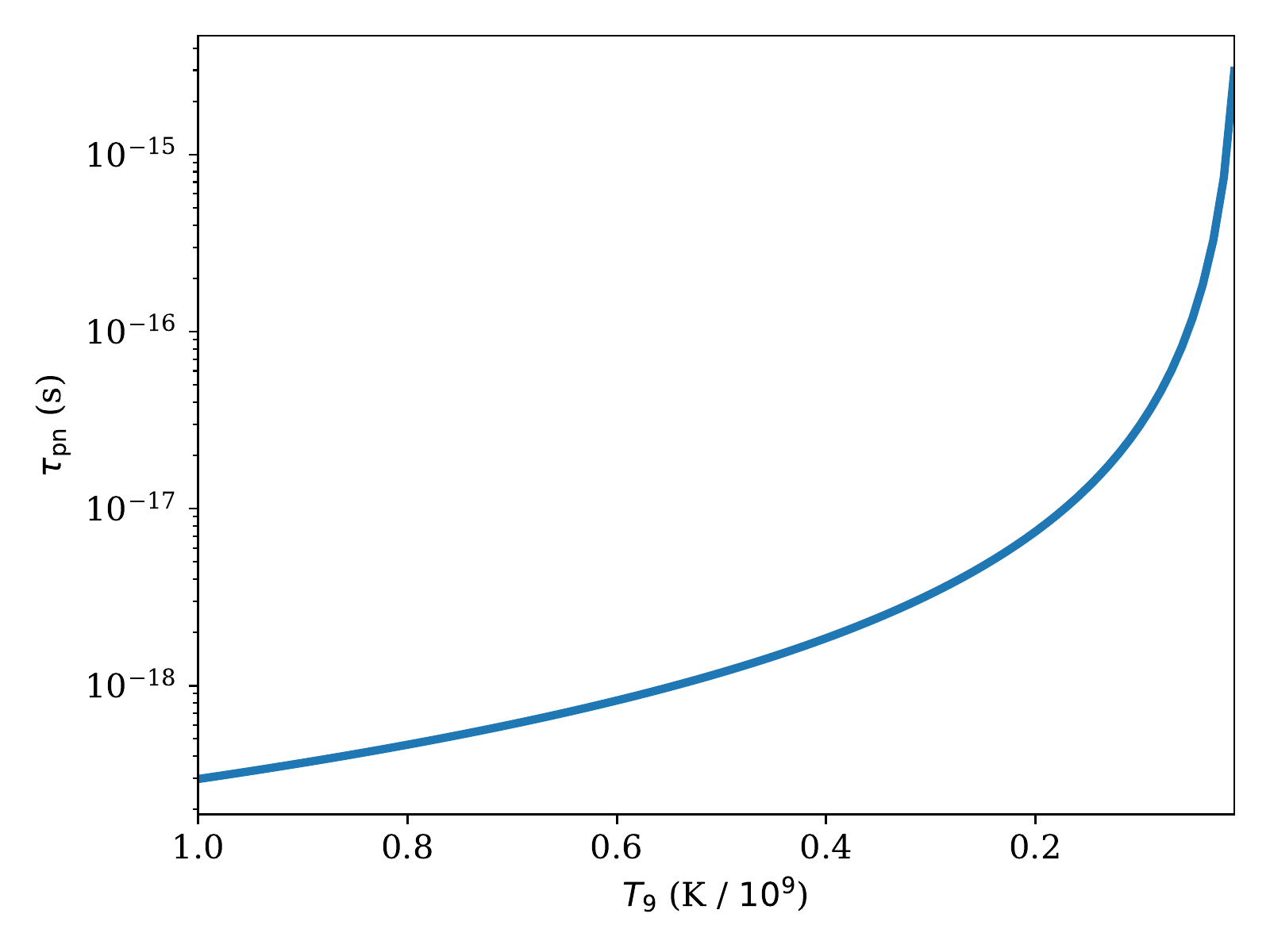}
    \end{minipage}
    \begin{minipage}{0.49\linewidth}
    \includegraphics[width=0.99\columnwidth]{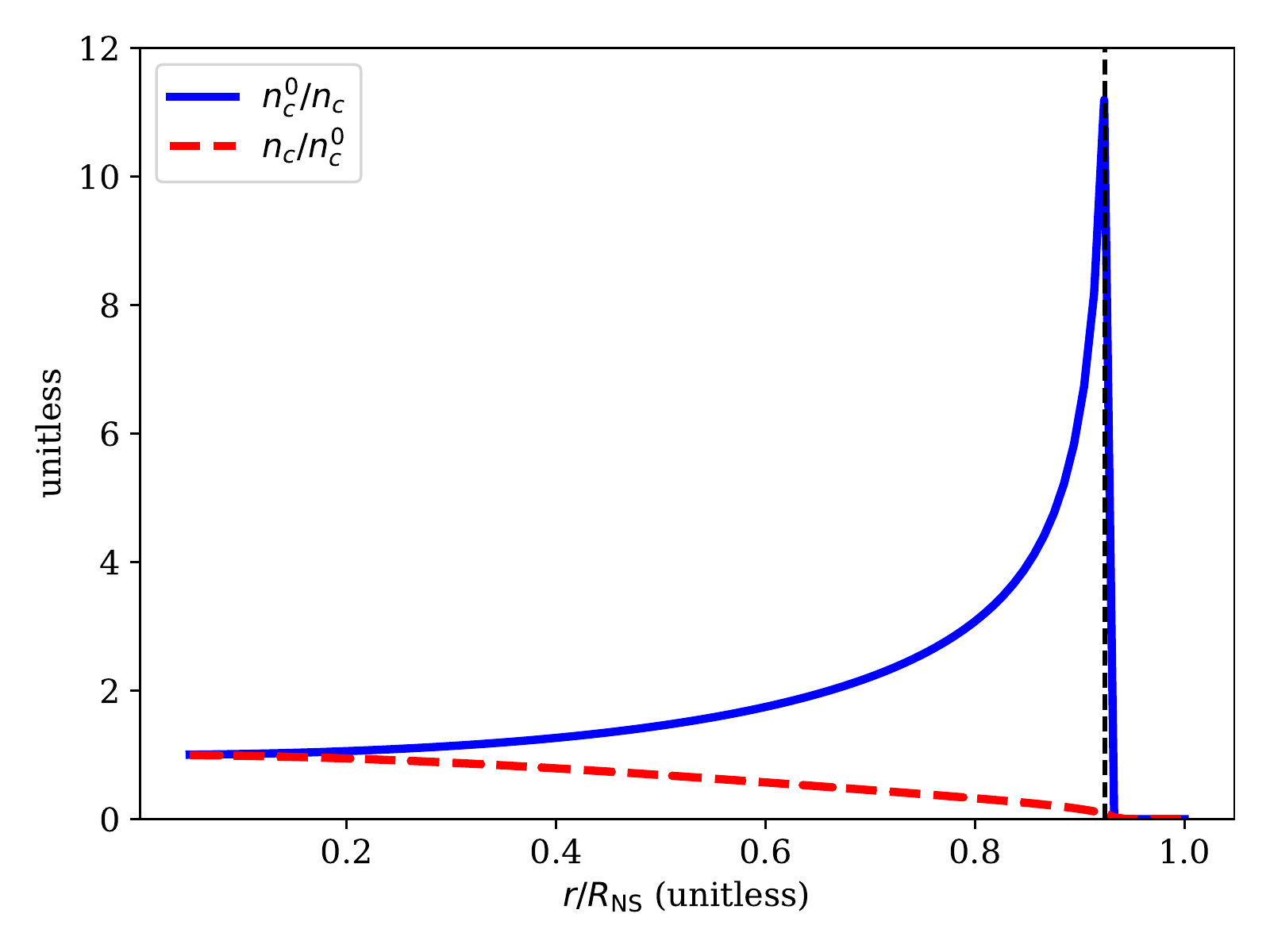}
    \end{minipage}    
    \caption{Left panel: $\taupn$ as a function of temperature. The temperature axis is reversed to highlight the fact that NS cools down with time and so moves from left to right.  Right panel: profile of number density of charged particles $\nnc$ and its inverse in the NS core. Dashed black line shows the crust-core boundary. }
    \label{fig:nc_profile}
\end{figure*}

The natural timescale to consider in this problem is given by \citep{Goldreich1992}:
\begin{equation}
t_\mathrm{amb} = \frac{R_\mathrm{ns}}{\langle v_\mathrm{amb}\rangle} = \frac{4 \pi \nnc m_p^* R_\mathrm{ns}^2}{\taupn B^2}.
\label{e:tamb}
\end{equation}
In order to estimate this timescale for some typical values we have to fix temperature ($\taupn$ is very sensitive to temperature, see Figure~\ref{fig:nc_profile}) and magnetic field strength. For typical parameters summarised in Table~\ref{tab:coeff} we obtain $t_\mathrm{amb}\approx 90$~Myr. To simulate magnetars we thus introduce dimensionless time $t' = t / t_0$ where $t_0 = 10^7$~years. It is worth noting that we estimate $t_\mathrm{amb}$ for the core centre, while for more external regions this timescale is significantly shorter due to decreasing $\nnc$, see Figure~\ref{fig:nc_profile}. During the first 10~Kyr of magnetar evolution the timescale of ambipolar diffusion is large because $\taupn$ is very small due to high interior temperature. For normal radio pulsars with magnetic field $B\approx 5\times 10^{12}$~G, the timescale for ambipolar diffusion is $t_\mathrm{amb}\approx 32$~Gyr.

\begin{table*}
    \centering
    \begin{tabular}{cclrl}
    \hline
    \multicolumn{5}{c}{Fixed values} \\
    \hline
    Symbol          & Eq. &  Meaning & Fixed value & Units \\
    \hline
    $B_0$           &  & Magnetic field strength                 & $1.00\times 10^{14}$   & G $=$ g$^{1/2}$ cm$^{-1/2}$ s$^{-1}$ \\
    $T_0$           &  & Core temperature                        & $1.00\times 10^8$      & K \\
    $R_\mathrm{NS}$ &  & NS radius                               & $1.26\times 10^6$      & cm \\
    $R_\mathrm{core} / R_\mathrm{NS}$ &  & Fraction of core to total NS radius          & 0.925 & Dimensionless \\
    $t_0$           &  & Timescale to make equations dimensionless & $10.00$              & Myr \\
    $m_\mathrm{p}^*$&  & Effective proton mass                   & $1.25\times 10^{-24}$  & g \\
    $s$             &  & Relative conductivity of the core       & $5.00\times 10^{-3}$   & Dimensionless \\
    $t_\mathrm{crust}$ & & Timescale of Ohmic decay in the crust & $30.00$                & Myr\\
    $\rho_0 = \rho (0)$  &  & Central NS density                 & $7.64\times 10^{14}$   & g cm$^{-3}$ \\
    $\nnc^0 = \nnc (0)$  &  & Number density of charged particles in NS centre & $3.30\times 10^{37}$ & cm$^{-3}$  \\
    $\rho_1$        &  & Typical NS density          & $2.8\times 10^{14}$ & g cm$^{-3}$ \\
    \hline
    \multicolumn{5}{c}{Intermediate and diagnostic values} \\
    \hline
    $\taupn^0$      & \ref{e:taupn} & Relaxation time for collisions        & $2.97\times 10^{-17}$  & s \\
    $\lambda_0$     & \ref{e:lmb}   & mUrca reaction rates                  & $9.76\times 10^{27}$    & erg$^{-1}$ cm$^{-3}$ s$^{-1}$\\
    $\mu_0$         & \ref{e:mu0}   & Chemical potential                    & $1.38\times 10^{-8}$   & erg \\
    $t_\mathrm{amb}$& \ref{e:tamb}  & Timescale of ambipolar diffusion      & $87.61$     & Myr \\
    $v_{\mathrm{amb},0}$ & \ref{e:vamb0} & Speed of ambipolar diffusion     & $0.14$   & km Myr$^{-1}$ \\
    $\sigma_0$    & \ref{eq:sigma0}  & Electrical resistivity for $t_0$         & $1.43\times 10^{22}$ & s$^{-1}$ \\
    $\epsilon_0^v$    & \ref{eq:eps0v}   & Volumetric energy release rate & $2.52\times 10^{12}$ & erg cm$^{-3}$ s$^{-1}$ \\
    $\epsilon_0$      & \ref{eq:eps00} & Thermal luminosity & $5\times 10^{30}$ & erg s$^{-1}$ \\
    \hline
    \multicolumn{5}{c}{Dimensionless numerical coefficients of partial differential equations} \\
    \hline
    Am              & \ref{eq:am}  &  & $0.11$  & Dimensionless \\
    $K$             & \ref{eq:k}   &  & $573.0$ & Dimensionless \\
    $d_1$           & \ref{e:d1} &   & $1.97\times 10^{-5}$ &  Dimensionless \\  
    \hline
    \end{tabular}
    \caption{Value of numerical coefficients involved in the problem. Top part of the table is for fixed coefficients and constants while the bottom part contains derived coefficients. }
    \label{tab:coeff}
\end{table*}

\begin{figure*}
    \centering
    \begin{minipage}{0.49\linewidth}
    \includegraphics[width=0.99\columnwidth]{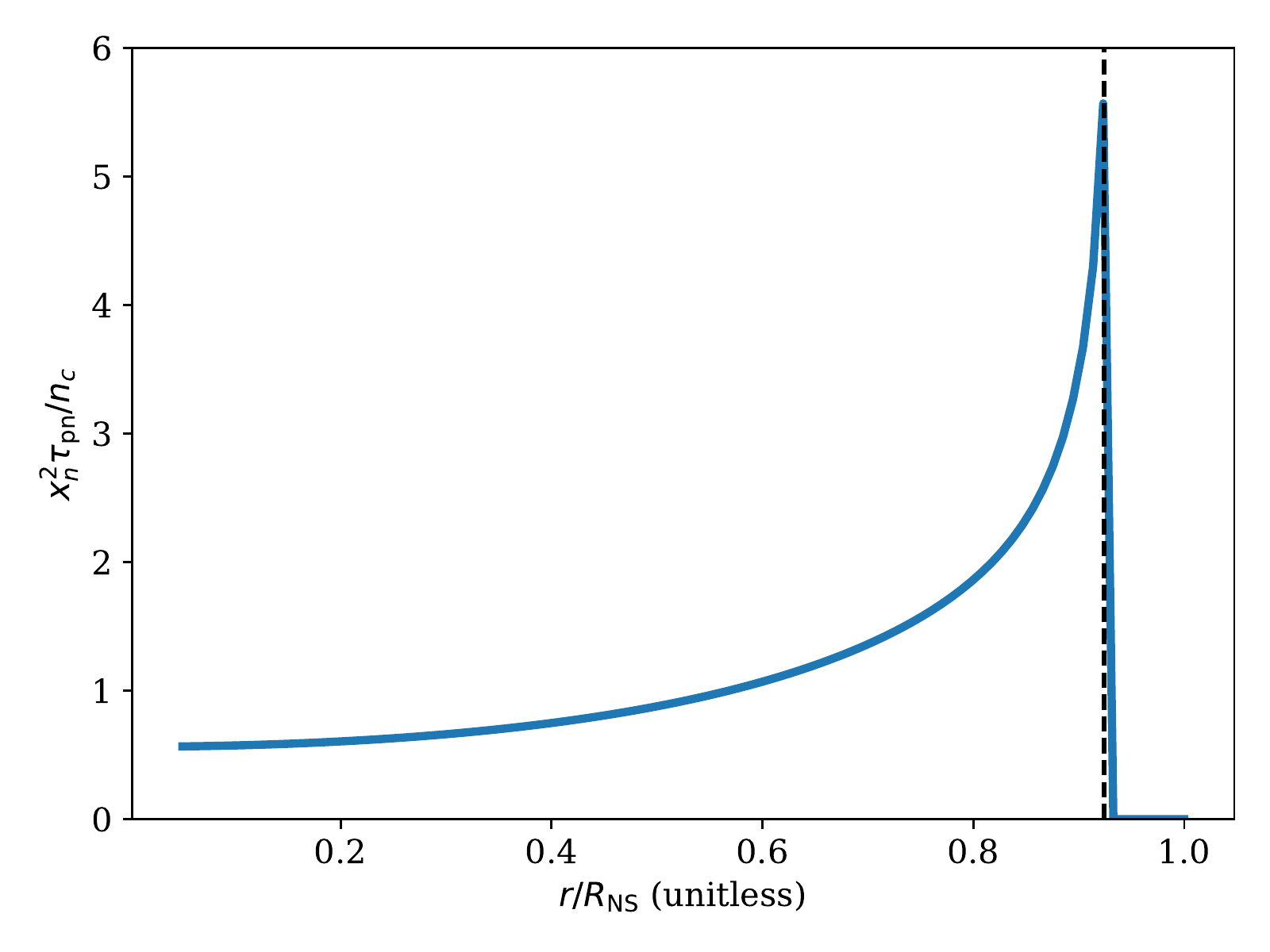}
    \end{minipage}
    \begin{minipage}{0.49\linewidth}
    \includegraphics[width=0.99\columnwidth]{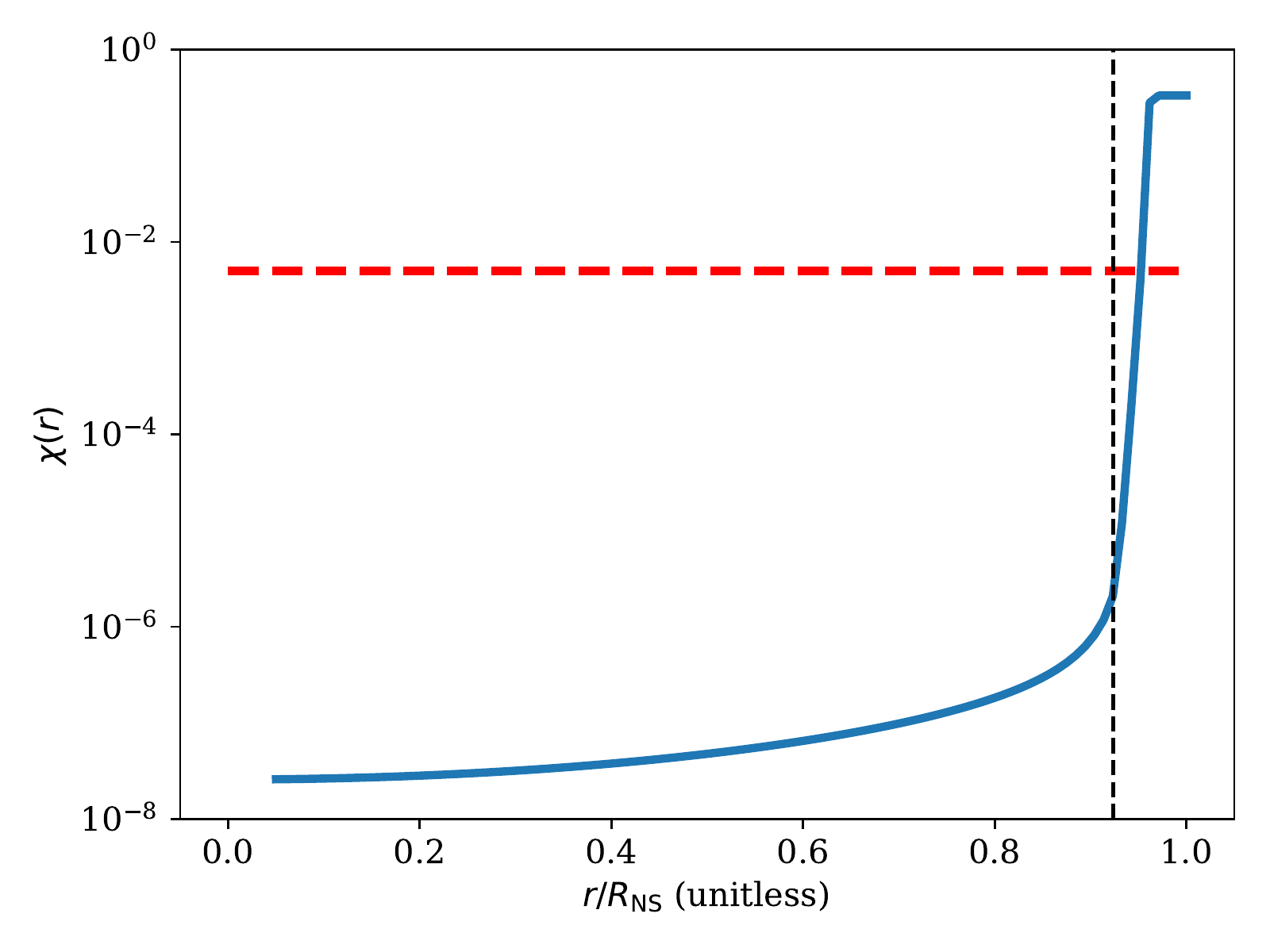}
    \end{minipage}    
    \caption{Left panel: radial profile of numerical coefficient determining the speed of ambipolar diffusion. Right panel: radial profile for relative conductivity $\chi$ (solid blue line) and additional conductivity $s$ (dashed red line). Dashed black line shows the crust-core boundary.}
    \label{fig:chi}
\end{figure*}

We write the equations in dimensionless form beginning with eq. (\ref{eq:vamb_orig}). The ambipolar velocity $\vamb$ is one of the key diagnostic variables which shows how ambipolar diffusion proceeds. In axisymmetric simulations \cite{Castillo2020MNRAS} it was found that $\vamb$ decreases by many orders of magnitude at $t_\mathrm{amb}$ because the configurations approach an equilibrium. To write eq. (\ref{eq:vamb_orig}) in dimensionless form, we introduce three auxiliary variables:
\begin{equation}
v_{\mathrm{amb}, 0} = \frac{\taupn^0B_0^2}{4\pi \nnc^0 m_\mathrm{p}^* R_\mathrm{NS}},
\label{e:vamb0}
\end{equation}
\begin{equation}
\mu_0 = k_B T_0,
\label{e:mu0}
\end{equation}
\begin{equation}
\mathrm{K} = \frac{4\pi \mu_0 \nnc^0}{B_0^2}.
\label{eq:k}
\end{equation}
Thus eq. (\ref{eq:vamb_orig}) becomes:
\begin{equation}
\vamb = v_{\mathrm{amb}, 0}  \frac{x_n^2 \taupn}{\nnc}  \left[ \left\{ \curl (\curl \vec A) \right\} \times (\curl \vec A) - \mathrm{K} \nnc \vec \nabla (\Delta \mu) \right].
\label{eq:vamb}
\end{equation}
The current dimension of this quantity is cm~s$^{-1}$, but for convenience it could also be computed in km Myr$^{-1}$ as in Table~\ref{tab:coeff}. Truly dimensionless velocity is:
\begin{equation}
v = v_{\mathrm{amb}, 0} \frac{t_0}{R_\mathrm{NS}}.
\end{equation}
At the next stage we consider eq. (\ref{eq:mf_evol_beginning}). The dimensionless version of this equation will contain:
\begin{equation}
\Am = \frac{\taupn^0 t_0 B_0^2}{4\pi n_c^0 m_\mathrm{p}^* R_\mathrm{NS}^2} = v_{\mathrm{amb},0} \frac{t_0}{R_\mathrm{NS}},
\label{eq:am}
\end{equation}
i.e. our new dimensionless coefficient has meaning of dimensionless velocity of ambipolar diffusion. If we assume that our $t_0$ corresponds to some fictional conductivity $\sigma_0$ via:
\begin{equation}
t_0 =  \frac{4\pi \sigma_0 R_\mathrm{NS}^2}{  c^2},
\label{eq:sigma0}
\end{equation}
we can also simplify the first term on the right side of eq. (\ref{eq:mf_evol_beginning}) by introducing a coefficient with profile:  
\begin{equation}
\chi(r) = \frac{\sigma_0}{\sigma (r)}.
\end{equation}
Using these variables we rewrite eq. (\ref{eq:mf_evol_beginning}) as:
\begin{equation}
\frac{\partial \vec A}{\partial t} = - \chi(r) \curl (\curl \vec A) + \mathrm{Am} \frac{x_n^2\taupn}{\nnc} \left[ \left\{\curl (\curl \vec A) \right\} \times (\curl \vec A) - \mathrm{K} \nnc \vec \nabla (\Delta \mu) \right] \times (\curl \vec A).
\label{eq:mf_evol}
\end{equation}
We also take into account that $A_0 = R_\mathrm{NS} B_0$. In order to solve this equation with a spectral method we further split it into linear (left side) and nonlinear (right side) parts:
\begin{equation}
\frac{\partial \vec A}{\partial t} + s \curl (\curl \vec A) = - \chi(r) \curl (\curl \vec A) + \mathrm{Am} \frac{x_n^2\taupn}{\nnc} \left[ \left\{\curl (\curl \vec A) \right\} \times (\curl \vec A) - \mathrm{K} \nnc \vec \nabla (\Delta \mu) \right] \times (\curl \vec A),
\label{eq:mf_evol_final}
\end{equation}
where $s$ is a constant dimensionless conductivity chosen at level $5\times 10^{-3}$. This level is well below the conductivity of the crust but still above the conductivity of the core for temperature $T_9 = 0.2$. We show the comparison of these conductivities in Figure~\ref{fig:chi} (right panel). For the conductivity of the crust we limit it at a value which corresponds to an Ohmic decay timescale of $t_\mathrm{crust} = 3\times 10^7$~years, using lower limit following the analysis by \cite{Igoshev2019MNRAS}:
\begin{equation}
\sigma_\mathrm{crust} = \frac{c^2 t_\mathrm{crust}}{4\pi R_\mathrm{NS}}.
\end{equation}
In the same Figure~\ref{fig:chi} we show the radial profile of the coefficient $x_n^2 \taupn / \nnc$ which determines the variation of the ambipolar diffusion speed over the NS core.

In the second equation we make the following replacements:
\begin{equation}
\vec \nabla^2 (\Delta \mu) - d_1 \xi_1 (r) \Delta \mu = \frac{1}{K} \vec \nabla \cdot \left(\frac{(\vec \nabla \times (\curl \vec A))\times (\curl \vec A)}{n_c} \right) - \frac{1}{K} \xi_3 (r) ((\vec \nabla \times (\curl \vec A))\times (\curl \vec A))\cdot \hat r + \xi_4 (r) \frac{\partial \Delta \mu}{\partial r}.
\label{eq:delta_mu_xi}
\end{equation}
The numerical coefficient $d_1$ is necessary to make the equation dimensionless, while $\xi_1 (r)$, $\xi_3 (r)$ and $\xi_4 (r)$ are radial profiles within the NS core. The coefficient $d_1$ is computed  as the following:
\begin{equation}
d_1 = \frac{m_p^* \lambda_0 R_\mathrm{NS}^2}{\nnc^0 \taupn^0}.
\label{e:d1}
\end{equation}

\begin{figure}
    \centering
    \includegraphics[width=0.99\columnwidth]{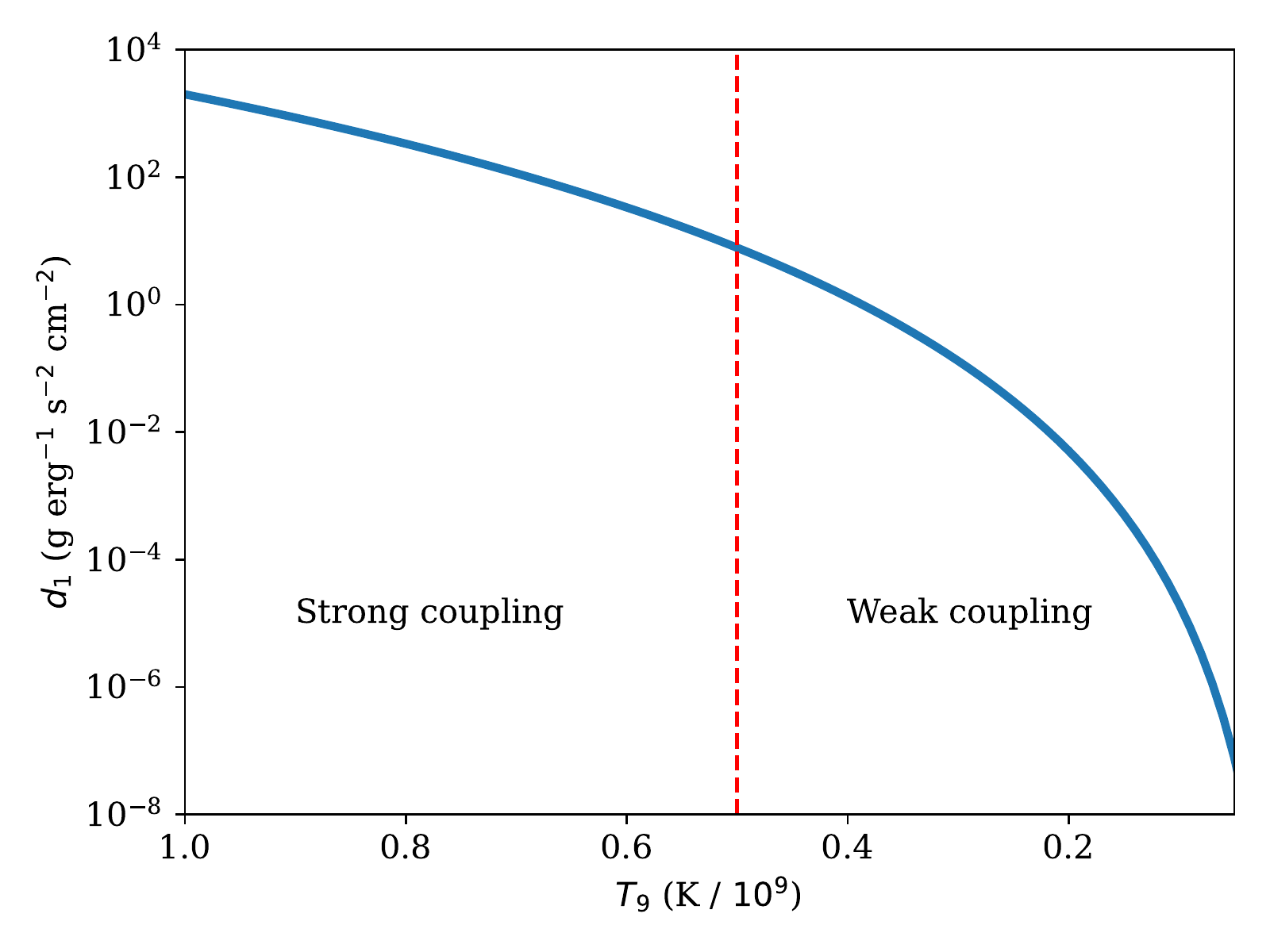}
    \caption{Dependence of $d_1$ on NS core temperature. The temperature axis is reversed to highlight the fact that NS cools down with time and thus moves from left to right.
    }
    \label{fig:xis}
\end{figure}

We show the dependence of $d_1$ on core temperature in Figure~\ref{fig:xis}. 
The radial profiles are given as:
\begin{align}
\xi_1 &= \frac{\lambda}{x_n^2 \nnc \taupn}, \\
\xi_3 &= x_n\taupn \frac{\partial }{\partial r}\left(\frac{1}{x_n\taupn \nnc}  \right), \\
\xi_4 &= x_n\taupn \nnc \frac{\partial }{\partial r}\left(\frac{1}{x_n\taupn \nnc}  \right).
\end{align}
In this second equation we assumed that $x_n\taupn \nnc$ varies only along the radial direction.

In comparison to work by \cite{Passamonti2017MNRAS} our coefficients translate to theirs as:
\begin{equation}
a = \frac{R_\mathrm{NS}}{\sqrt{d_1 \xi_1}},
\end{equation}
\begin{equation}
b = \frac{R_\mathrm{NS}}{ \xi_4}.
\label{e:b}
\end{equation}

\begin{figure*}
    \centering
    \begin{minipage}{0.49\linewidth}
    \includegraphics[width=0.99\columnwidth]{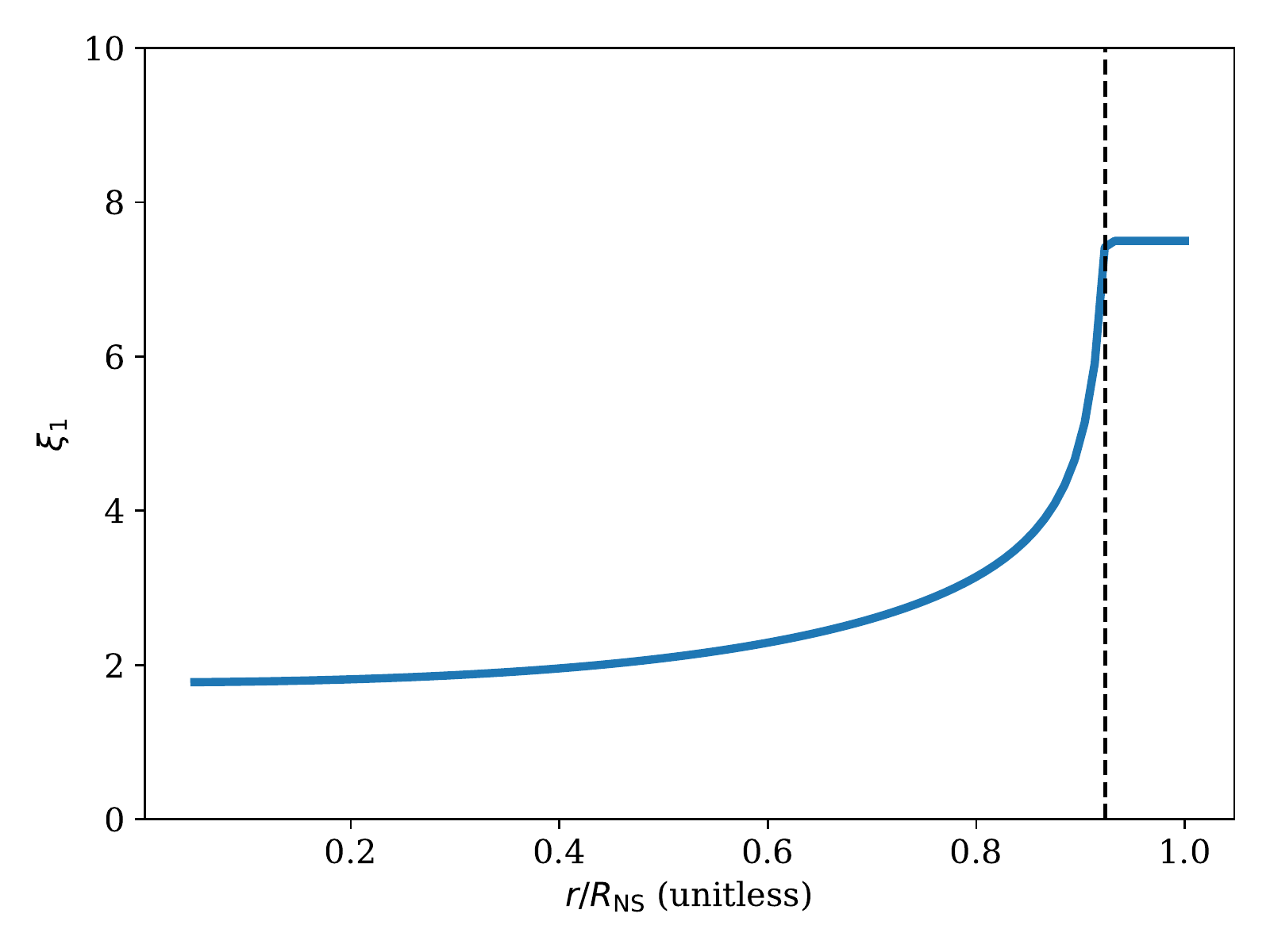}
    \end{minipage}
    \begin{minipage}{0.49\linewidth}
    \includegraphics[width=0.99\columnwidth]{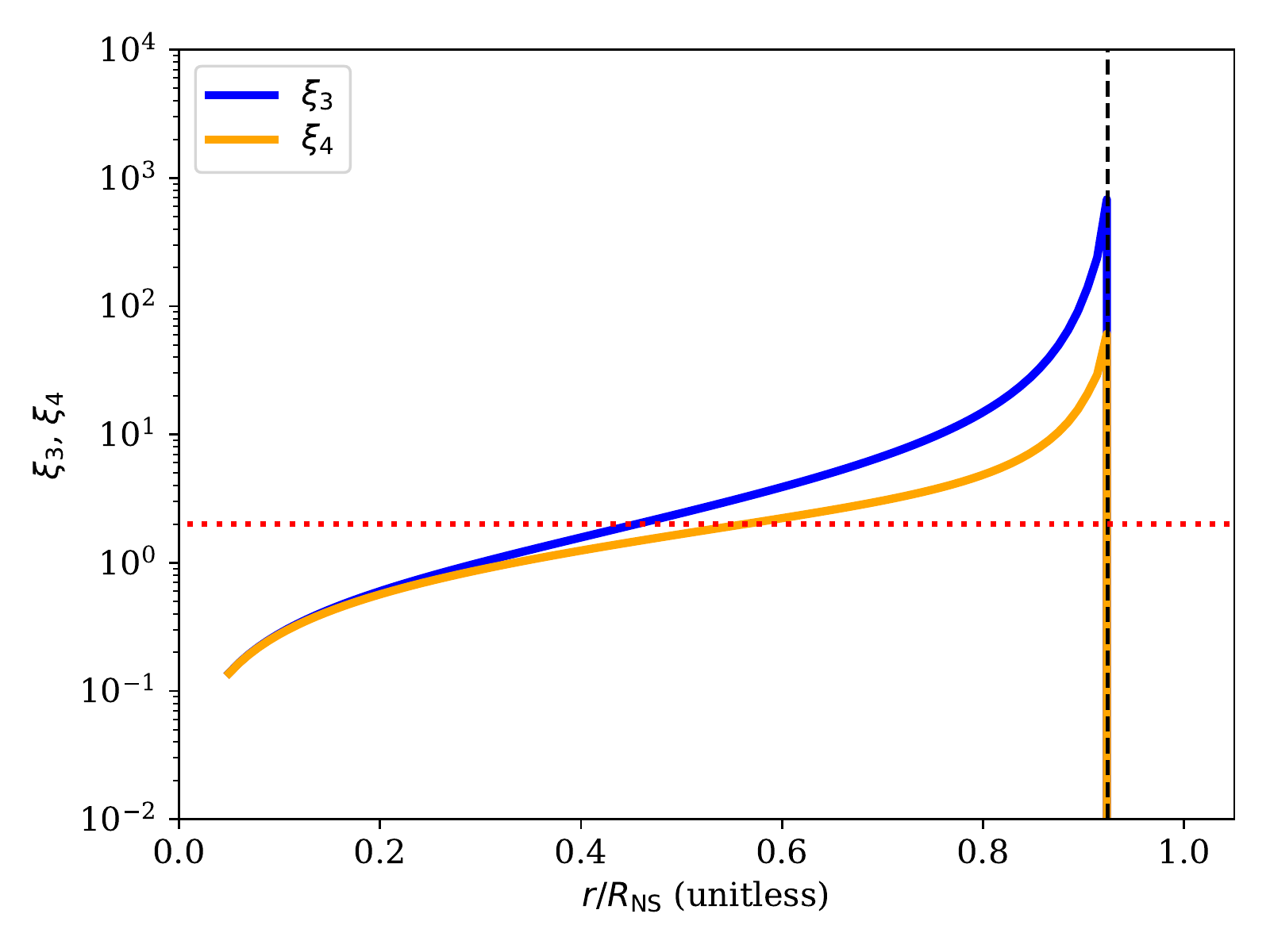}
    \end{minipage}    
    \caption{Profile of numerical coefficients $\xi_1$ (left panel) and $\xi_3, \xi_4$ (right panel). In the right panel we show the physical $\xi_3, \xi_4$ (solid blue and orange lines) and the value $r_\mathrm{cut} = 2$ at which we restrict our numerical profiles (horizontal red dotted line). Vertical dashed black line shows the crust-core boundary. }
    \label{fig:xi134_profile}
\end{figure*}

We show the radial profile for coefficients $\xi_1$, $\xi_3$ and $\xi_4$ in Figure~\ref{fig:xi134_profile}. We set the values of the $\xi_3$ and $\xi_4$ coefficients in the crust to be zero. It is not essential if $\Delta \mu \neq 0$ in the crust because we set the value of $n_c$ (see Figure~\ref{fig:nc_profile}) to zero in the crust, so ambipolar diffusion does not proceed there. The values of $\xi_3$ and $\xi_4$ change over a few orders of magnitude in the core. To avoid numerical difficulties for these realistic values of $\xi_3$ and $\xi_4$ we introduce a parameter $r_\mathrm{cut}$ which determines the maximum value which $\xi_3$ and $\xi_4$ can reach within the core. We provide more details about this $r_\mathrm{cut}$ parameter in Appendix~\ref{appendix:code_verification}.

\subsection{Strong and weak couplings}

Because the value of $d_1$ is very sensitive to the NS core temperature, the behaviour of eq. (\ref{eq:delta_mu_xi}) changes when the NS cools down. Two different asymptotic cases are called strong coupling ($T_9 \sim 1$ and $d_1 \gg 1/K$) and weak coupling ($T_9 \sim 0$ and $d_1 \approx 1/K$ ). 

In the strong coupling regime ($T > 5\times 10^8$~K) the eq. (\ref{eq:delta_mu_xi}) transforms into:
\begin{equation}
\Delta \mu \approx 0.0,
\end{equation}
because all coefficients in the original equation are dwarfed in comparison to $d_1 \approx 10^3$. Therefore, the material is in $\beta$-equilibrium. 
Alternatively, in a weak coupling regime the coefficient $d_1$ becomes comparable or smaller than the remaining terms in this equation.


\subsection{Complete system of equations and boundary conditions}

Overall, our system of equations includes also the Coulomb gauge $\nabla \cdot \vec A = 0$:
\begin{align}
\frac{\partial \vec A}{\partial t} - s\nabla^2 \vec A + \vec \nabla \Phi &=  \chi(r) \nabla^2 \vec A + \mathrm{Am} \frac{x_n^2\taupn}{\nnc} \left[ \left\{\curl (\curl \vec A) \right\} \times (\curl \vec A) - \mathrm{K} \nnc \vec \nabla (\Delta \mu) \right] \times (\curl \vec A),  \label{eq:complete_system_beginning} \\
\vec \nabla \cdot \vec A &= 0, \\
\vec \nabla^2 (\Delta \mu) - \xi_4 (r) \frac{\partial \Delta \mu}{\partial r}  - d_1 \xi_1 (r) \Delta \mu &= \frac{1}{K} \vec \nabla \cdot \left(\frac{(\vec \nabla \times (\curl \vec A))\times (\curl \vec A)}{n_c} \right) - \frac{1}{K} \xi_3 (r) ((\vec \nabla \times (\curl \vec A))\times (\curl \vec A))\cdot \hat r, 
\label{eq:complete_system_end}
\end{align}
where $\Phi$ is the scalar potential.

The vector potential $\vec A$ is subject to the potential boundary condition:
\begin{equation}
\frac{\partial \vec A_l}{\partial r} + \frac{l+1}{r} \vec A_l \; \Big|_{r = R_\mathrm{core}} = 0,
\end{equation}
where $l$ is the spherical harmonic degree.
For the deviation from the $\beta$-equilibrium we use the same boundary condition as \cite{Passamonti2017MNRAS}:
\begin{equation}
K \nnc \frac{\partial \Delta \mu}{\partial r} = \left(\vec \nabla \times (\curl A) \right) \times (\curl A ) \cdot \hat r \;  \Big|_{r = R_\mathrm{core}}.
\end{equation}
This boundary condition means that $v_\mathrm{amb} = 0$ at the crust-core boundary, see eq. (\ref{eq:vamb}).



\subsubsection{Solution procedure}
We solved the coupled partial differential equations (\ref{eq:complete_system_beginning}-\ref{eq:complete_system_end}) using the publicly available spectral code \texttt{Dedalus} v.3 \citep{DedalusPaper,dedalusSphereI,dedalusSphereII} in spherical coordinates. This code expands the solution using a combination of spherical harmonics for angular directions and Jacobi polynomials for the radial direction. 
We propagate the simulation in time using the second-order implicit-explicit Runge-Kutta integrator \citep{ASCHER1997151}. We rewrite the system of differential equations (\ref{eq:complete_system_beginning}-\ref{eq:complete_system_end}) to satisfy requirements of the \texttt{Dedalus} code as the following:
\begin{align}
\frac{\partial \vec A}{\partial t} - s\nabla^2 \vec A + \vec \nabla \Phi + \tau P(A) &=  \chi(r) \nabla^2 \vec A + \mathrm{Am} \frac{x_n^2\taupn}{\nnc} \left[ \left\{\curl (\curl \vec A) \right\} \times (\curl \vec A) - \mathrm{K} \nnc \vec \nabla (\Delta \mu) \right] \times (\curl \vec A), \\
\vec \nabla \cdot \vec A + \tau_\phi &= 0, \\
\vec \nabla^2 (\Delta \mu) - \xi_4 (r) \frac{\partial \Delta \mu}{\partial r} - d_1 \xi_1 (r) \Delta \mu  + \tau P (\Delta \mu) + \tau_\mu &= \frac{1}{K} \vec \nabla \cdot \left(\frac{(\vec \nabla \times (\curl \vec A))\times (\curl \vec A)}{n_c} \right) - \frac{1}{K} \xi_3 (r) ((\vec \nabla \times (\curl \vec A))\times (\curl \vec A))\cdot \hat r.
\label{eq:complete_system_tau}
\end{align}
We introduced so-called tau-terms: $\tau P(A)$ and $\tau P (\Delta \mu)$ in order to apply the generalised tau-method. These terms allow additional degrees of freedom, so the problem can be solved exactly over polynomials when the boundary conditions are introduced as additional equations in the system. There are two terms $\tau_\phi$ and $\tau_\mu$ which are the same at every point in the grid. These are introduced to deal with uncertainty of the type $A' = A + C$ where $C$ is a constant. This uncertainty appears also for $\Delta \mu$ when $d_1$ is very small i.e. in the weak-coupling case.
Because we introduced these two additional degrees of freedom not covered by the current set of equations and boundary conditions we add two more equations:
\begin{align}
\int \Phi \; dV = 0, \\
\int \Delta \mu \; dV = 0. 
\end{align}
We store the results of simulations after every few thousand timesteps.

\begin{table}
    \centering
    \begin{tabular}{ccccc}
    \hline
    Name &  $L_\mathrm{max}$ & $M_\mathrm{max}$ & $N_\mathrm{max}$  & $\Delta t$ \\
    \hline
    A    &  32               & 64              & 24                & $2\times 10^{-5}$ \\
    B    &  64               & 128              & 64               & $10^{-6}$ \\
    C    &  64               & 128              & 128              & $2\times 10^{-7}$ \\
    D    &  256              & 512              & 128              & $2\times 10^{-7}$ \\
    \hline
    \end{tabular}
    \caption{Summary of the setup for numerical simulations. $L_\mathrm{max}$ and $M_\mathrm{max}$ indicate the maximum spherical harmonic degree and order; $N_\mathrm{max}$ denotes the number of Jacobi polynomials in the radial direction.}
    \label{tab:simulation_summary}
\end{table}

We run three-dimensional simulations using spherical harmonics with different resolutions. We summarise the resolutions in Table~\ref{tab:simulation_summary}.
The lowest resolution is mostly used for testing purposes.
The radial resolution is not uniform and grows toward the surface. In the setup B, the crust is covered with 15 collocation points in radial direction while the distance between consecutive collocation points near the NS centre is 0.024~$R_\mathrm{NS}$. The centre itself is not included as one of the radial grid points, since the coordinate singularity there would require it to be treated somewhat differently. The equatorial and meridional sections presented below therefore have a small empty circle at the centre. Depending on the strength of the magnetic field and the numerical resolution, the timestep is adjusted to ensure stability.

\subsection{Initial conditions}
For the initial magnetic field configuration we transform the analytical configuration by \cite{Akgun2013MNRAS} to vector potential form:
\begin{equation}
\begin{array}{ccc}
A_r     &=& b_t,\\
A_\theta &=& 0,\\
A_\phi   &=& \frac{f(r)}{r} \sin \theta,
\end{array}
\end{equation}
where
\begin{equation}
f(x) = \frac{35}{8} x^2 - \frac{21}{4} x^4 + \frac{15}{8} x^6.
\end{equation}
The exact form of $b_t$ together with derivations can be found in Appendix~\ref{a:vect_pot}. We add a Gaussian noise with amplitude of $10^{-5}$ to all components of $\vec A$. We are then in a position to study the stability of axisymmetric configurations to non-axisymmetric disturbances.

\begin{figure*}
    \centering
    \begin{minipage}{0.43\linewidth}
    \includegraphics[width=0.99\columnwidth]{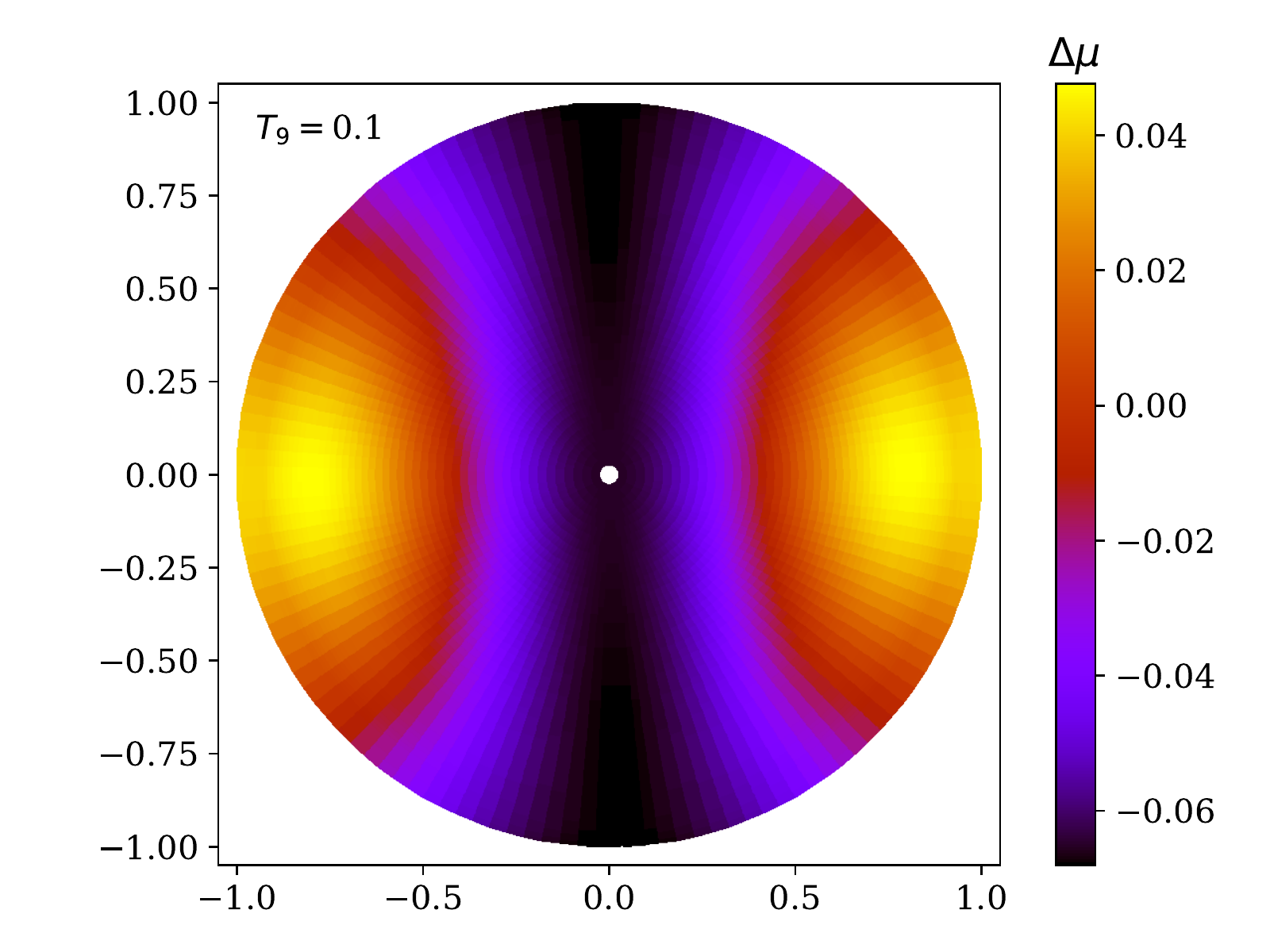}
    \end{minipage}
    \begin{minipage}{0.43\linewidth}
    \includegraphics[width=0.99\columnwidth]{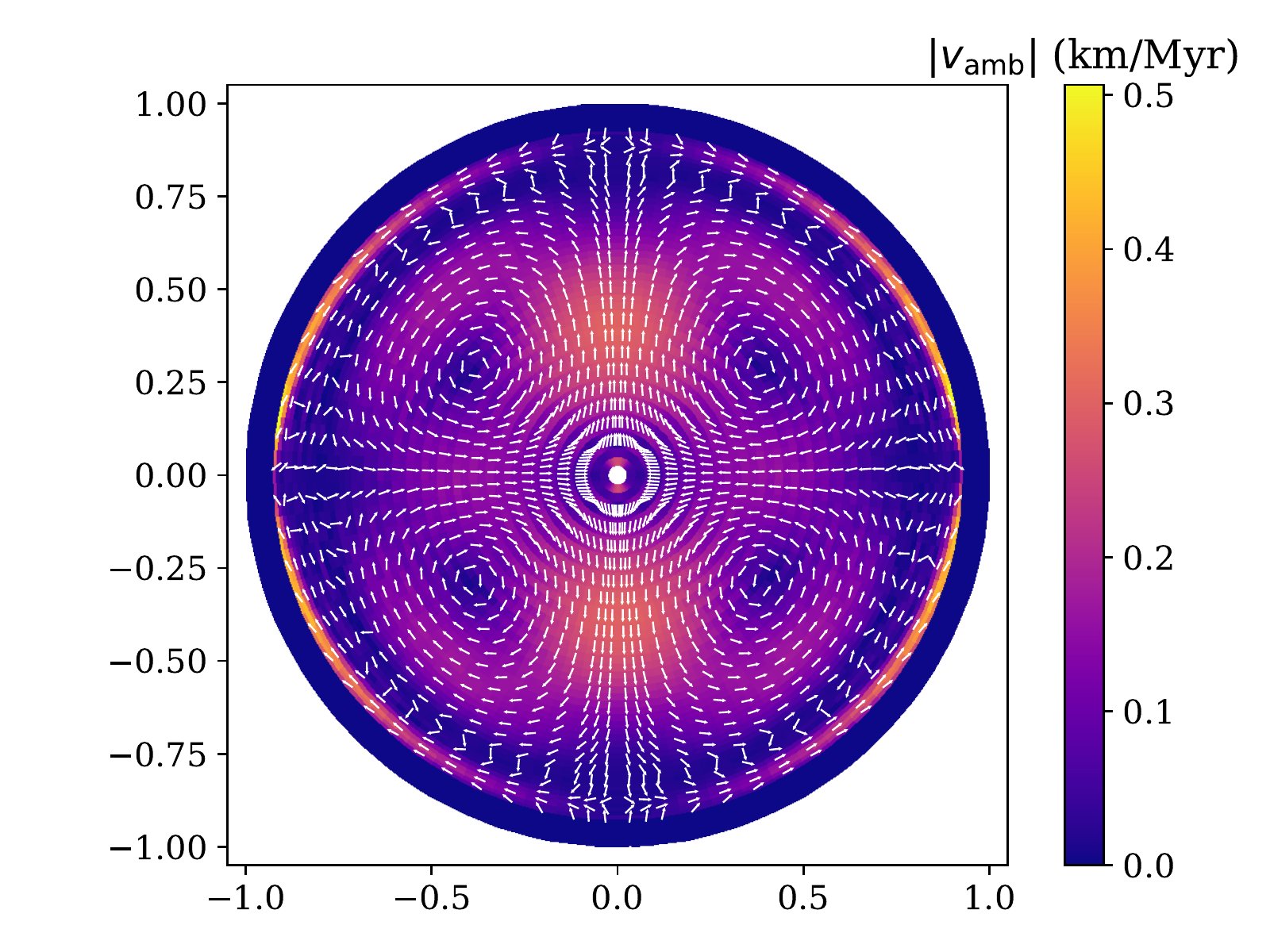}
    \end{minipage}
    \caption{Left panel: deviation from the chemical equilibrium. Right panel: speed of ambipolar diffusion. These values are computed at the beginning of the simulation for $T_9 = 0.1$. }
    \label{fig:v_amb_temp}
\end{figure*}


\section{Results}
\label{s:results}

In order to check if our code works correctly we compute the deviation from the chemical equilibrium and the speed of ambipolar diffusion for a range of temperatures studied by \cite{Passamonti2017MNRAS}. We show the results of a short simulation with $T_9 = 0.1$ in Figure~\ref{fig:v_amb_temp}. The velocity field is purely solenoidal. Our equation for the deviation from chemical equilibrium guarantees that the system relaxed to the magneto-hydrostatic quasi-equilibrium state at every timestep.

We make a few more verification runs. The goal of these runs is to check if the code correctly reproduces the previous results found in two-dimensional simulations by \cite{Passamonti2017MNRAS} and if our resolution is adequate to follow the long-term evolution of magnetic field. More details can be found in Appendix~\ref{appendix:code_verification}. The main conclusions of these technical investigations are as follows: (1) we reproduce the results of  \cite{Passamonti2017MNRAS} with the exception of precise $\Delta \mu$ amplitudes, (2) our numerical resolutions presented in Table~\ref{tab:simulation_summary} are enough to resolve physics with $r_\mathrm{cut}=2$, and (3) the exact choice of parameter $r_\mathrm{cut}$ does not seem to affect the development of the azimuthal field. It is worth noting important differences between our work and \cite{Passamonti2017MNRAS}: (1) our model includes the crust with finite conductivity, thus boundary conditions for magnetic field are written at the top of crust and not at the crust-core interface as was done by previous authors, (2) in our work we propagate the evolution of magnetic field in time while \cite{Passamonti2017MNRAS} only solved for $\Delta \mu$ for different fixed temperatures.

We present our results in the following order. We start with discussing the basic physical variables such as speed of ambipolar diffusion, magnetic energy, azimuthal magnetic field and electric currents. At this stage we identify a development of non-axisymmetric instability. We characterise properties of this instability in Section~\ref{s:instability}. Further we summarise the astrophysical implications in Section~\ref{s:astro_res}. Our basic run is computed with resolution B for 40~Myr and is numerically expensive. To cover the long-term behaviour we also run simulations with resolution A for 160~Myr. 

\subsection{Basic physical variables}
In this section we describe how basic quantities evolve. We compute the speed of ambipolar diffusion using eq. (\ref{eq:vamb}). In order to be more quantitative while characterising the evolution of ambipolar velocity we introduce the mean ambipolar speed $\langle v_\mathrm{amb} \rangle$. 

\subsubsection{Speed of ambipolar diffusion}

We plot the ambipolar diffusion velocities for temperature $T_9=0.1$ in Figure~\ref{fig:v_amb_temp}. The maximum velocities reached within the NS are $\approx 0.5$~km~Myr$^{-1}$. These velocities are a few times larger than the value we initially estimated in Table~\ref{tab:coeff}. The reason for this is a mismatch between our $B_0 = 10^{14}$~G in Table~\ref{tab:coeff} and the maximum magnetic field reached within the NS core, which is $B_\mathrm{max}\approx 8\times 10^{14}$~G. Since $v_\mathrm{amb}$ depends on magnetic field strength as $B^2$, our velocities could be $\approx 64$ times faster. This motion is partially cancelled by deviation from the chemical equilibrium. That is why maximum velocities are not $\approx 6$~km/Myr but much slower. It also means that the timescale of ambipolar diffusion is $\approx 5$ times shorter, i.e. $t_\mathrm{amb}\approx 17$~Myr, or $\approx 2$ in dimensionless time. 

\begin{figure*}    
    \begin{minipage}{0.49\linewidth}
    \includegraphics[width=0.99\columnwidth]{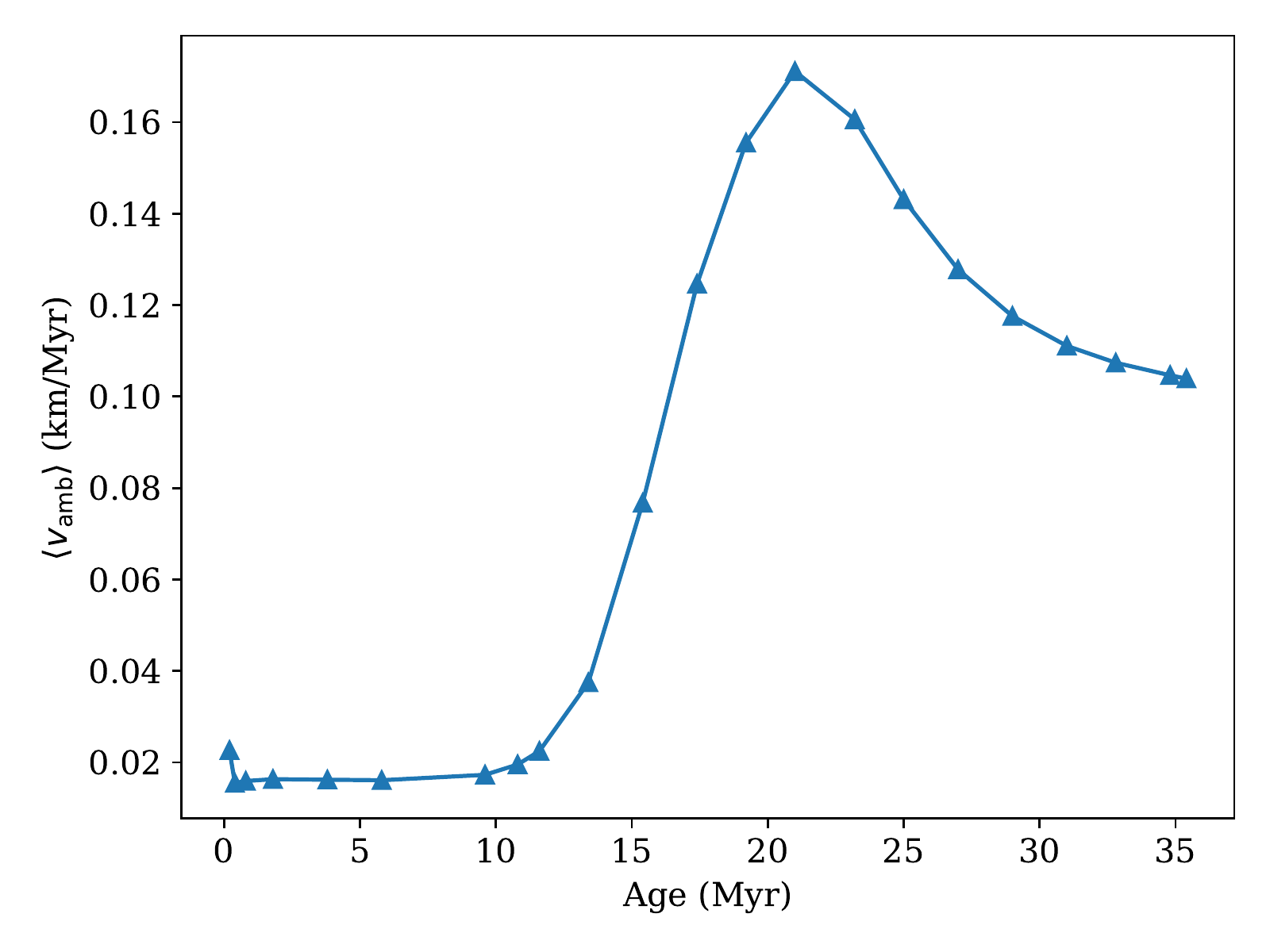}
    \end{minipage}
    \begin{minipage}{0.49\linewidth}
    \includegraphics[width=0.99\columnwidth]{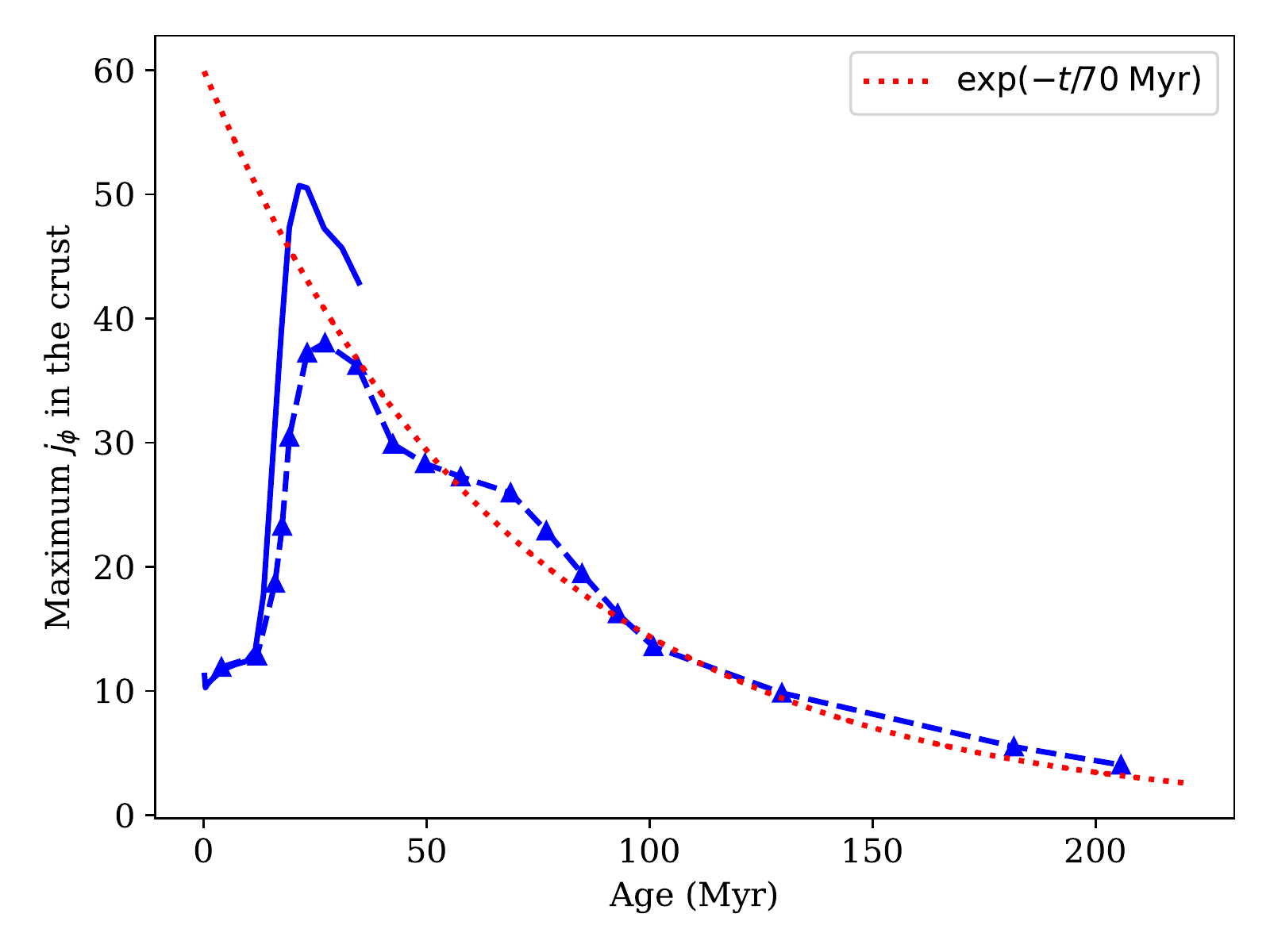}
    \end{minipage}
    \caption{Left panel: evolution of mean ambipolar velocity inside the neutron star. Right panel: evolution of maximum azimuthal electric current in NS crust between radial distance of 0.924~$R_\mathrm{NS}$ and 0.955~$R_\mathrm{NS}$.
    In both panels we show simulations with $r_\mathrm{cut}=2$. Solid lines correspond to simulations with resolution B while dashed lines correspond to similar calculations with resolution A.}
    \label{fig:v_amb_jphi}
\end{figure*}

Here we introduce a mean speed of ambipolar diffusion as:
\begin{equation}
\langle v_\mathrm{amb} \rangle = \frac{1}{V}\,{\int |v_\mathrm{amb}| d^3 V},
\end{equation}
where $|v_\mathrm{amb}|$ is the amplitude of the velocity vector and $V$ is the total NS volume including the crust. There is a small caveat related to this definition. The speed of ambipolar diffusion in the crust is zero; our mean is thus slightly less in comparison to that we would obtain if we only integrated over the NS core. The mean speed of ambipolar diffusion initially decays with time, see Figure~\ref{fig:v_amb_jphi}. This speed starts growing again after $\approx 10$~Myr, which corresponds to development of an instability.
The initial decrease of ambipolar velocity is related to the decay of magnetic fields generated due to the noise added to the simulations. The mean speed reaches its maximum around $20$~Myr, i.e. on the timescale of ambipolar diffusion.

\begin{figure*}
    \centering
    \begin{minipage}{0.42\linewidth}
    \includegraphics[width=0.99\columnwidth]{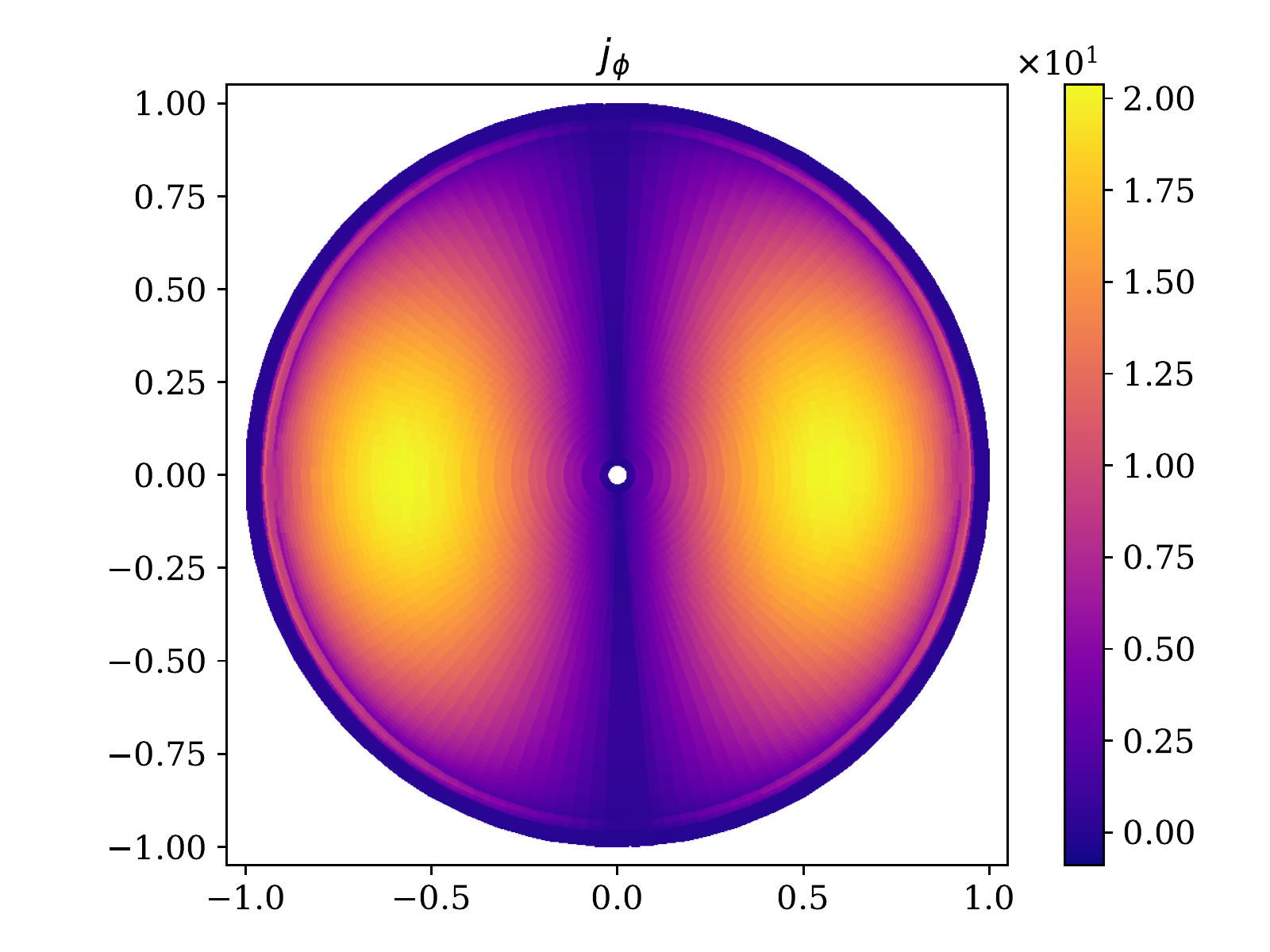}
    \end{minipage}
    \begin{minipage}{0.42\linewidth}
    \includegraphics[width=0.99\columnwidth]{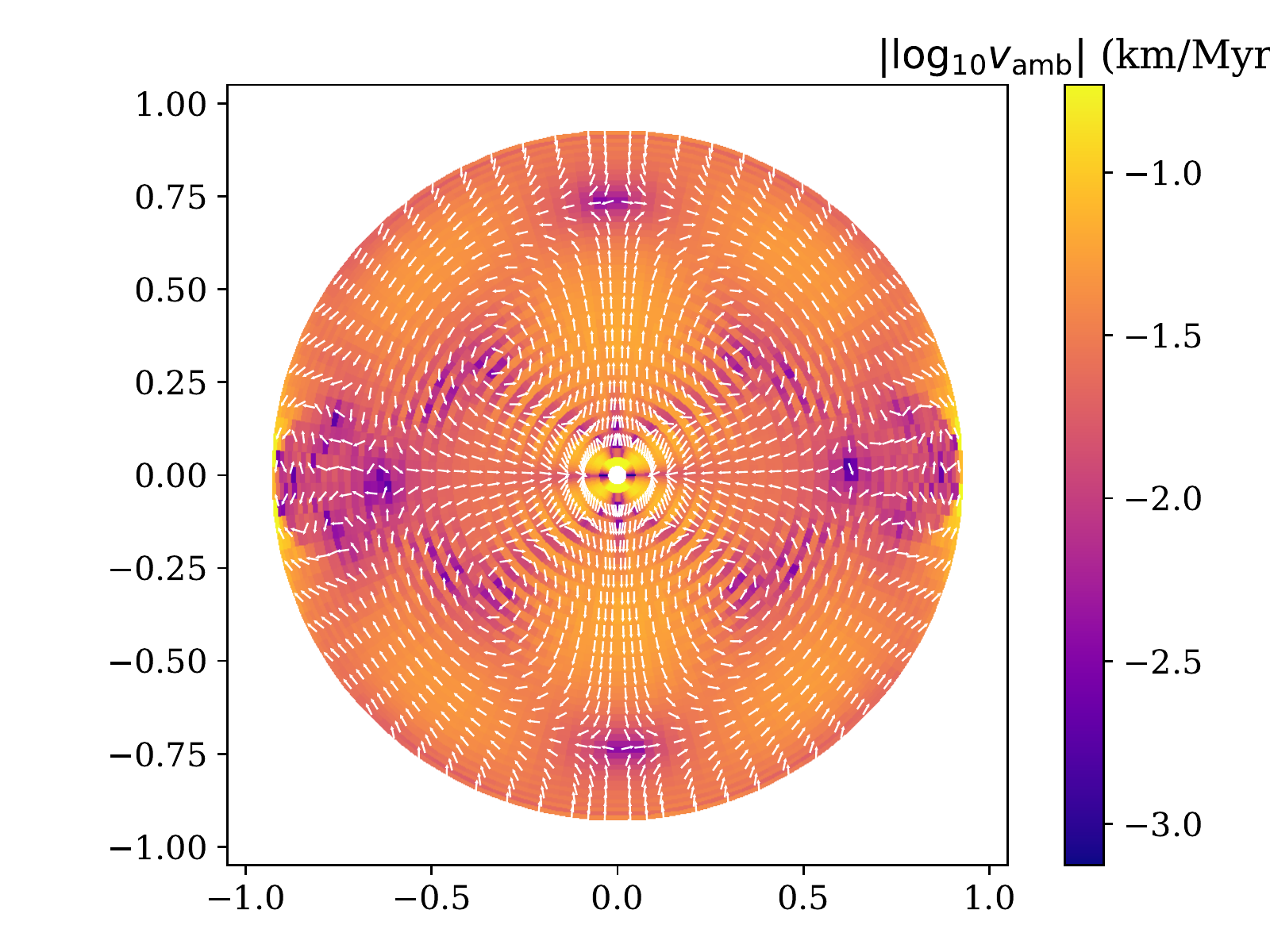}
    \end{minipage}
    \begin{minipage}{0.42\linewidth}
    \includegraphics[width=0.99\columnwidth]{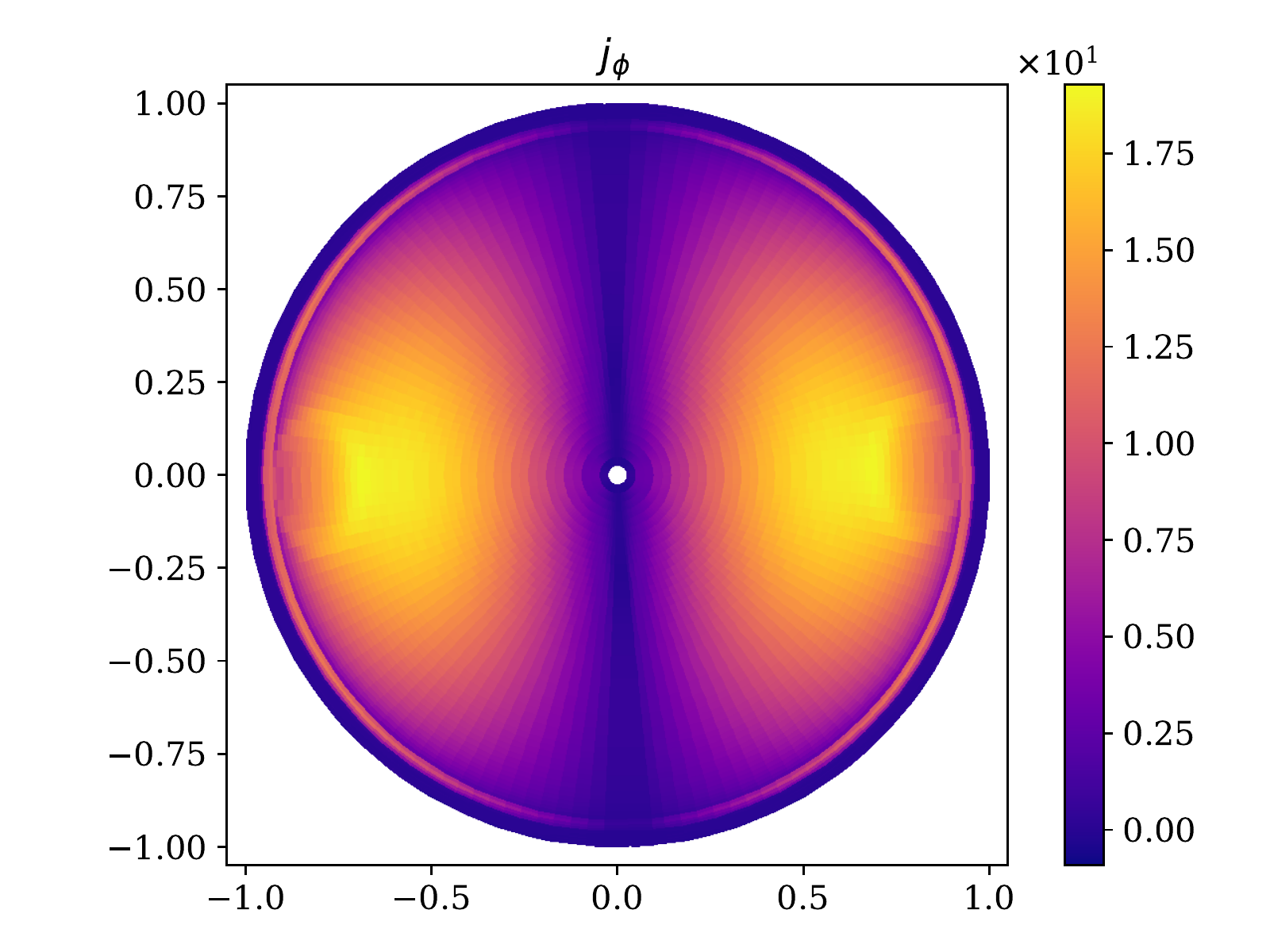}
    \end{minipage}
    \begin{minipage}{0.42\linewidth}
    \includegraphics[width=0.99\columnwidth]{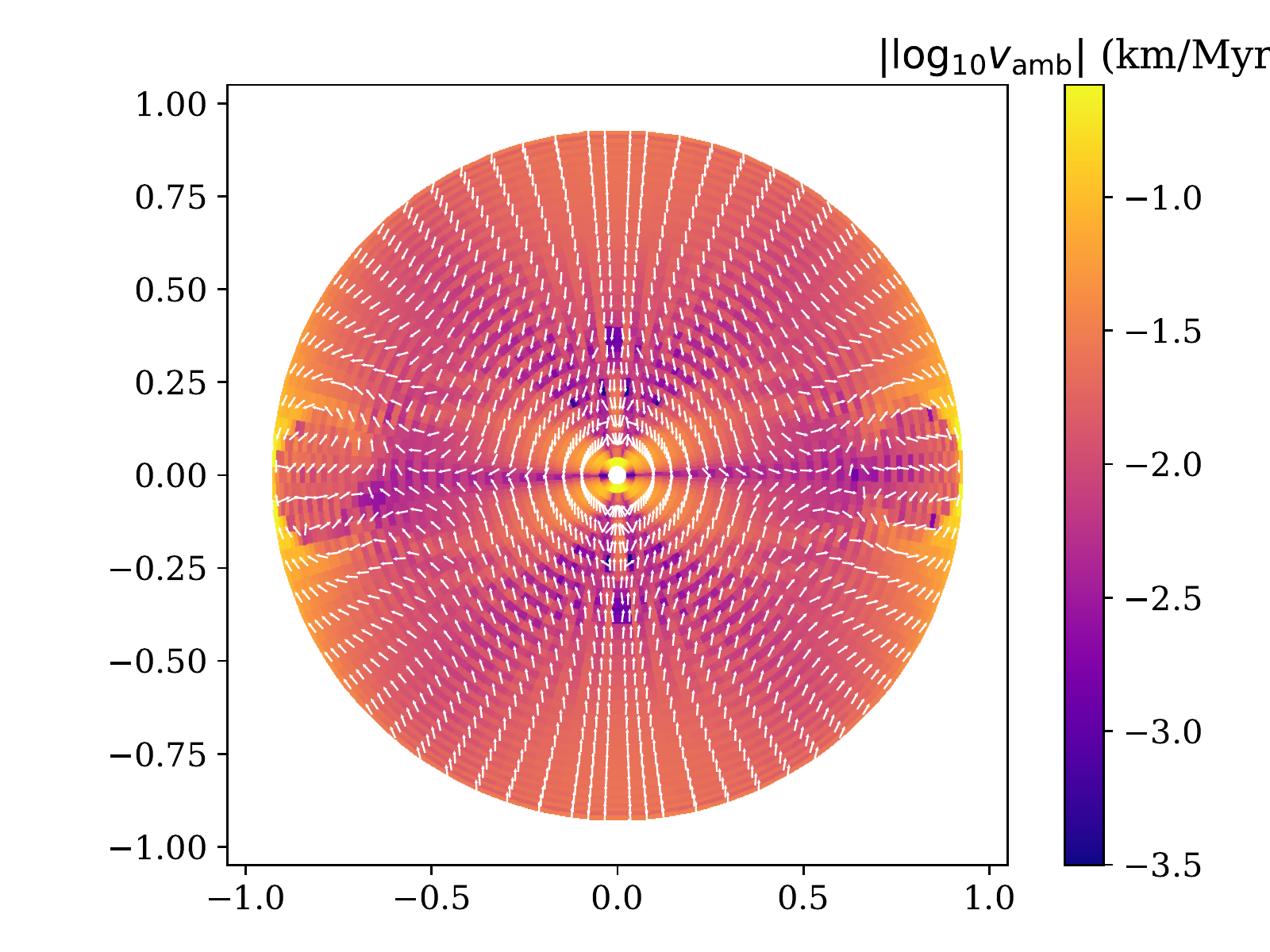}
    \end{minipage}
    \begin{minipage}{0.42\linewidth}
    \includegraphics[width=0.99\columnwidth]{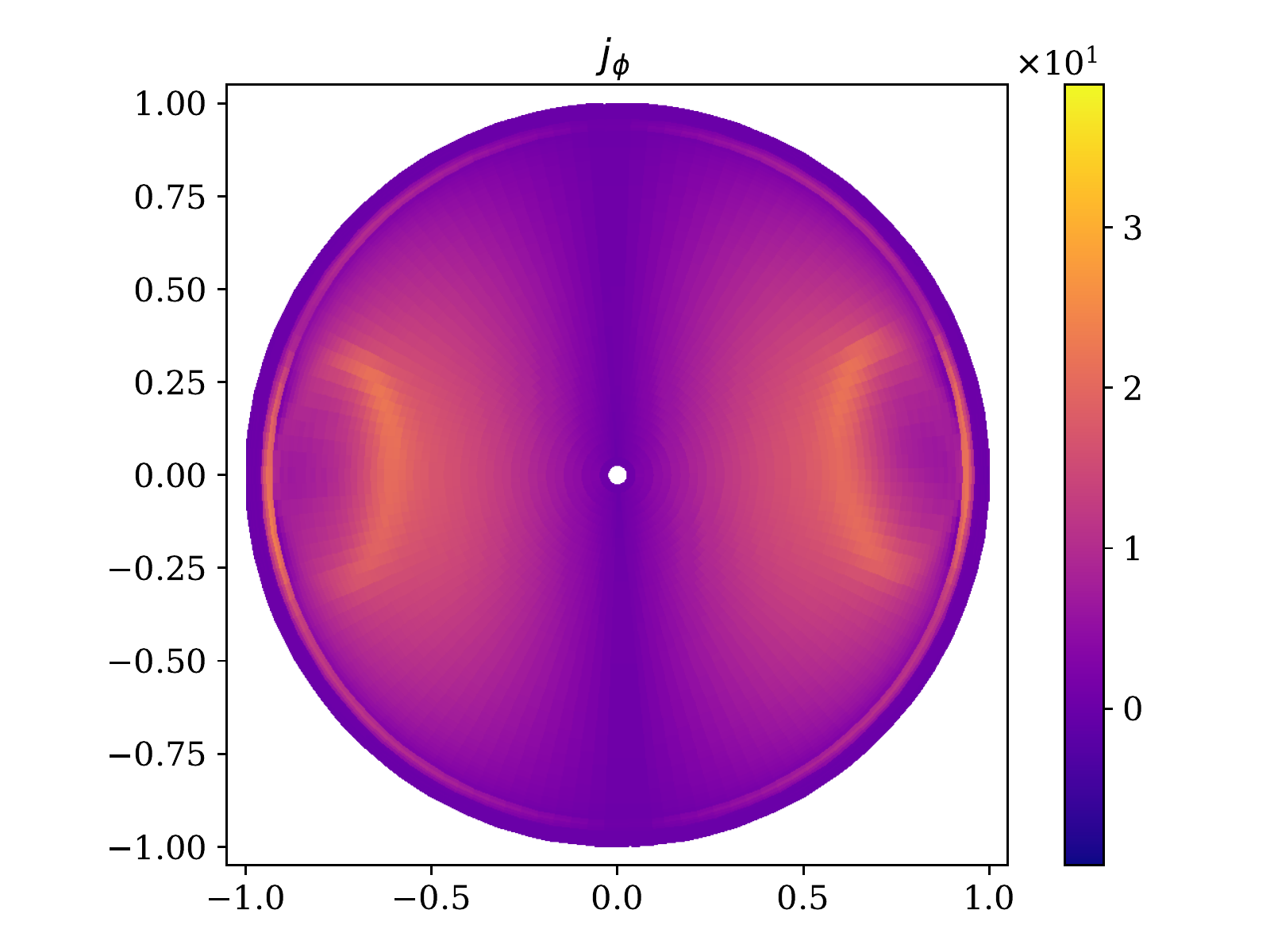}
    \end{minipage}
    \begin{minipage}{0.42\linewidth}
    \includegraphics[width=0.99\columnwidth]{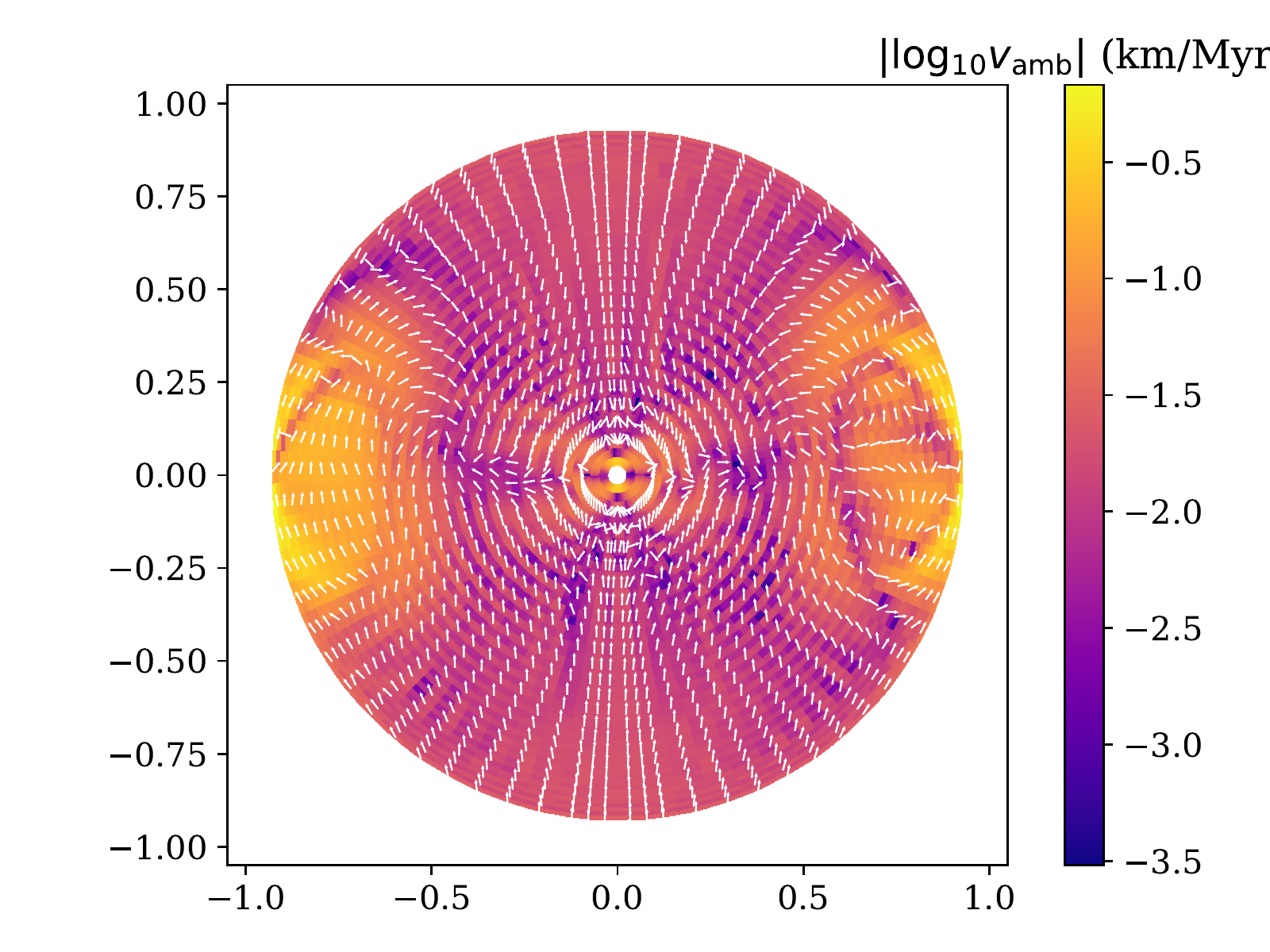}
    \end{minipage}
    \begin{minipage}{0.42\linewidth}
    \includegraphics[width=0.99\columnwidth]{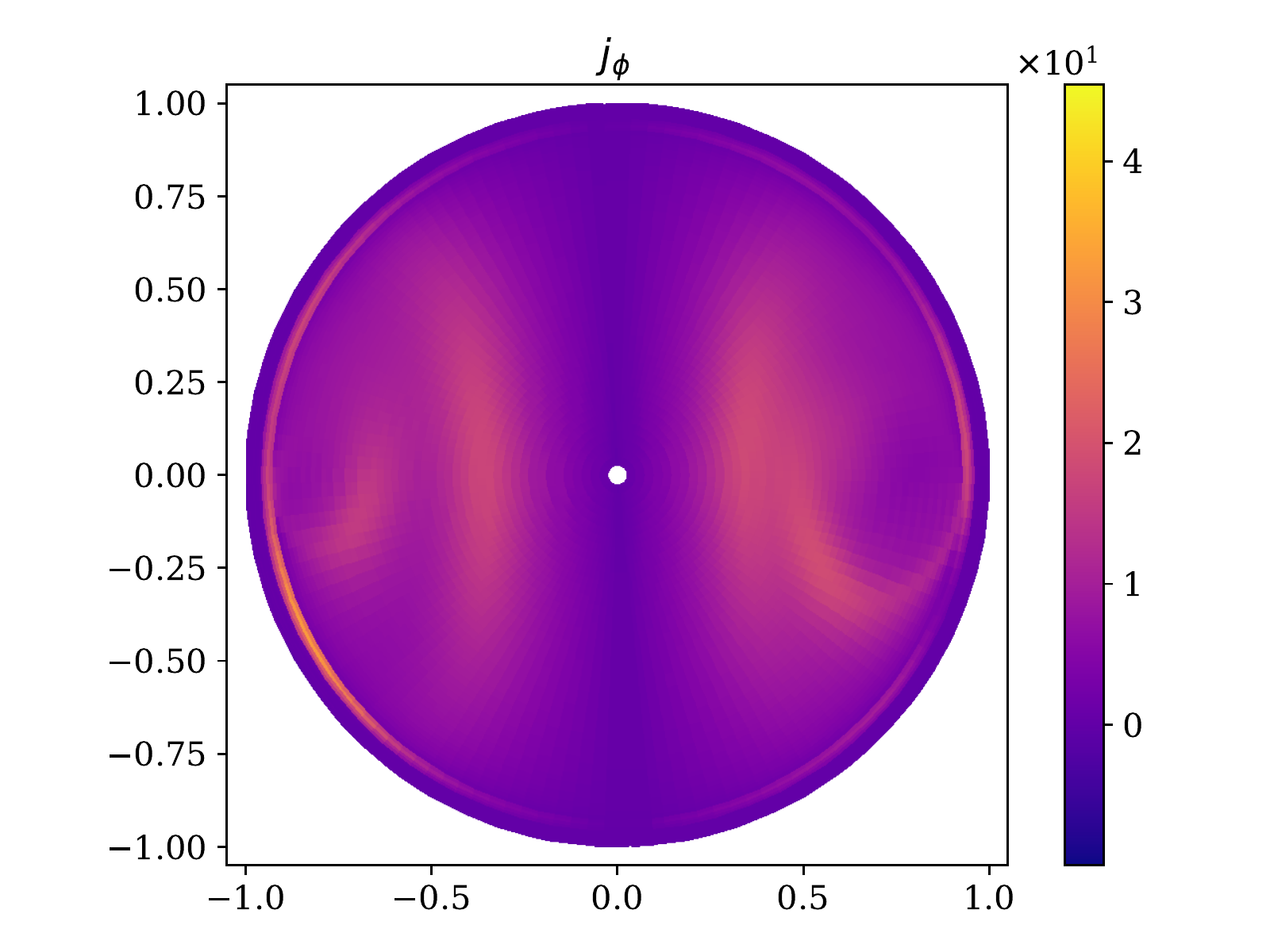}
    \end{minipage}
    \begin{minipage}{0.42\linewidth}
    \includegraphics[width=0.99\columnwidth]{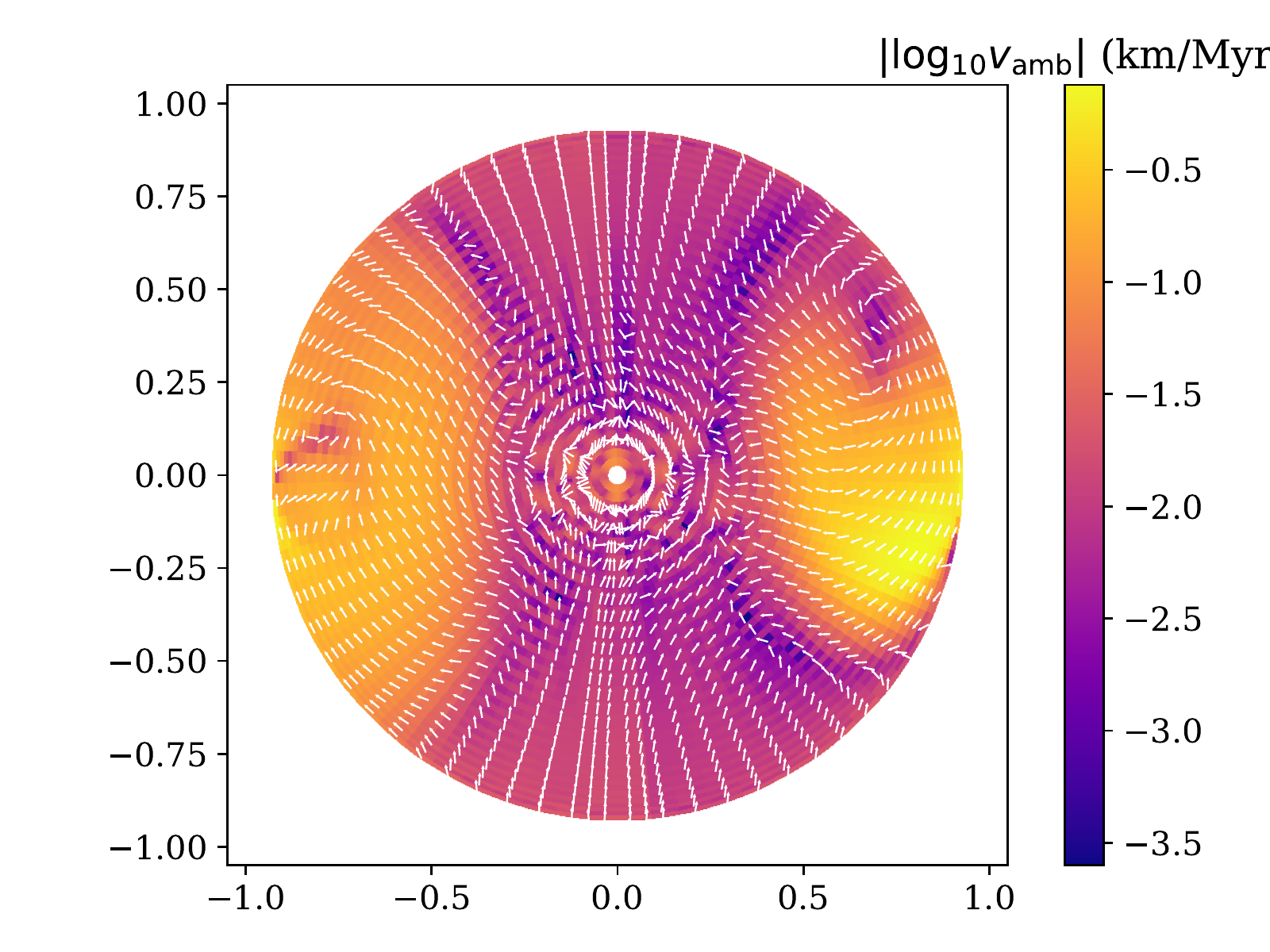}
    \end{minipage}
    \caption{The left column shows the electric current $j_\phi$, and the right column the speed of ambipolar diffusion $v_\mathrm{amb}$, computed for resolution B and $r_\mathrm{cut} = 2$. From top to bottom the four rows are at times $t=0.02,\ 0.96,\ 1.73,\ 3.14$.}
    \label{fig:evol_merid_cut_time}
\end{figure*}

We show the extended evolution of ambipolar velocity in Figure~\ref{fig:evol_merid_cut_time}. Around dimensionless time 1, north-south symmetry is broken. The velocity field in the northern hemisphere behaves differently than the velocity field in the southern hemisphere. After dimensionless time 2, the axial symmetry is broken, i.e. the velocity field in the right part of the figure does not look the same as the velocity field in the left part of the same figure. 

\begin{figure*}
    \centering
    \begin{minipage}{0.49\linewidth}
    \includegraphics[width=0.99\columnwidth]{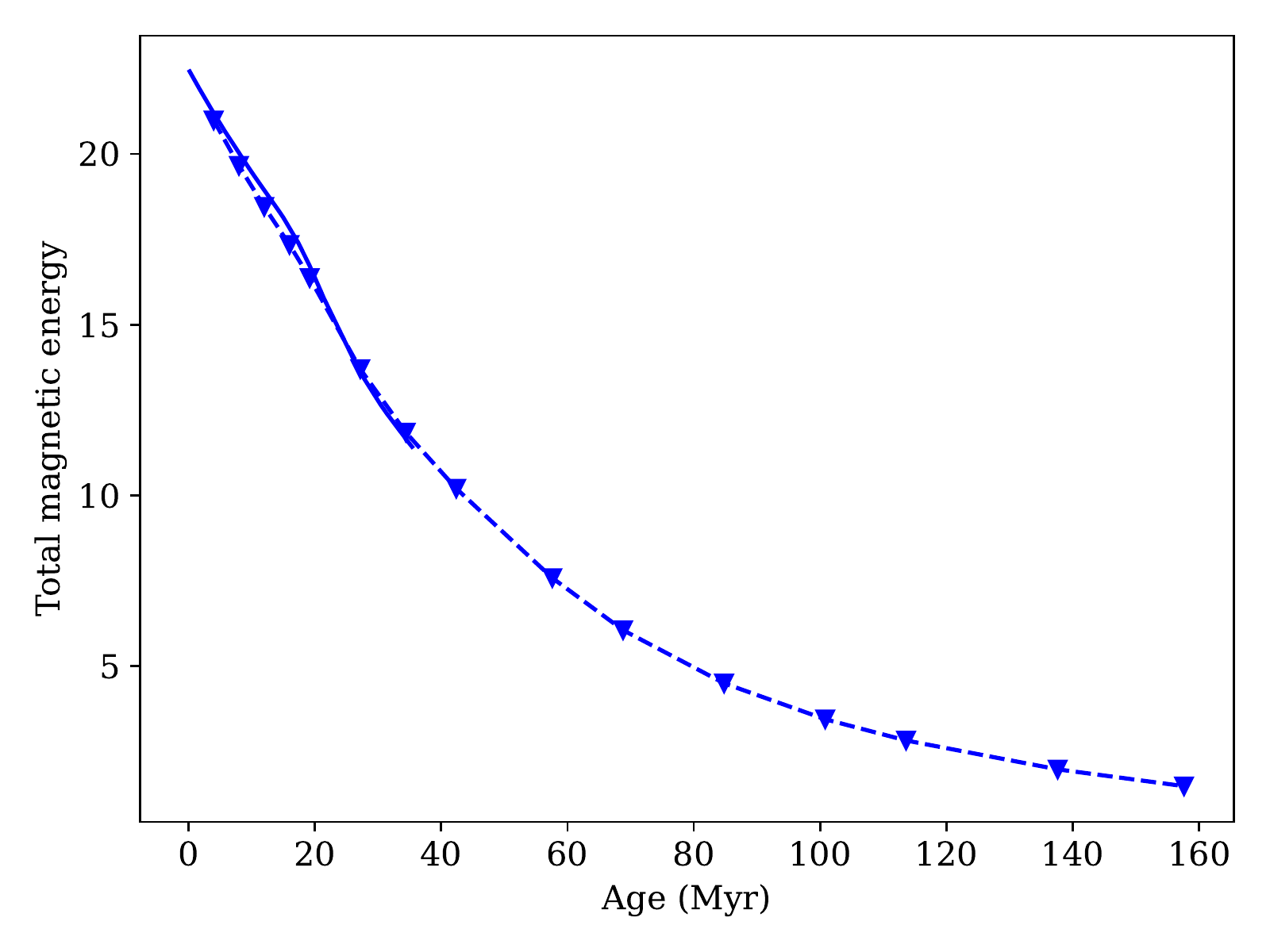}
    \end{minipage}
    \begin{minipage}{0.49\linewidth}
    \includegraphics[width=0.99\columnwidth]{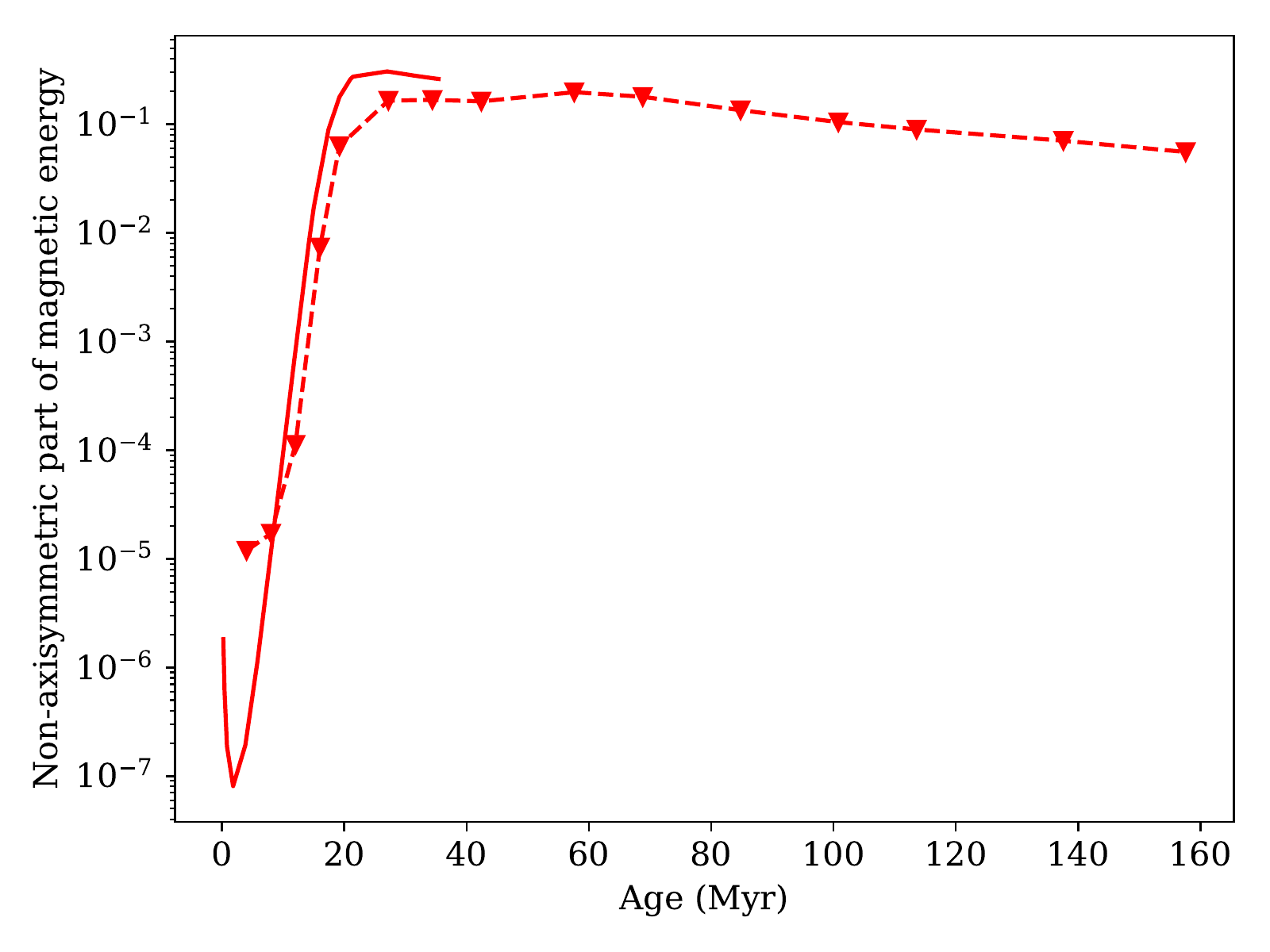}
    \end{minipage}
    \caption{Left panel: evolution of total magnetic energy $E_t$. Right panel: evolution of non-axisymmetric part $E_n$. On both panels we show simulations with $r_\mathrm{cut}=2$. Solid lines correspond to simulations with resolution B while dashed lines correspond to similar calculations with resolution A.}
    \label{fig:e_tot}
\end{figure*}

\subsubsection{Magnetic energy}
We compute the total magnetic energy in our simulations as:
\begin{equation}
E_t = \frac{1}{2} \int \left(\curl \vec A \right)^2 d^3 V.
\end{equation}
It decays slowly during the simulations, see Figure~\ref{fig:e_tot}. In the shorter simulations the energy seems to decay nearly linearly with time. During the first 35~Myr half of the total magnetic energy is released from the system. This decay rate is surprising since the Ohmic decay timescale in the core is chosen to be fixed at 6~Gyr. This decay timescale is comparable with the timescale of ambipolar diffusion ($\approx 20$~Myr) and the Ohmic timescale in the crust ($30$~Myr). We can estimate the timescale for energy decay in high-resolution simulations using the exponential model as:
\begin{equation}
t_\mathrm{decay} = - \frac{t}{\log E_t(t) - \log E_t (0)} \approx 50\; \mathrm{Myr}.
\end{equation}

In our first attempt to model the non-axisymmetric evolution of magnetic fields, we add random values to all components of the initial vector potential $\vec A$ with amplitude of $10^{-5}$ in dimensionless units. We further check how these perturbations evolve, tracking the non-axisymmetric part of the total magnetic energy. In order to compute this quantity, we compute first the axisymmetric part of the magnetic energy by averaging the field over the $\phi$-coordinate:
\begin{equation}
\vec B_\mathrm{axi} = \frac{1}{2\pi} \int_0^{2\pi} \vec B (r, \theta, \phi) d\phi.
\end{equation}
Then we find the energy as:
\begin{equation}
E_\mathrm{axi} = \pi \int \vec B^2_\mathrm{axi} d^2 V.
\end{equation}
The non-axisymmetric part of magnetic energy is then:
\begin{equation}
E_\mathrm{non} = E_t - E_\mathrm{axi}.
\end{equation}
We plot the evolution of $E_\mathrm{non}$ in Figure~\ref{fig:e_tot}. We notice that until $t=2$~Myr it decays much faster than the total magnetic energy, so initially the field becomes more axisymmetric. However, once the random initial conditions have adjusted themselves, after 2~Myr the non-axisymmetric part of the energy grows, reaching values of $0.3$, several orders of magnitude greater than the initial perturbations. That is, the large-scale axisymmetric field is unstable to the presence of the small-scale non-axisymmetric noise that was added. The non-axisymmetric magnetic energy peaks around time 2.7, and thereafter decays slower than the total magnetic energy. This means that the instability continues to operate, and a fraction of non-axisymmetric energy is constantly regenerated from the large-scale axisymmetric field.

\subsubsection{Azimuthal magnetic field}

\begin{figure*}
    \centering
    \begin{minipage}{0.42\linewidth}
    \includegraphics[width=0.99\columnwidth]{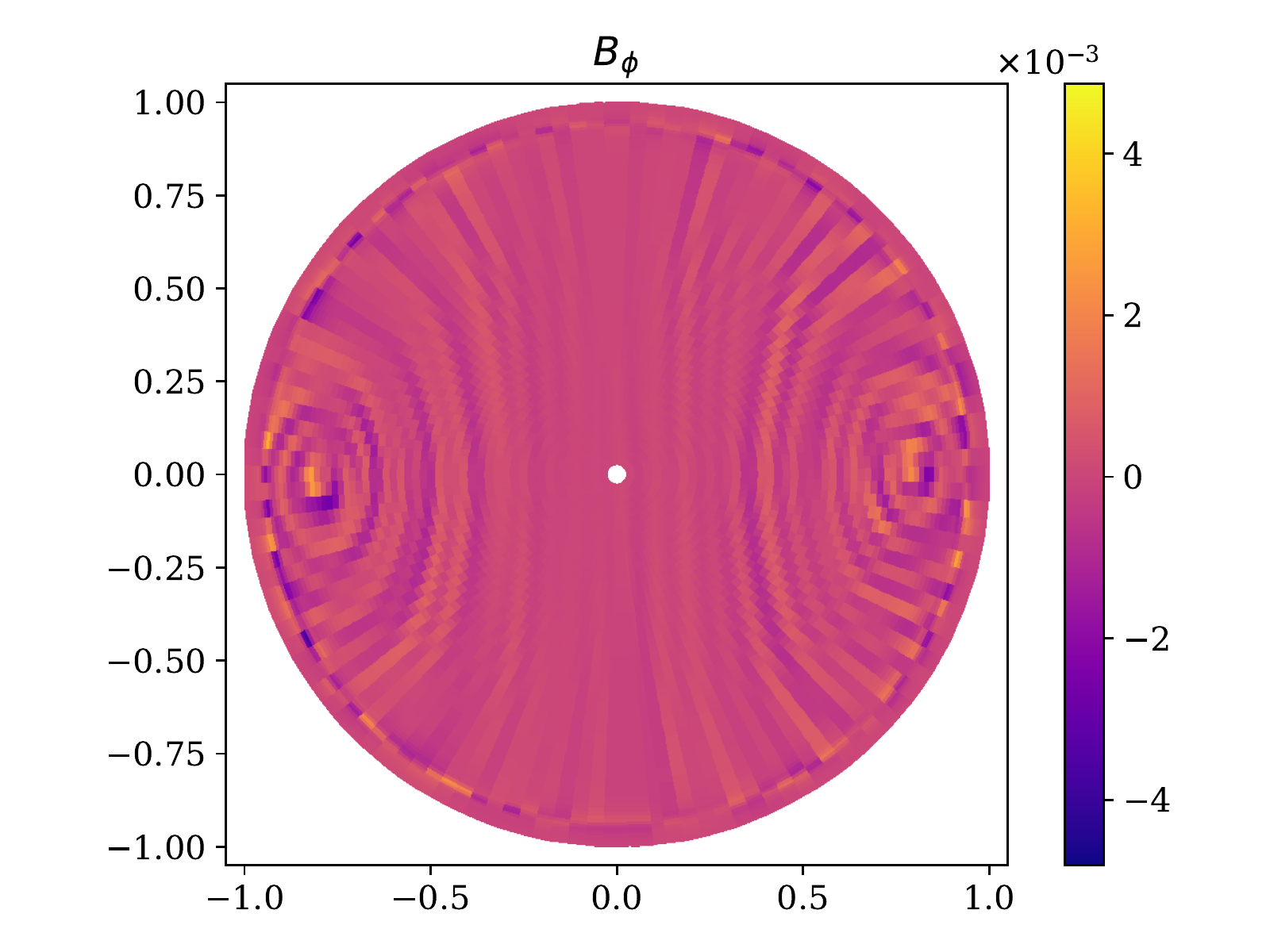}
    \end{minipage}
    \begin{minipage}{0.42\linewidth}
    \includegraphics[width=0.99\columnwidth]{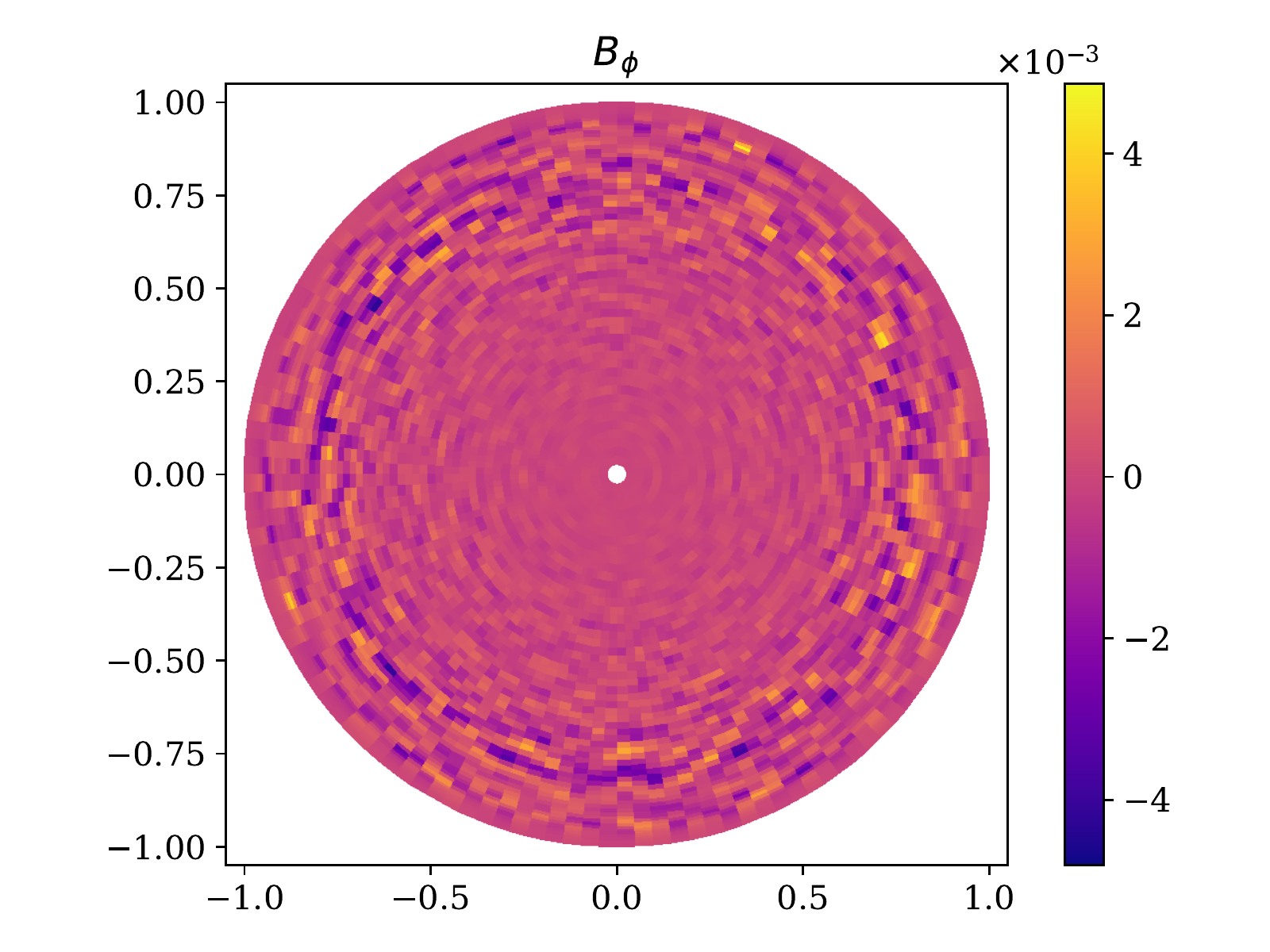}
    \end{minipage}
    \begin{minipage}{0.42\linewidth}
    \includegraphics[width=0.99\columnwidth]{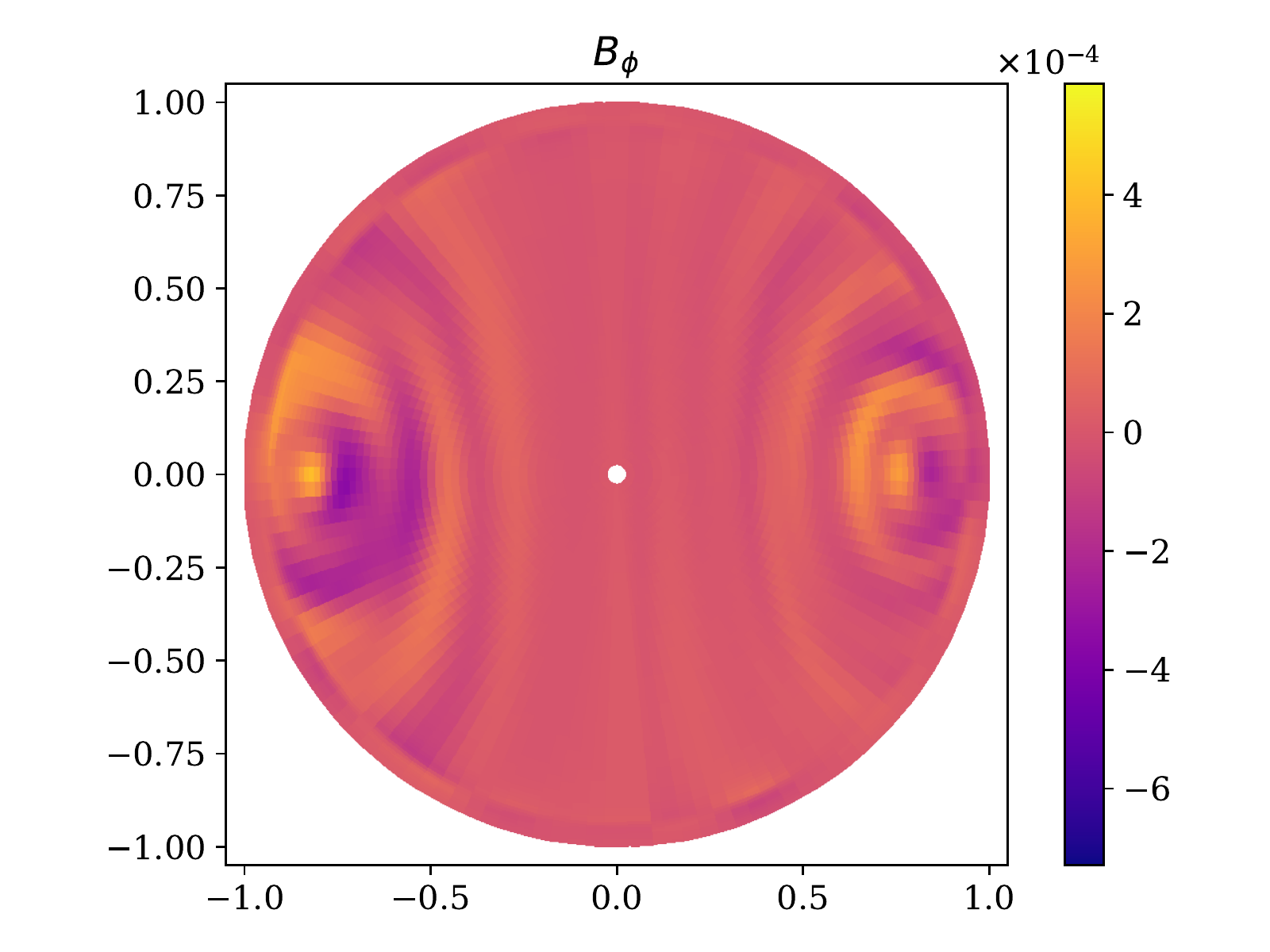}
    \end{minipage}
    \begin{minipage}{0.42\linewidth}
    \includegraphics[width=0.99\columnwidth]{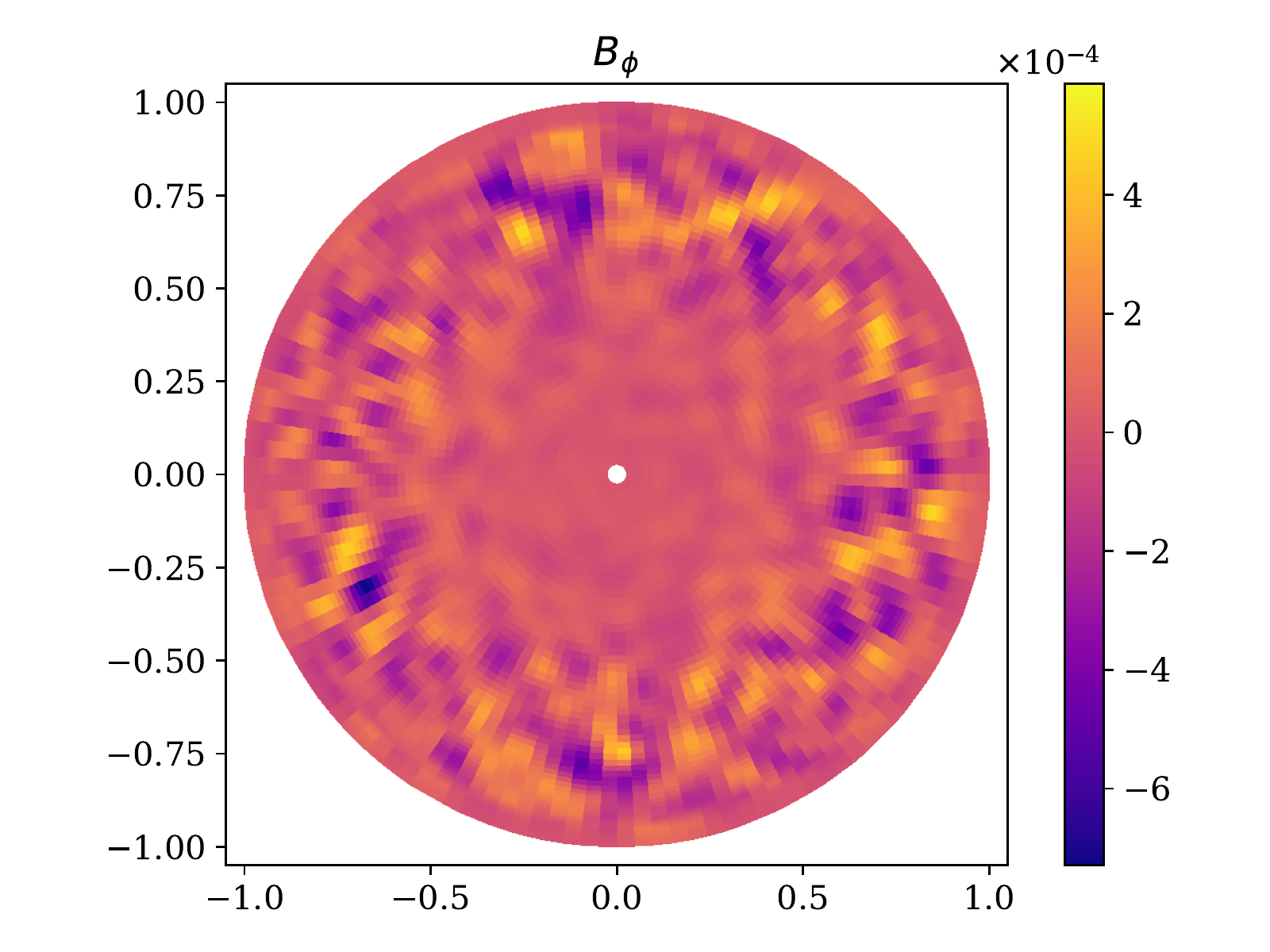}
    \end{minipage}
    \begin{minipage}{0.42\linewidth}
    \includegraphics[width=0.99\columnwidth]{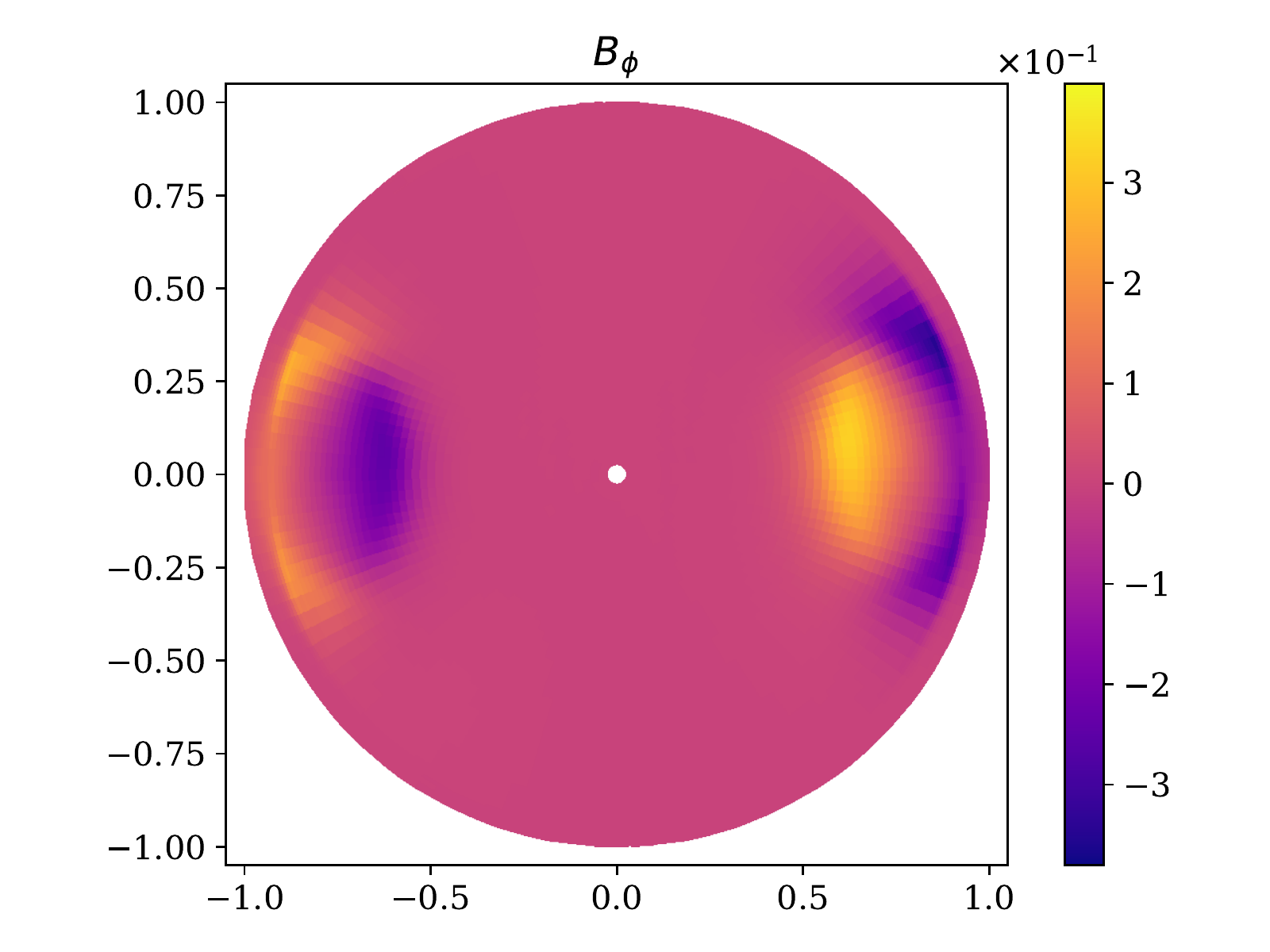}
    \end{minipage}
    \begin{minipage}{0.42\linewidth}
    \includegraphics[width=0.99\columnwidth]{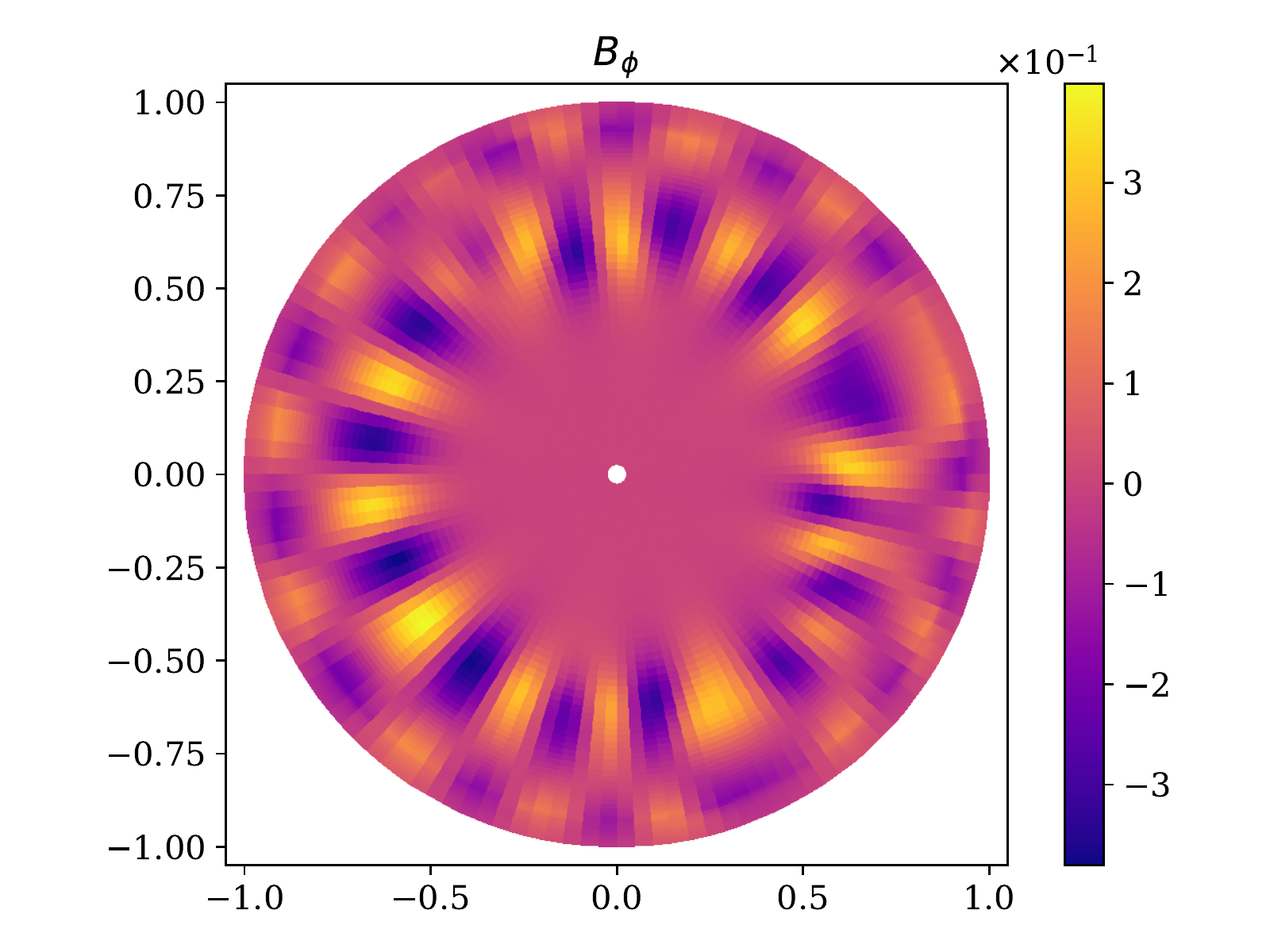}
    \end{minipage}
    \begin{minipage}{0.42\linewidth}
    \includegraphics[width=0.99\columnwidth]{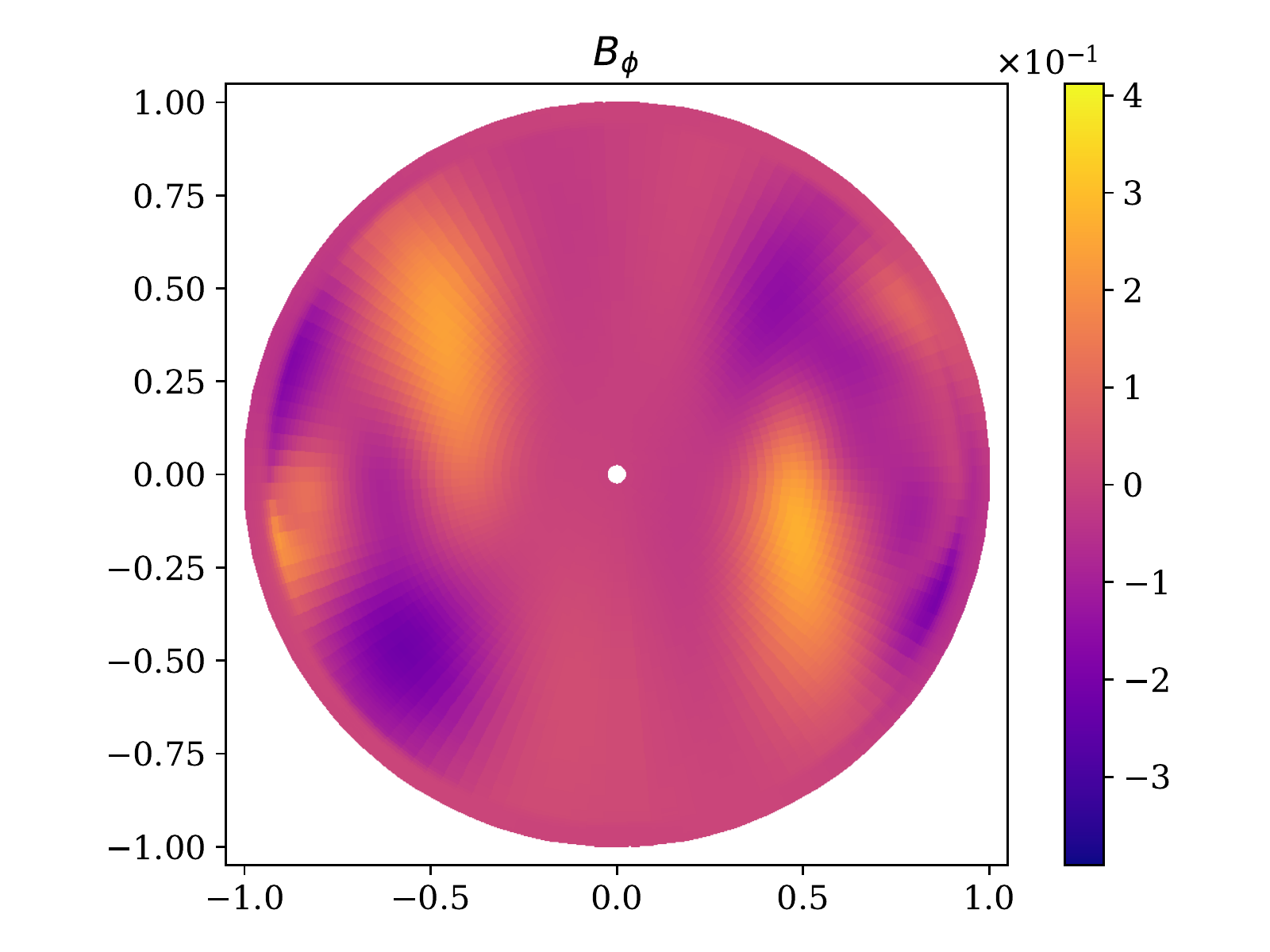}
    \end{minipage}
    \begin{minipage}{0.42\linewidth}
    \includegraphics[width=0.99\columnwidth]{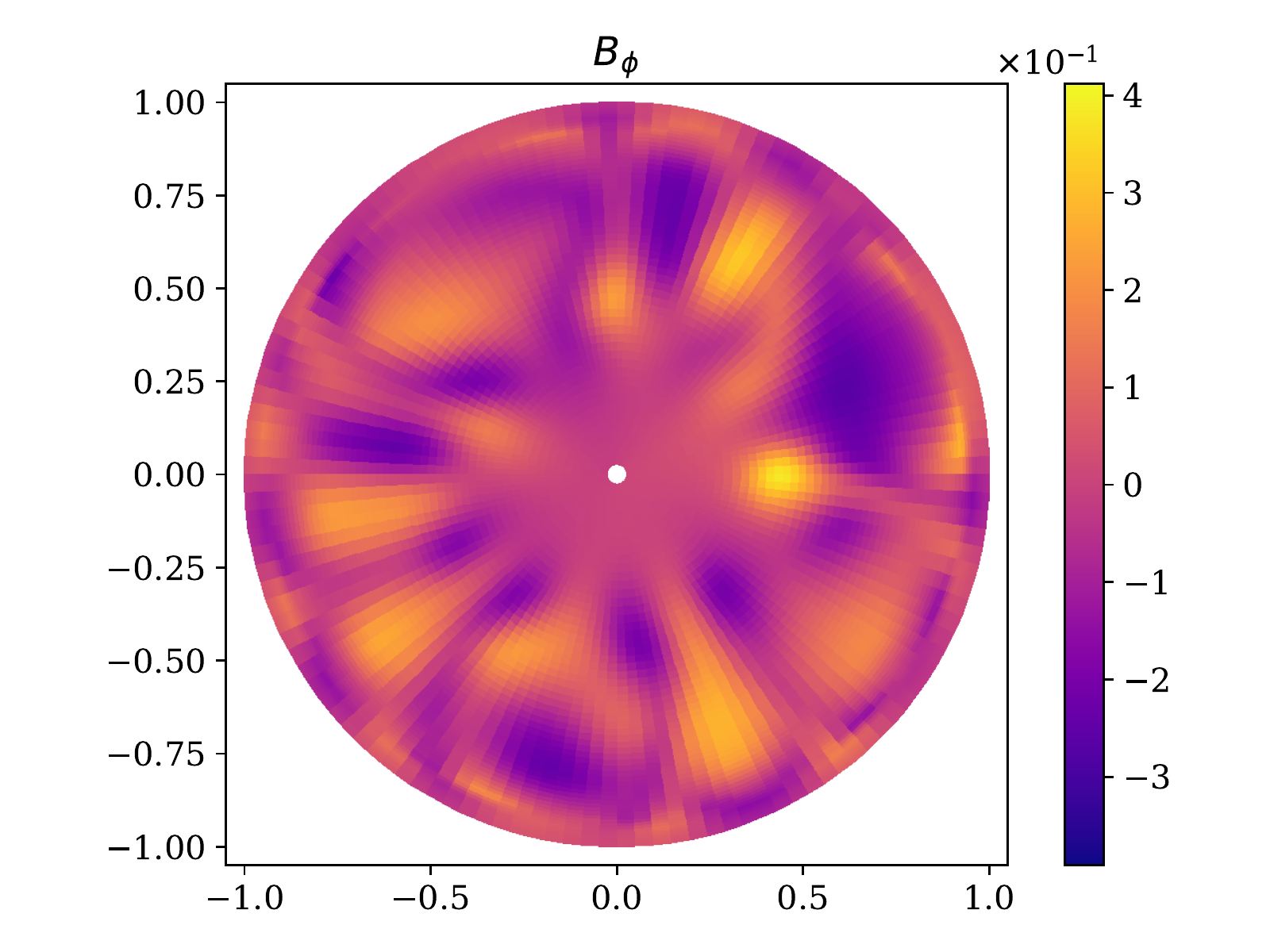}
    \end{minipage}
    \caption{Evolution of $B_\phi$ with time. The left column shows meridional cuts through the NS while the right column shows equatorial cuts. The resolution is B, and $r_\mathrm{cut} = 2$. From top to bottom the four rows are $t=0.02,\ 0.18,\ 1.74,\ 3.54$.}
        \label{fig:evol_toroidal}
\end{figure*}

The evolution of the non-axisymmetric part of the energy is easy to track if we examine the $B_\phi$ component of the field, which roughly corresponds to toroidal magnetic field because our initial conditions are nearly axisymmetric, see Figure~\ref{fig:evol_toroidal}. In our initial conditions we have very limited $B_\phi$ caused by the random perturbations of $A_r$ and $A_\theta$ because we assume regular magnetic field only for $A_\phi$. During the simulations these perturbations merge, forming complicated semi-regular large-scale structures inside the NS core. Initially these structures are elongated along the magnetic field lines with width $300-600$~m and length comparable to $R_\mathrm{NS}$. 

Over time these elongated structures decay and merge, forming much larger regions with regular $B_\phi$. After a few million years the strength of magnetic fields in these structures starts growing, reaching values of $3\times 10^{12}$~G.  When we examine the equatorial cut we observe that $B_\phi$ reaches positive and negative values 14 times, that is, the azimuthal wavenumber $m=14$. This is well within our numerical resolution (recall Table \ref{tab:simulation_summary}), so we believe that these structures correspond to true physical instabilities rather than numerical ones. We further test issues related to resolution and the influence of $r_\mathrm{cut}$ in Appendix~\ref{appendix:code_verification}.

\begin{figure*}
    \centering
    \begin{minipage}{0.42\linewidth}
    \includegraphics[width=0.99\columnwidth]{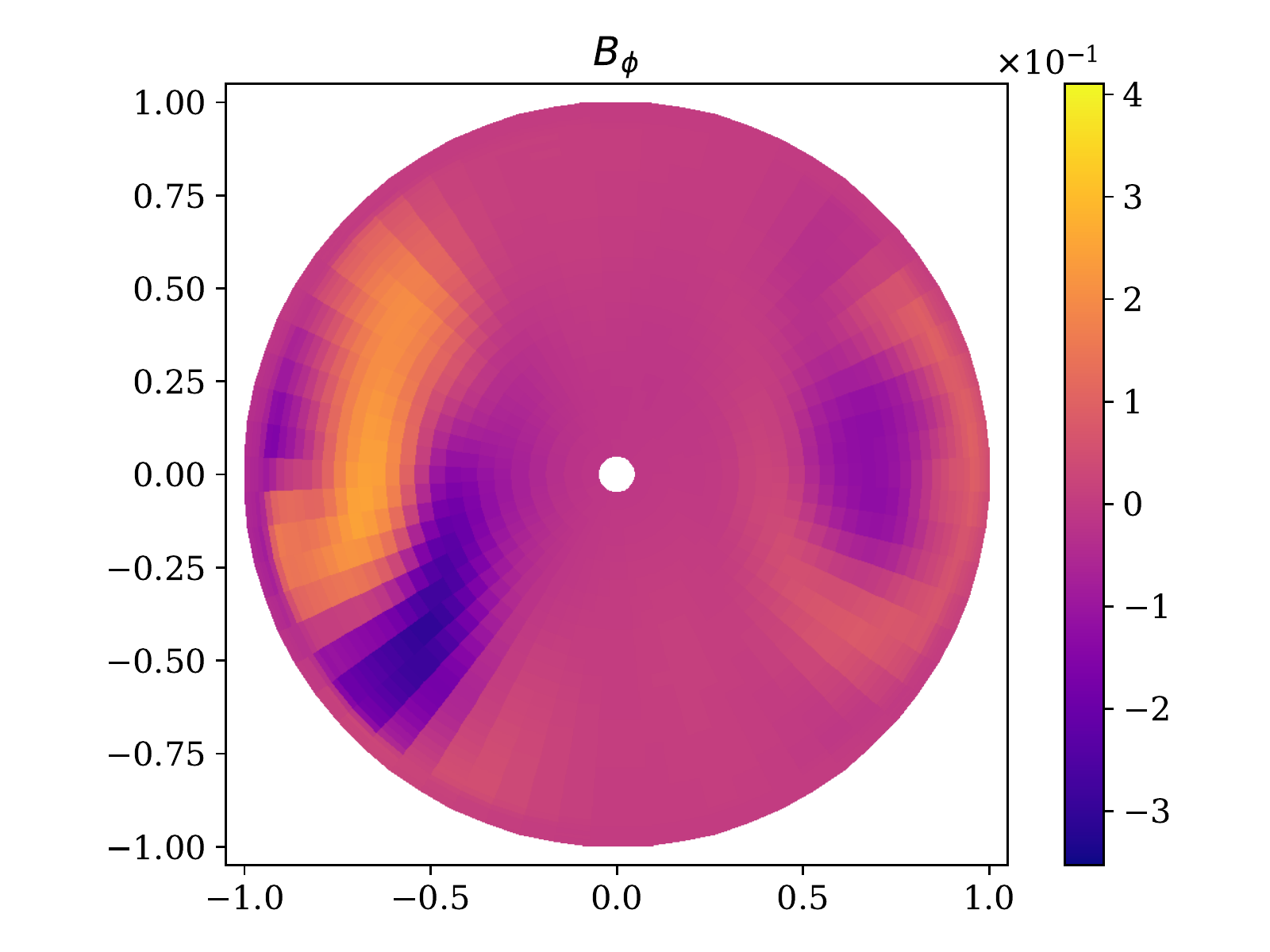}
    \end{minipage}
    \begin{minipage}{0.42\linewidth}
    \includegraphics[width=0.99\columnwidth]{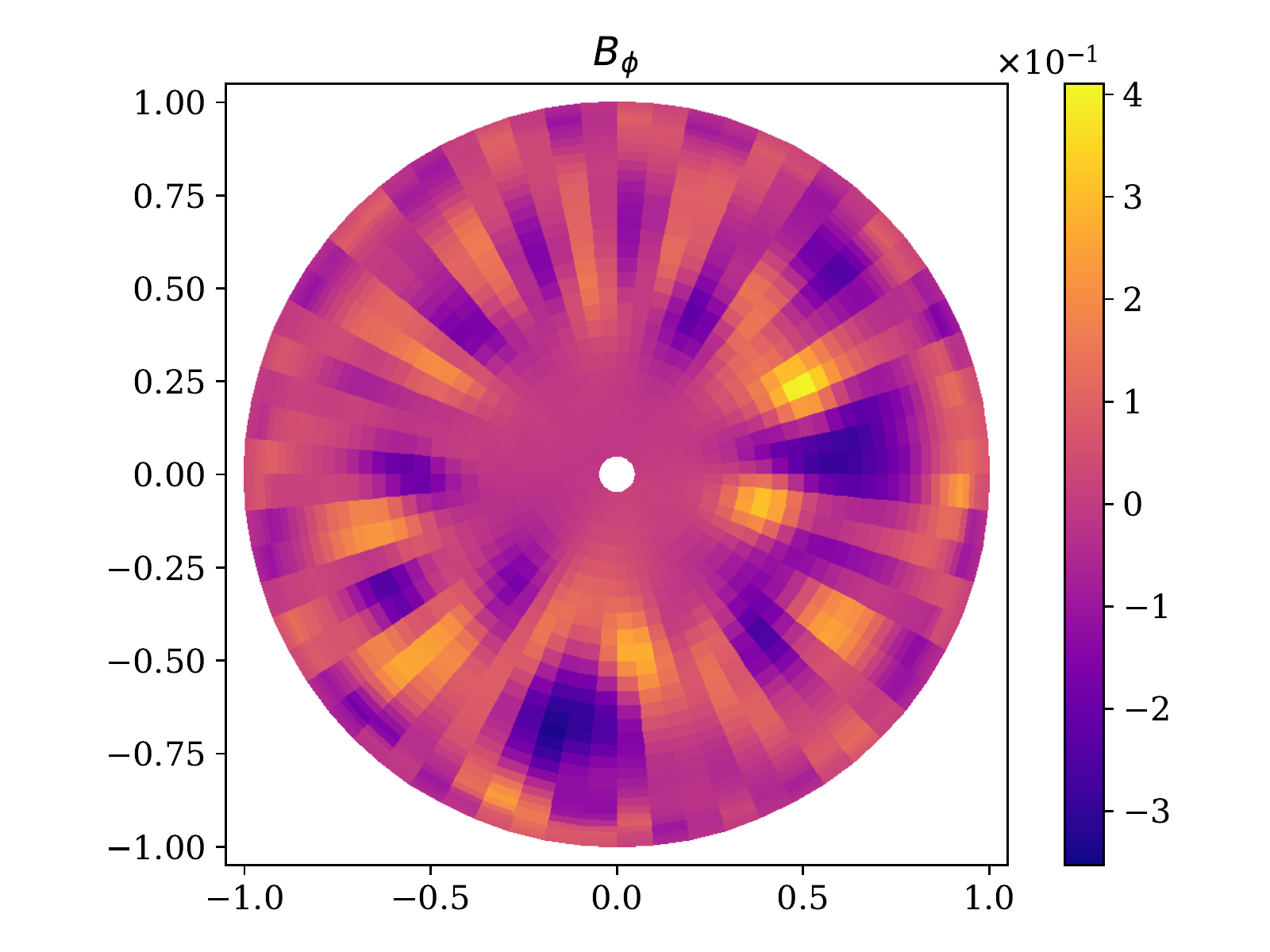}
    \end{minipage}
    \begin{minipage}{0.42\linewidth}
    \includegraphics[width=0.99\columnwidth]{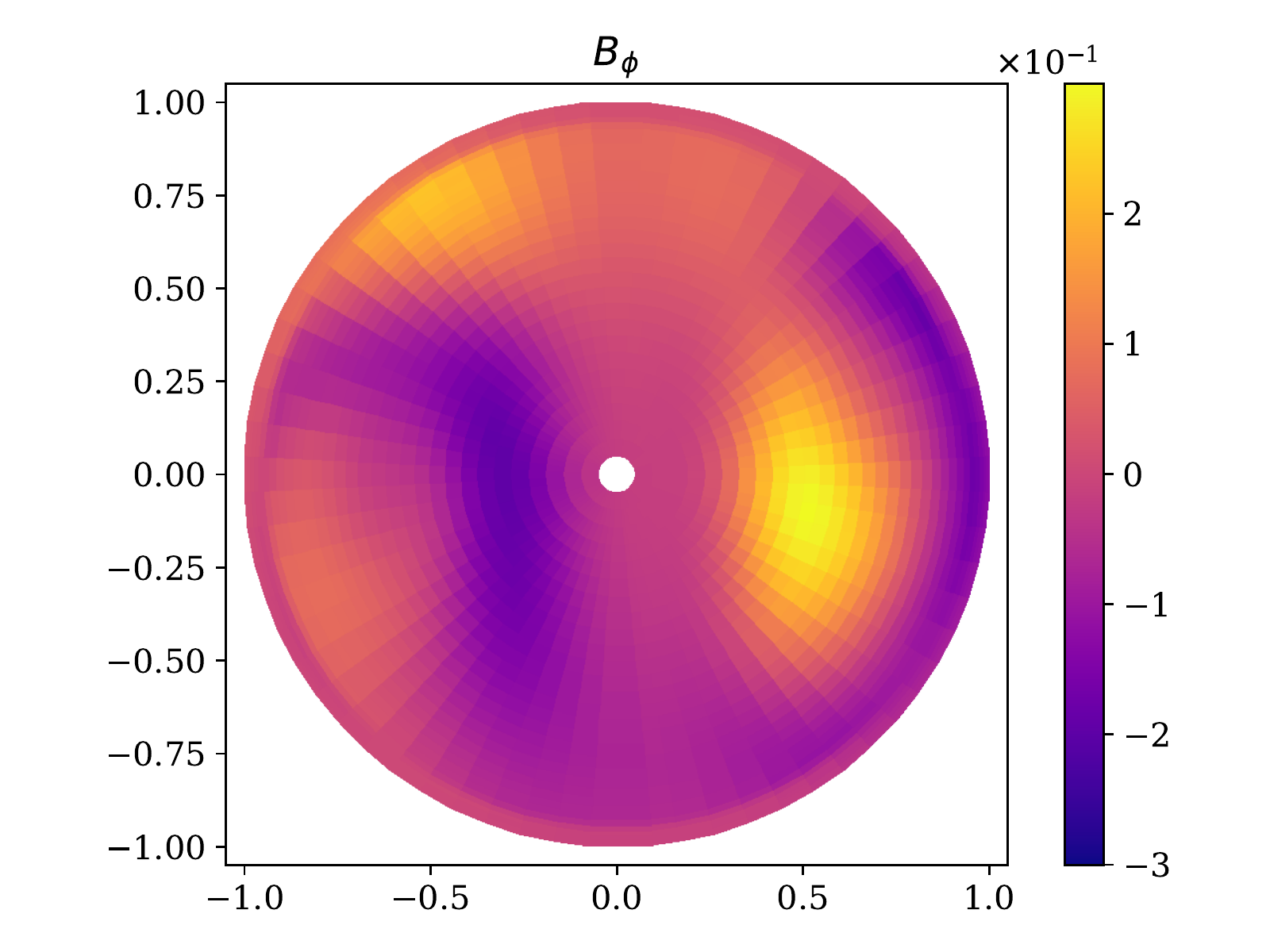}
    \end{minipage}
    \begin{minipage}{0.42\linewidth}
    \includegraphics[width=0.99\columnwidth]{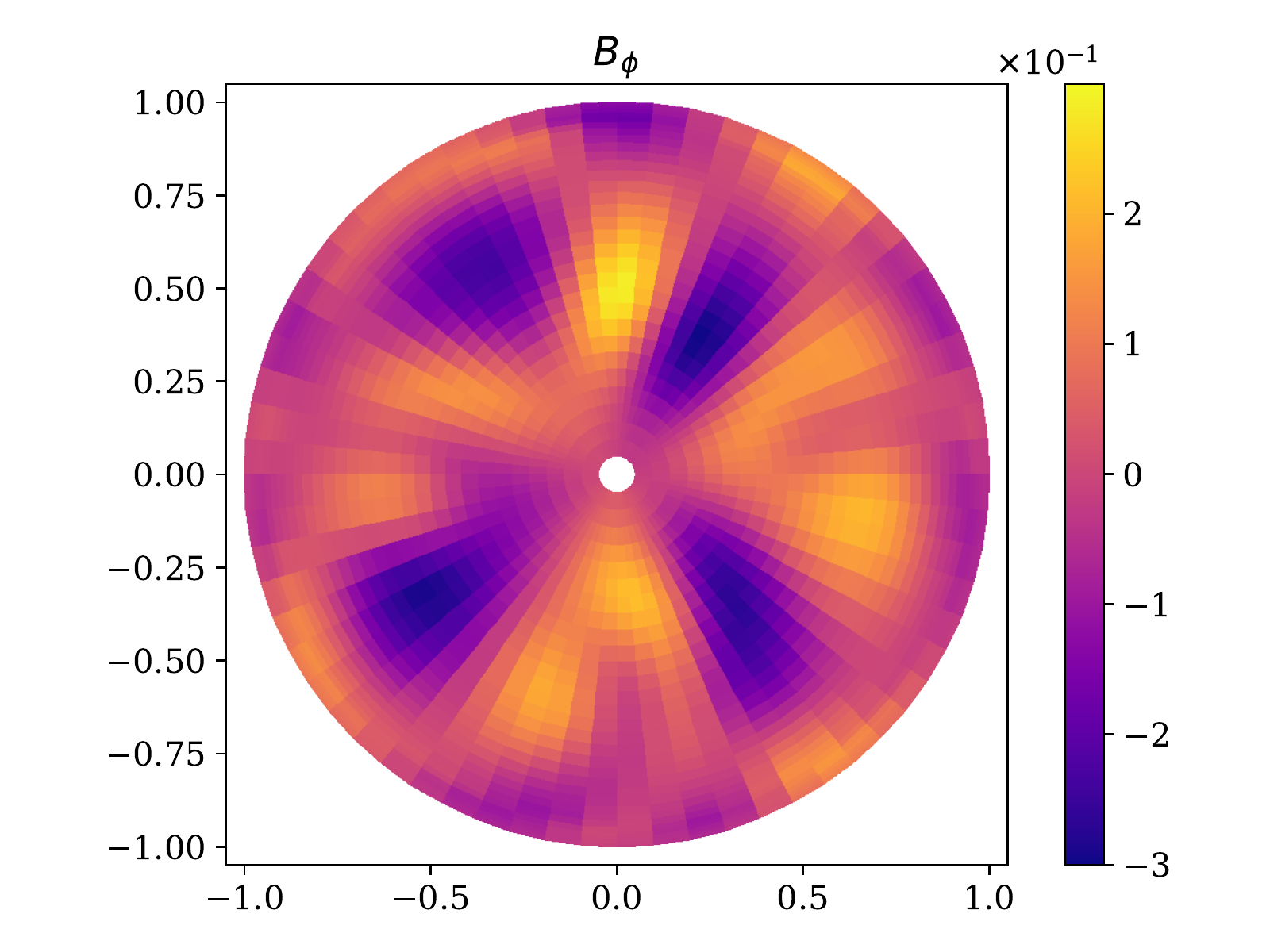}
    \end{minipage}
    \caption{Evolution of $B_\phi$ with time. The left column shows meridional cuts through the NS while the right column shows equatorial cuts. The resolution is A, and $r_\mathrm{cut} = 2$. The top row is $t = 3.52$ and the bottom row $t = 10.08$.}
    \label{fig:evol_toroidal_32}
\end{figure*}

The long calculations (above 35~Myr) with resolution B are numerically challenging. Nevertheless, we are still interested in later stages of this simulations, so we revert back to resolution A for extended calculations. We show some results of these calculations in Figure~\ref{fig:evol_toroidal_32}. In these simulations the structure of magnetic field stays quite similar to more detailed simulations; compare the last panel of Figure~\ref{fig:evol_toroidal} with the first panel of Figure~\ref{fig:evol_toroidal_32}. On longer timescales the fine structures of $B_\phi$ continue merging so by 100~Myr the $m=4$ mode becomes dominant.

\subsubsection{Electric currents}
Among other quantities we track the evolution of electric current. In our simulation setup electric currents in the crust are expected to decay on a 30~Myr timescale. Soon after we start the simulations, we see the formation of electric current near the core-crust boundary, see Figure~\ref{fig:evol_merid_cut_time}.
In our simulations this current is localised between the crust-core boundary (0.924~$R_\mathrm{NS}$) and radial distance $\approx 0.955$~$R_\mathrm{NS}$ in the region where the resistivity grows rapidly, see Figure~\ref{fig:chi}. In this region resistivity still does not reach its maximum. Some electric current also flows through the outer crust where the resistivity is fixed.
We resolve the current above the crust-core boundary relatively well since this region is covered with 4 collocation points in numerical simulations with resolution B and with 8 collocation points in simulations with resolution D. In both these simulations the radial extent of the current is the same, confirming that this current sheet is adequately resolved.

Under the influence of ambipolar diffusion the electric currents in the core start evolving, forming arcs reaching from the core-crust boundary to distances $R = 0.3-0.5$~$R_\mathrm{NS}$ inside the NS, see Figure~\ref{fig:evol_merid_cut_time}. The electric current in the crust evolves as well, see Figure~\ref{fig:v_amb_jphi}. In this figure we show the time evolution of maximum $j_\phi$ between radial distance  0.924~$R_\mathrm{NS}$ and 0.955~$R_\mathrm{NS}$.  Ambipolar diffusion induces strong electric current in the NS crust after 15~Myr. The  current reaches a maximum around 20~Myr and decays nearly exponentially after this. This current decays on approximately twice the Ohmic timescale (60-70~Myr) of the crust, leading to global decay of magnetic energy on a timescale comparable to the Ohmic timescale in the crust.
At later stages (t > 1.5) most of the crust current concentrates between $\pi/4<\theta < 3\pi/4$, see Figure~\ref{fig:evol_merid_cut_time}. 
At advanced stages of evolution the arcs of electric current separate regions with fast ambipolar diffusion speed ($v_\mathrm{amb}$ > 0.1~km/Myr) from regions with slow ambipolar diffusion speed.

\subsection{Instability of poloidal magnetic field}
\label{s:instability}

As we noted in previous sections the non-axisymmetric part of total magnetic energy begins growth at 2-3~Myr and reaches maximum around 21~Myr, see Figure~\ref{fig:e_tot}. Simultaneously the speed of ambipolar diffusion starts growing and reaches maximum around the same time. 
We also noted that azimuthal magnetic field with $m\approx 14$ start emerging from the initial fluctuations. The azimuthal number $m=14$ is seen as the number of regions with negative values of $B_\phi$ in the equatorial cut in Figure~\ref{fig:evol_toroidal}.

\begin{figure*}
    \centering
    \begin{minipage}{0.49\linewidth}
    \includegraphics[width=0.99\columnwidth]{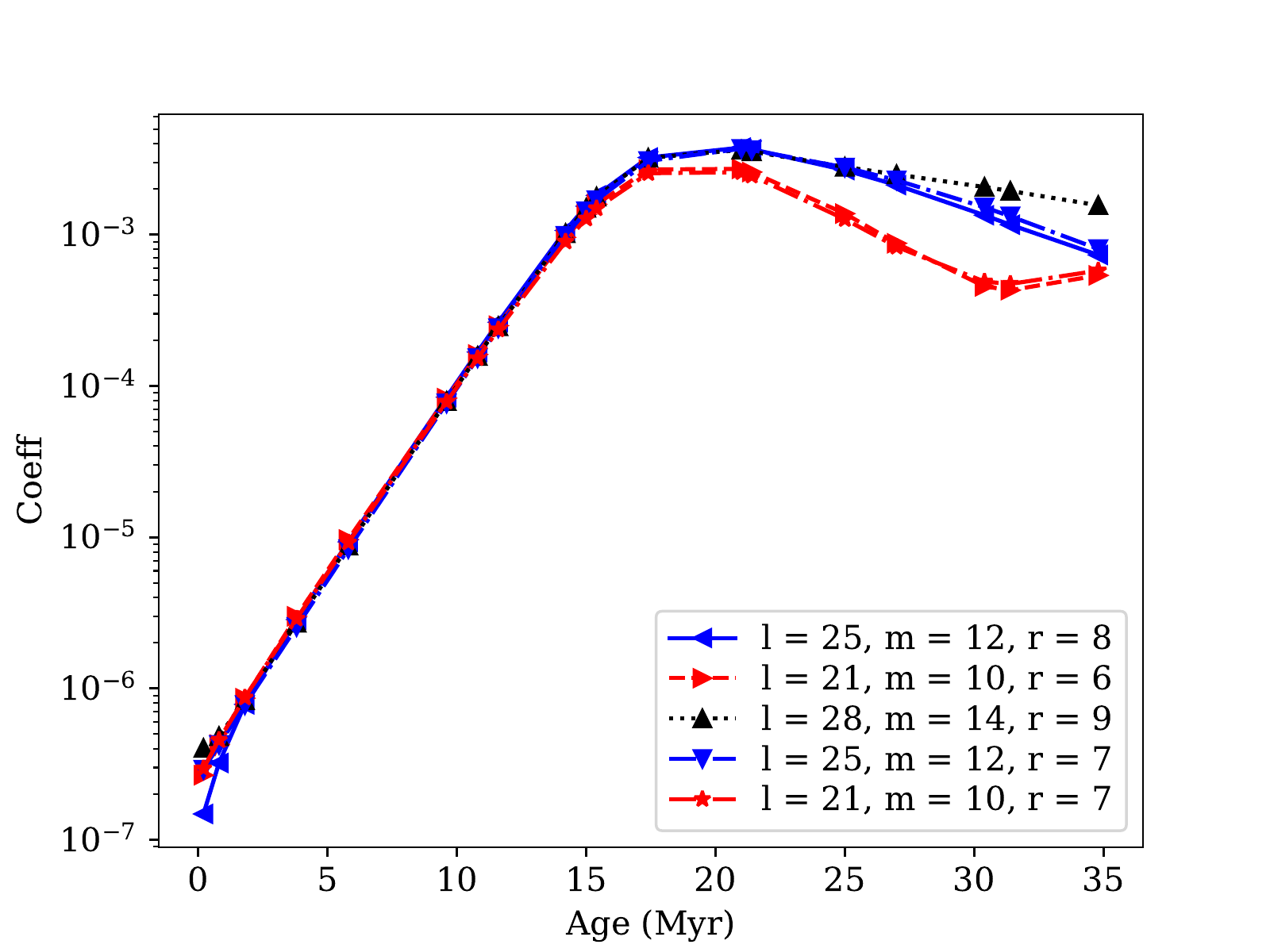}
    \end{minipage}
    \begin{minipage}{0.49\linewidth}
    \includegraphics[width=0.99\columnwidth]{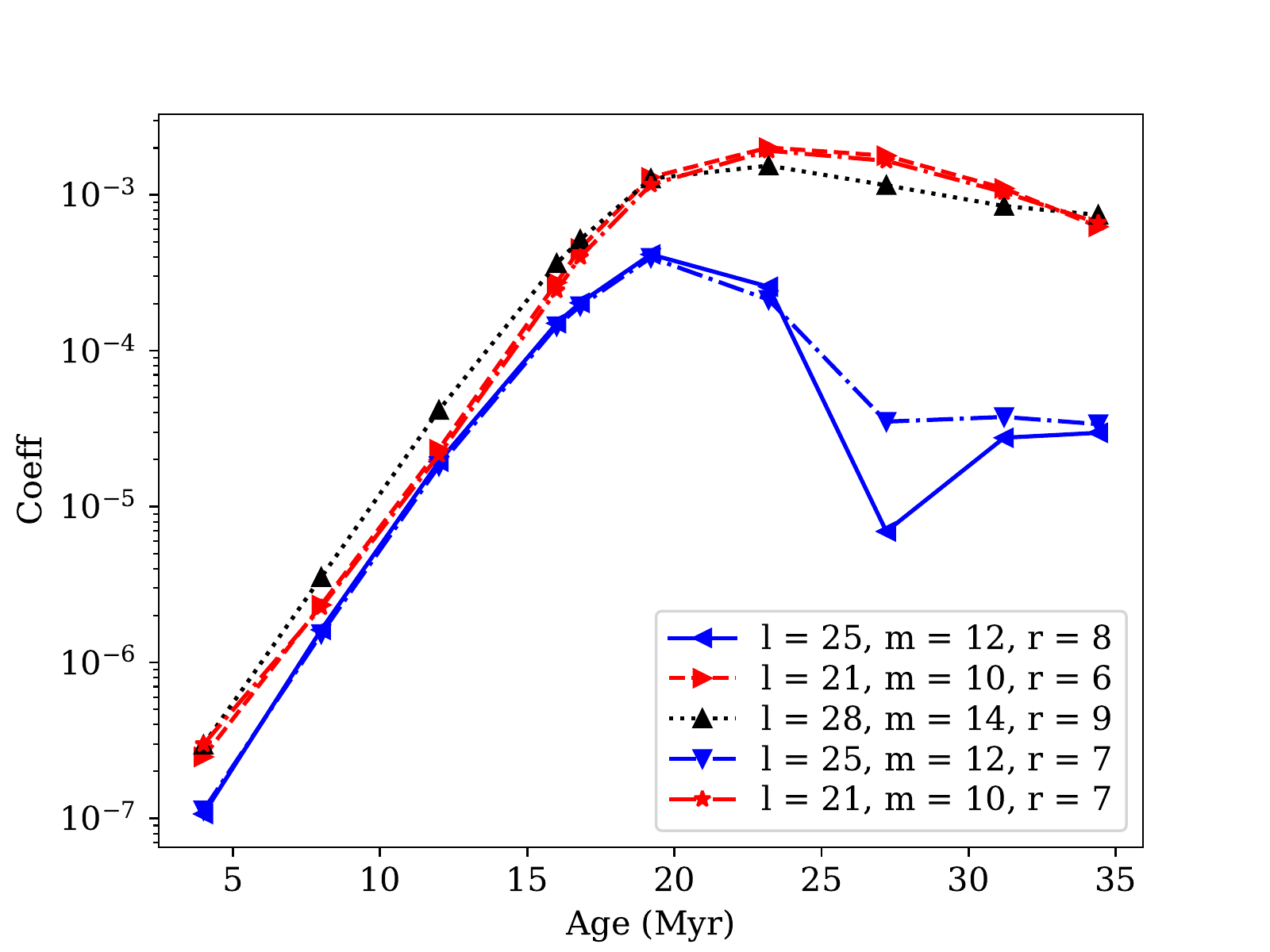}
    \end{minipage}
    \caption{Growth of selected coefficients of the magnetic field with time; $l,\ m$ correspond to spherical harmonic degree and order, while $r$ corresponds to degree of Jacobi polynomial. In the left panel we show the result of calculations with resolution B and in the right panel we show the truncated results computed with resolution A.
    }
    \label{fig:ll_evol}
\end{figure*}

In order to investigate growth of azimuthal magnetic field in more detail, we identify the coefficients of the spectral expansion with the largest absolute value at age $15$~Myr. From this selection we exclude the coefficients corresponding to the initial condition. We plot the evolution of this identified cluster of harmonics in  Figure~\ref{fig:ll_evol}  for numerical resolutions A and B. The contribution of these harmonics to the solution increases from $10^{-7}$ (roughly the level of noise perturbations added to the simulations) to $(2-3)\times 10^{-3}$ when these harmonics start affecting the ambipolar velocity field. Although the resolution A is only half of the resolution B, we see the growth of exactly the same harmonics during the first 15~Myr. The behaviour after the saturation is different, which means that after the saturation harmonics with number $l > 32$ start contributing to the evolution.  

The growth of well-resolved spectrally localised structures is an indication of instability. We have thus shown that an initial axisymmetric poloidal magnetic field is unstable under effects of ambipolar diffusion in three dimensions, giving rise to non-axisymmetric structures. Using Figure~\ref{fig:ll_evol} we conclude that the growth rate of this instability is $\approx 2$~Myr or 0.2 in dimensionless units. The instability is saturated around $15-20$~Myr when the selected harmonics reach maximum value.

\begin{figure}
    \centering
    \includegraphics[width=0.99\columnwidth]{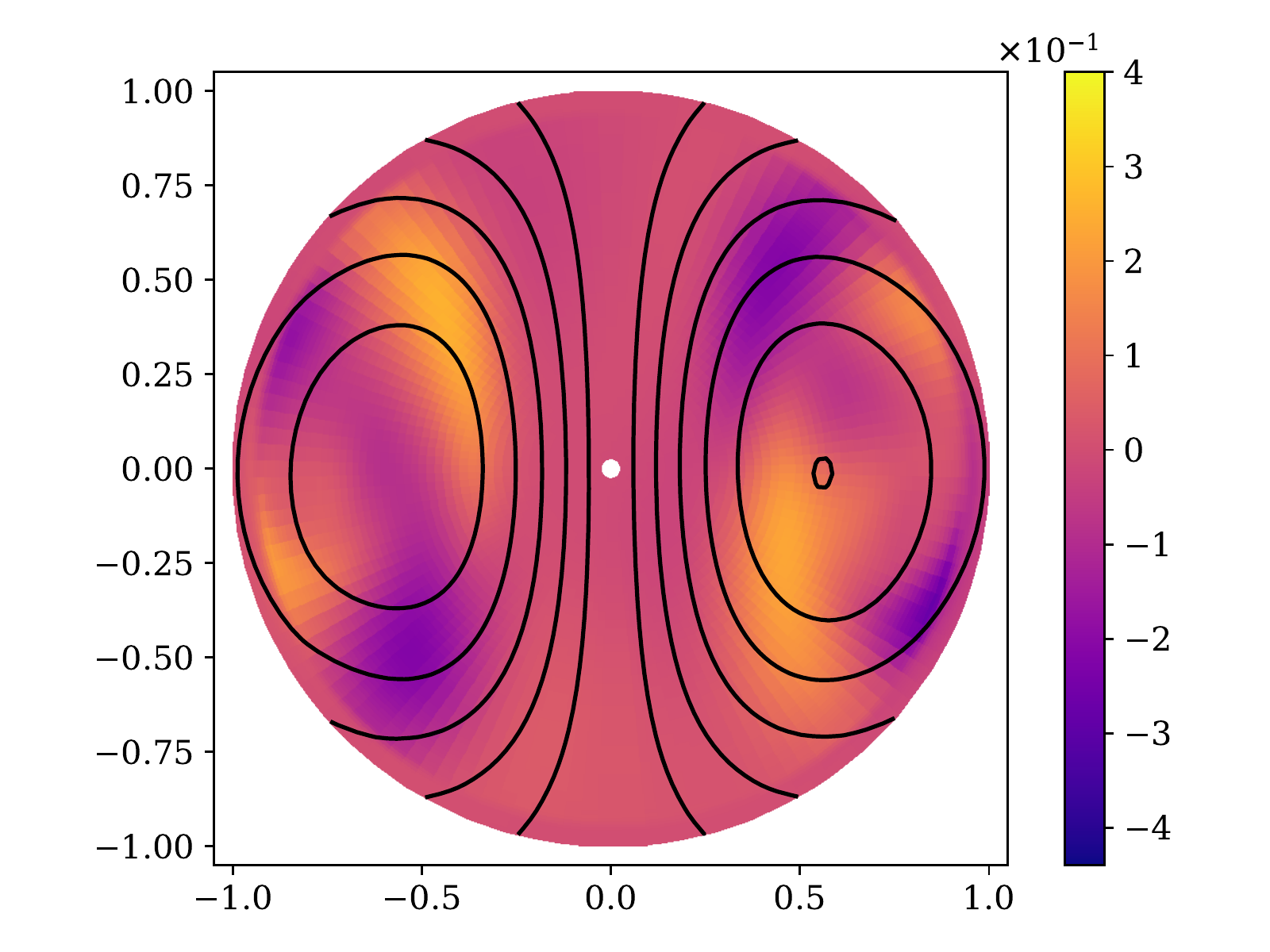}
    \caption{A meridional cut showing the $B_\phi$ component of magnetic field at 29~Myr, computed with resolution B. Colour shows the strength of $B_\phi$ while black solid lines correspond to field lines for $B_r$ and $B_\theta$ components. 
    }
    \label{fig:countor_Bphi}
\end{figure}

This instability is intrinsically three-dimensional and was not seen before in one and two-dimensional simulations. Earlier on, \cite{Castillo2017MNRAS} found a formation of toroidal magnetic field in two-dimensional, axisymmetric simulations. Our azimuthal field might be related to that one but has a complicated structure in the azimuthal direction. \cite{Castillo2017MNRAS} found that their newly generated toroidal magnetic field is bounded within the closed magnetic field lines of poloidal magnetic field. It is not the case in our simulations. In Figure~\ref{fig:countor_Bphi} we see that $B_\phi$ is generated also in regions of open field lines. This difference might be related to the following factors: (1) our simulations are not axisymmetric, (2) our $B_\phi$ is therefore also not a purely toroidal magnetic field, and (3) we added a crust with finite conductivity in our simulations.

We show evolution of magnetic energy computed in our low-resolution simulations (A resolution) in Figure~\ref{fig:e_tot}. The total magnetic energy decay is quite similar to that computed in high-resolution simulations (B resolution). There are some small differences around 20~Myr which might be related to a slight delay in development of non-axisymmetric magnetic field. On longer timescales it becomes apparent that magnetic energy decay is exponential with timescale of $\approx 60$~Myr, i.e. $\approx 3$ ambipolar diffusion timescales estimated using velocities derived from the numerical solution. The energy decays in our simulations much faster and much stronger than it was found in axisymmetric simulations by \cite{Castillo2017MNRAS}. We do not see any indications that magnetic energy stops decaying after some time. We could speculate that magnetic energy decay might slow down when total magnetic energy becomes comparable (i.e. 2-3 times stronger) to the non-axisymmetric part of the energy. It will require a decay of one order of magnitude more, i.e. on a timescale of another 150~Myr.

\subsection{Astrophysical implications}
\label{s:astro_res}
In this section we summarise the results of our simulations which could be probed in astronomical observations. These are the structure and evolution of surface magnetic field, surface temperature, and crust failure.

\subsubsection{Deep crustal heating}

\begin{figure*}
    \centering
    \begin{minipage}{0.49\linewidth}
    \includegraphics[width=0.99\columnwidth]{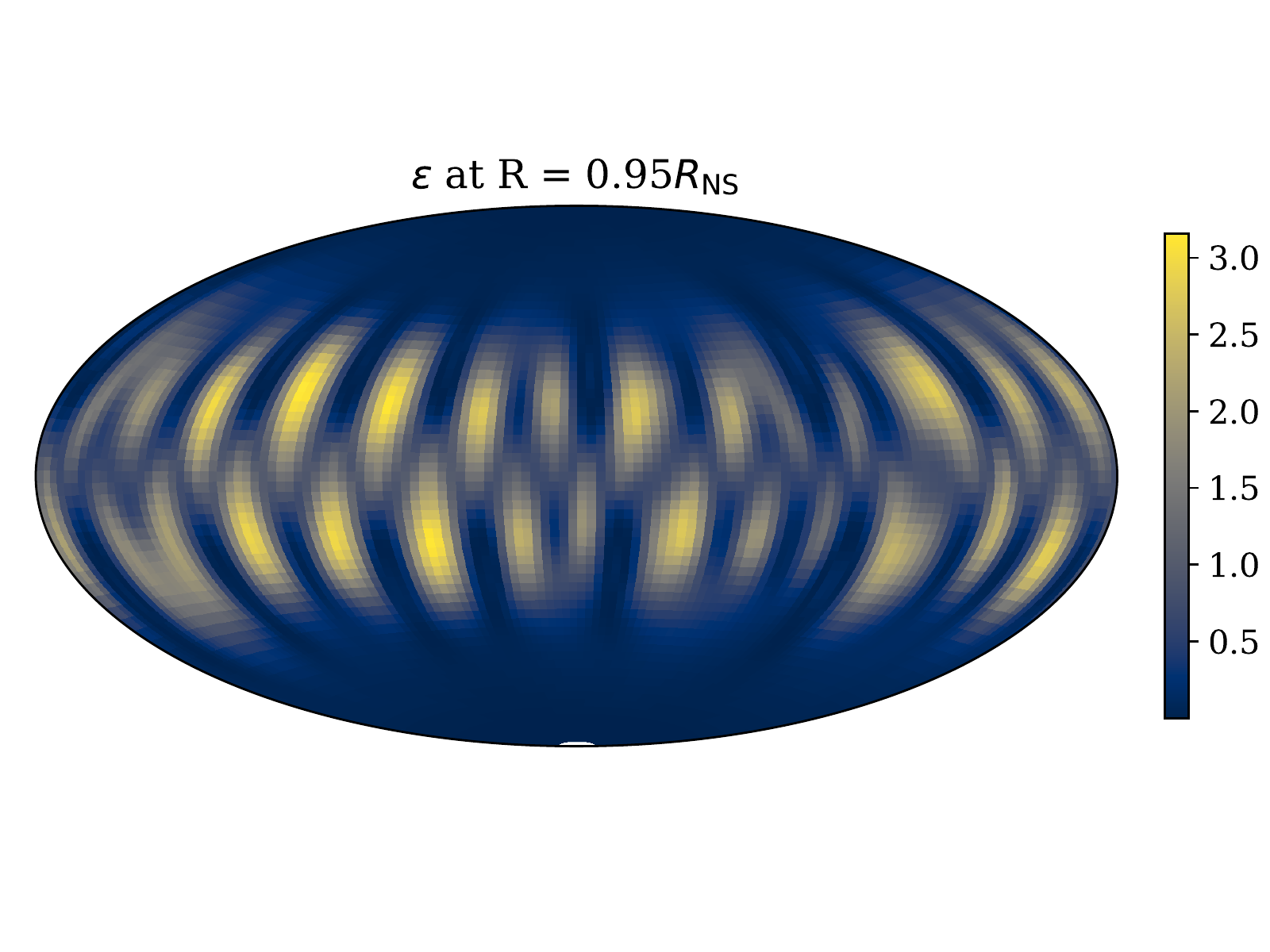}
    \end{minipage}
    \begin{minipage}{0.49\linewidth}
    \includegraphics[width=0.99\columnwidth]{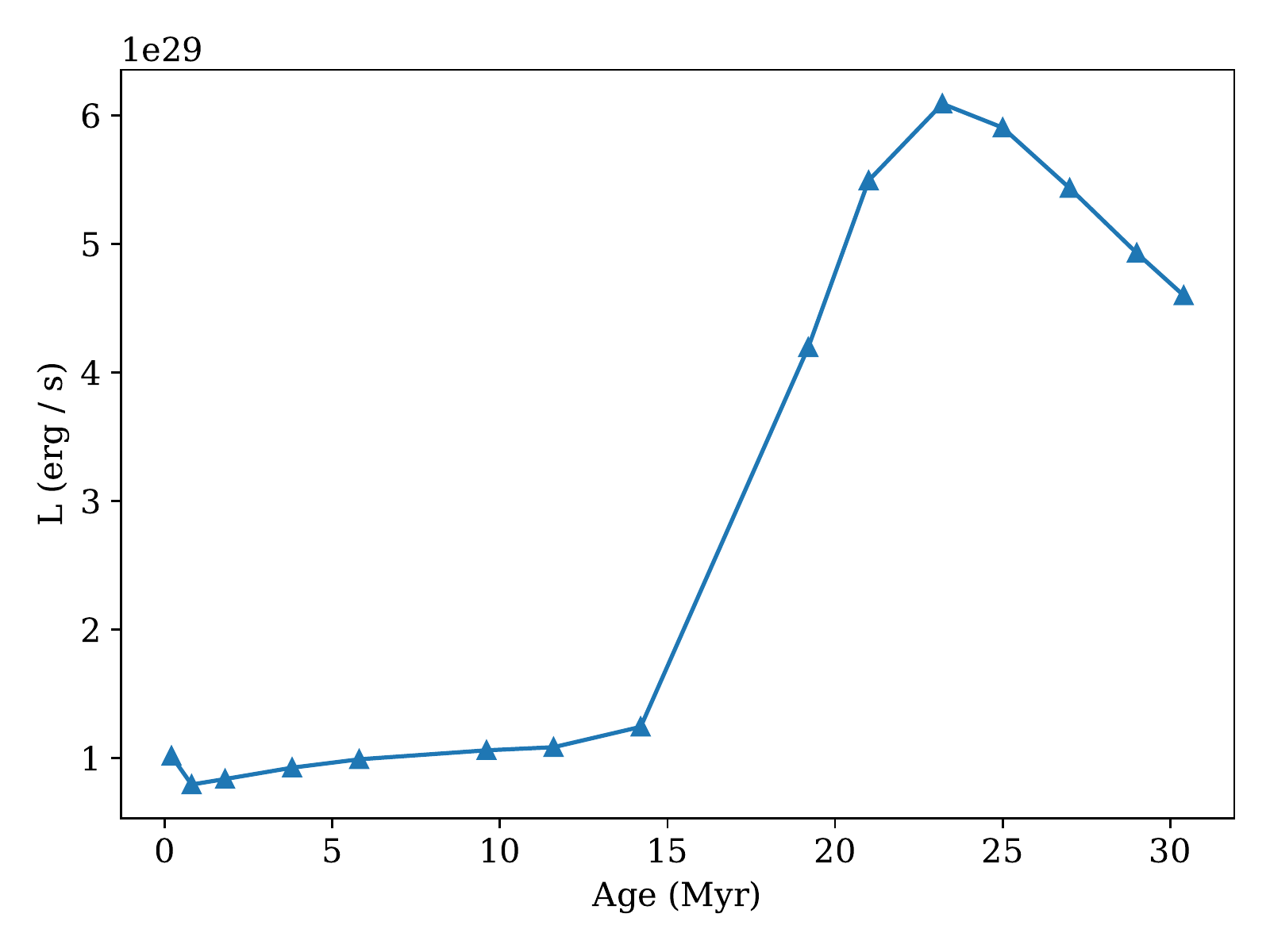}
    \end{minipage}
    \caption{Left panel: energy release in the NS crust rate due to the Ohmic decay at 23~Myr for $r_\mathrm{cut} = 2$ at depth $R = 0.95$~R$_\mathrm{NS}$. Right panel: luminosity as a function of time.}
    \label{fig:eps}
\end{figure*}

The presence of electric currents in the crust lead to its heating. We can estimate the rate of this heating as follows (see e.g. \citealt{DeGrandis2020ApJ}, \citealt{igoshev2021NatAs}):
\begin{equation}
\epsilon = \frac{1}{\sigma} \left[\frac{c}{4\pi} (\curl \vec B) \right]^2 \;\; \mathrm{erg} \; \mathrm{cm}^{-3} \; \mathrm{s}^{-1}.
\end{equation}
In our dimensionless system the energy release rate is:
\begin{equation}
\epsilon = \chi(r) \left[ \curl (\curl \vec A) \right]^2.
\end{equation}
We show this dimensionless quantity in Figure~\ref{fig:eps}. Most of the energy is released in the deep NS crust around $R = 0.95$~R$_\mathrm{NS}$. To convert the dimensionless energy release into cgs units, we use the conversion factor:
\begin{equation}
\epsilon_0^v = \frac{c^2 B_0^2}{16\pi^2 R_\mathrm{NS}^2 \sigma_0},
\label{eq:eps0v}
\end{equation}
which is the volumetric energy release rate. 
Thus each cm$^{3}$ in the deep crust releases up to $\approx 10^{12}$~erg~s$^{-1}$, see Table~\ref{tab:coeff}. If we numerically integrate this energy release over the whole NS we obtain:
\begin{equation}
\epsilon_t = \int \chi(r) \left[ \curl (\curl \vec A) \right]^2 dV.
\end{equation}
To convert this value into cgs we use the value:
\begin{equation}
\epsilon_0 = R_\mathrm{NS}^3  \epsilon_0^v.
\label{eq:eps00}
\end{equation}
We plot the evolution of the energy release rate in Figure~\ref{fig:eps}. During the first 10~Myr the energy release rate stays at the level of $\approx 10^{29}$~erg~s$^{-1}$. Since energy is released in the deep crust, a part of this energy could be emitted as neutrino radiation and cannot be detected. If a significant fraction of this energy reaches the NS surface it allows the NS to stay relatively hot with surface temperature given by:
\begin{equation}
T = \left(\frac{L}{4\pi R_\mathrm{NS}^2 \sigma_B}\right)^{1/4} \approx 10^5\; \mathrm{K}.
\end{equation}
The energy release starts growing after 1~Myr when the instability started developing. The heat release reaches its maximum around 20-25~Myr when the instability reaches its saturation. 

\begin{figure}
    \centering
    \includegraphics[width=0.99\columnwidth]{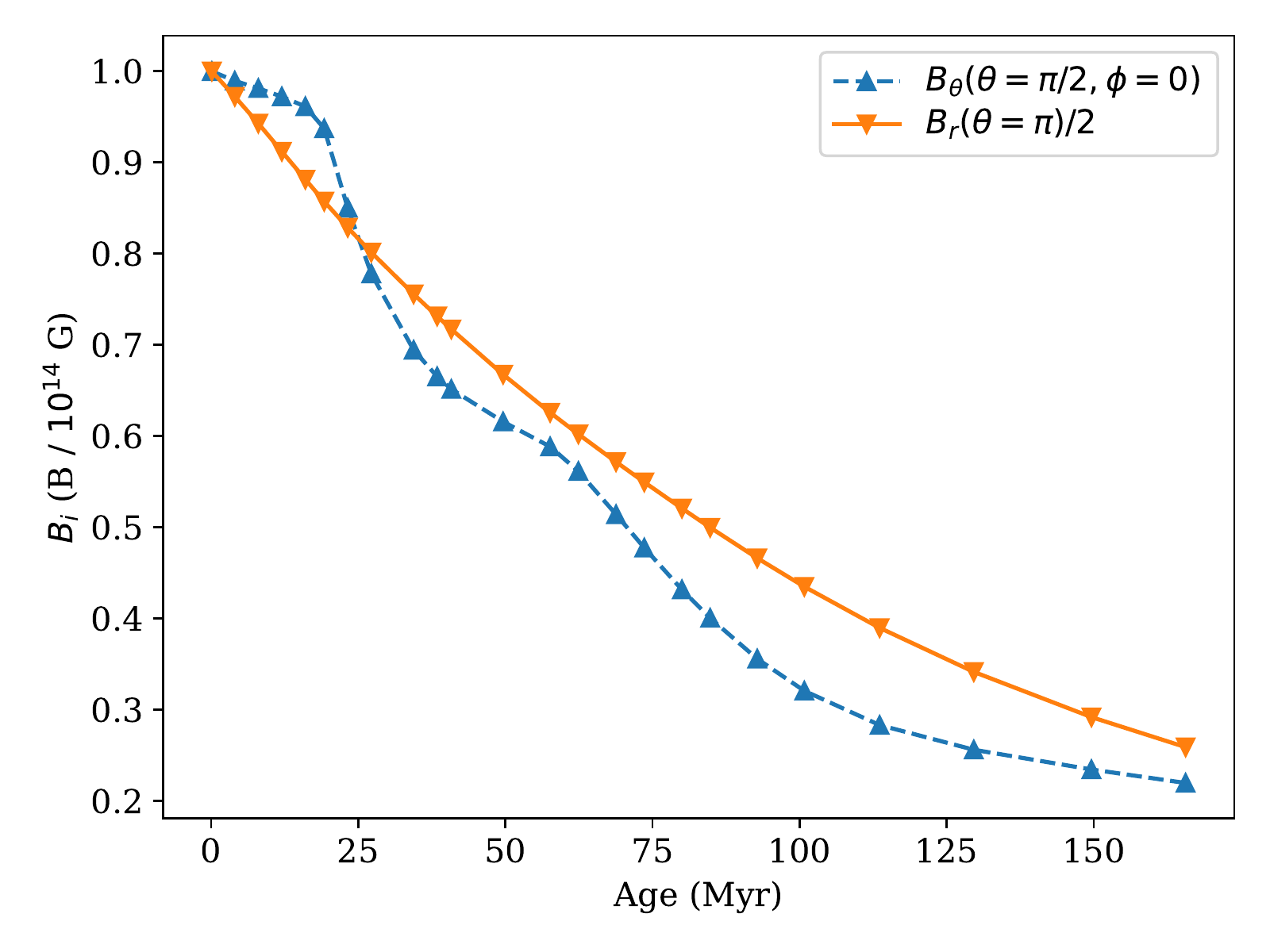}
    \caption{Evolution of surface magnetic field strength as a function of time.
    }
    \label{fig:B_evol}
\end{figure}

Spatially the energy release concentrates toward the magnetic equator where the crust thermal conductivity is limited. Further simulations of magneto-thermal evolution are required to understand what the surface map could look like. In any case the heat is not concentrated toward small-scale structures but instead forms a wide belt around the equator. This thermal emission might thus be detected as the bulk NS emission. It is known that some older NSs have bulk temperatures comparable to $10^5$~K, see e.g. \cite{Mignani2008A,Pavlov2009ApJ}. 

The magnetic energy decays and this energy is released from the system in the form of the deep crust heating which evolves with age. The energy release pattern in the deep crust also has azimuthal angular structure with $m=14$; that is, it is the same as the current structure. Thus, this pattern evolves with time and by 160~Myr simplifies to $m=4$.  

\subsubsection{Structure and evolution of surface magnetic field}

We show the evolution of magnetic field strength at the equator and pole in Figure~\ref{fig:B_evol}. While the surface field at the pole decays with the same rate, the field at the equator is affected by the growth of the small-scale field. Its decay thus proceeds with different rates. Overall, the decay of magnetic field proceeds on a timescale of $\approx 120$~Myr, i.e. on twice the timescale for decay of magnetic energy ($E\propto B^2$), and on the timescale of six ambipolar diffusion timescales estimated based on numerical velocity. We see no indications that magnetic field decay stops. It is possible to extrapolate that in our particular setup we could suggest that a magnetar-strength field decays to values of $10^8$~G on a timescale of 1.1~Gyr under the influence of ambipolar diffusion.

\begin{figure*}
    \centering
    \begin{minipage}{0.49\linewidth}
    \includegraphics[width=0.99\columnwidth]{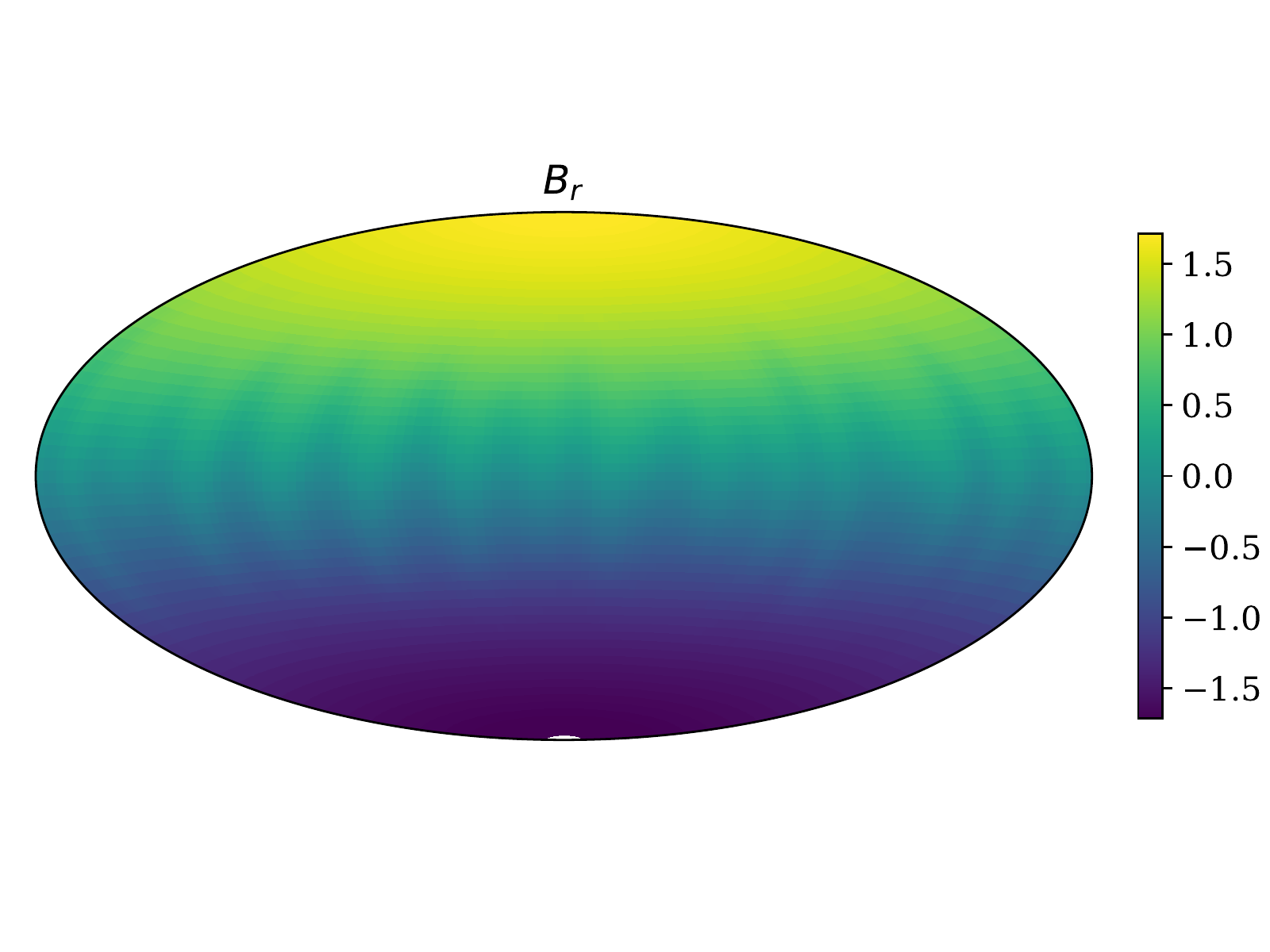}
    \end{minipage}
    \begin{minipage}{0.49\linewidth}
    \includegraphics[width=0.99\columnwidth]{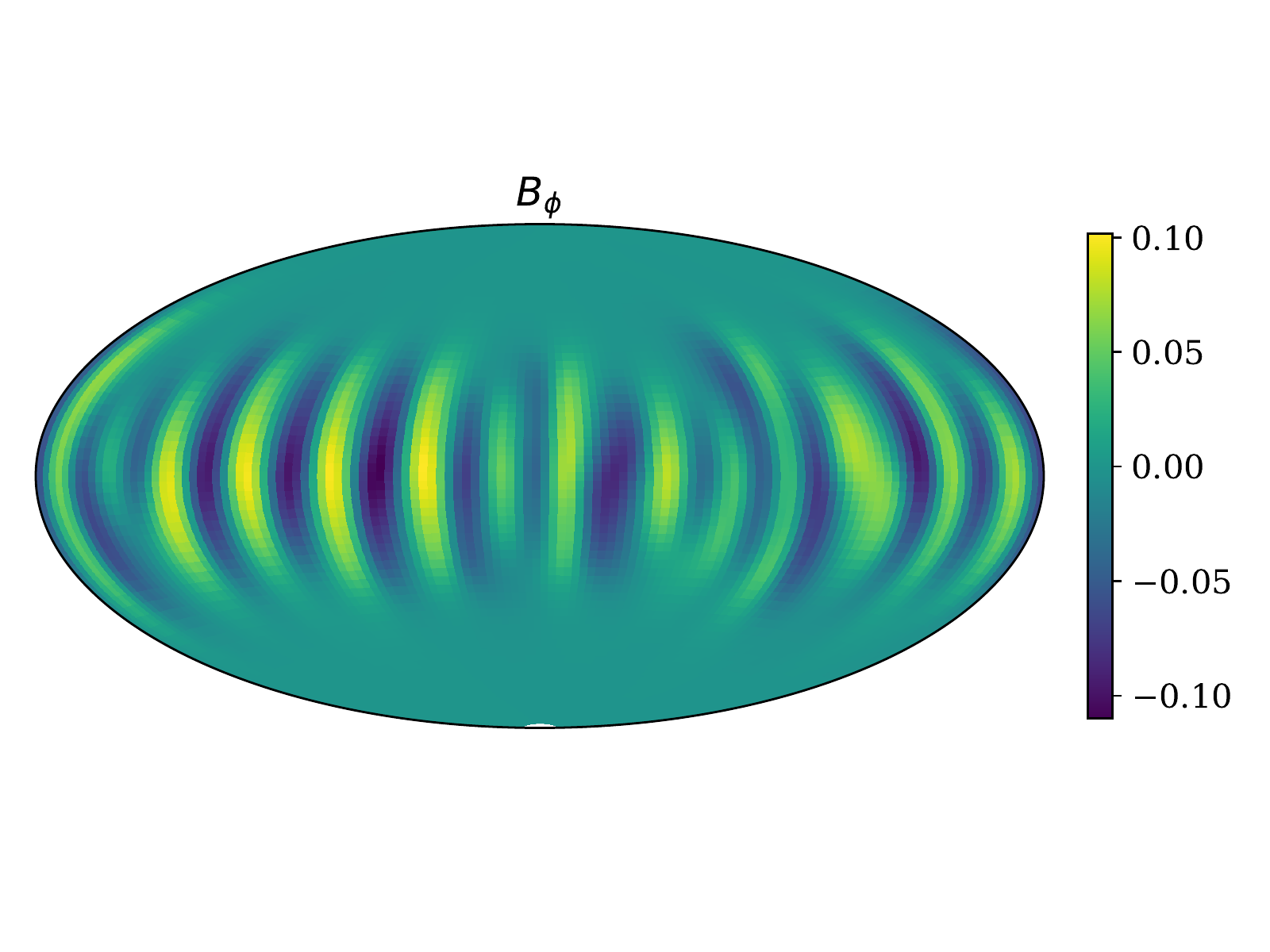}
    \end{minipage}
    \begin{minipage}{0.49\linewidth}
    \includegraphics[width=0.99\columnwidth]{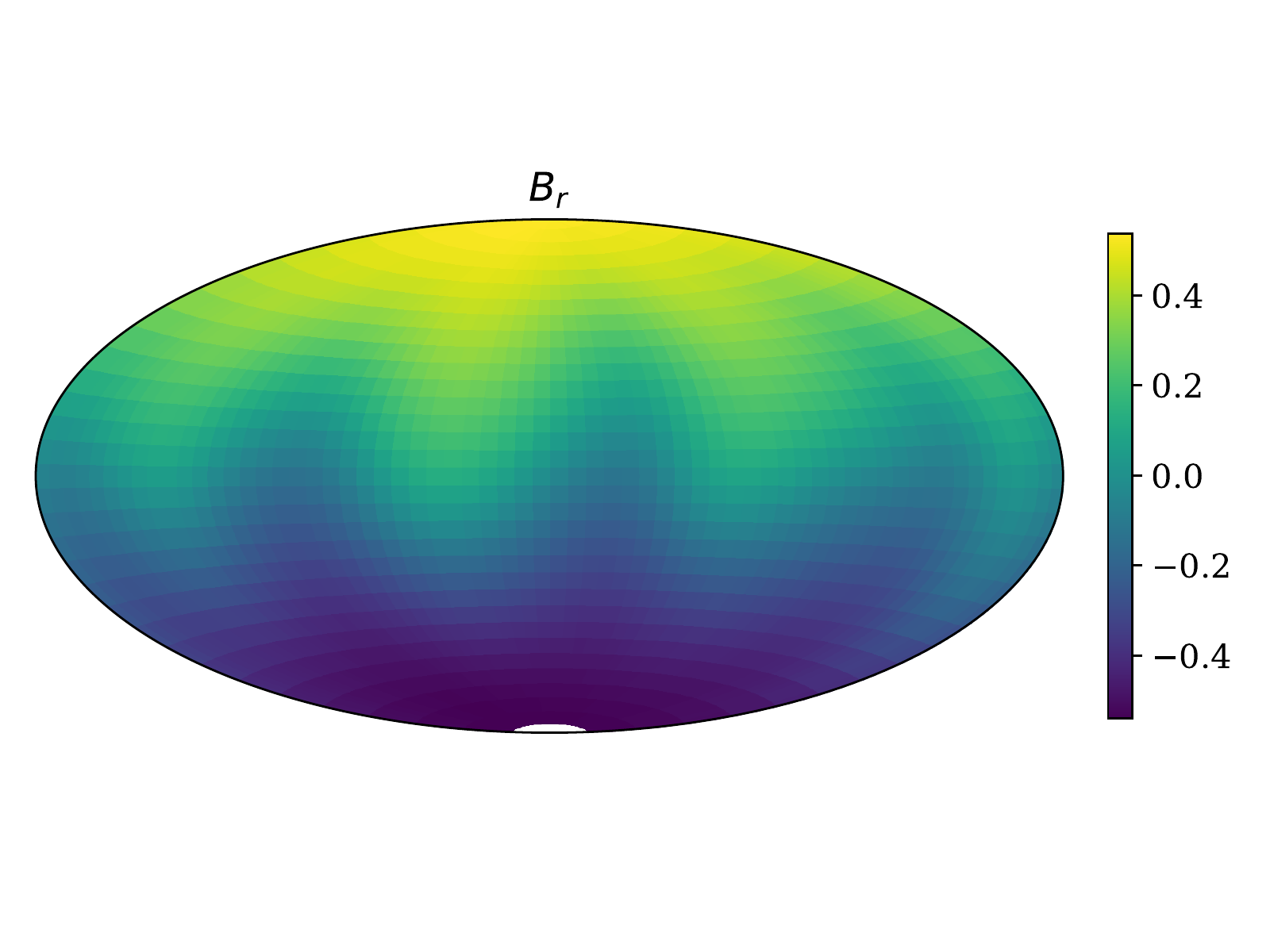}
    \end{minipage}
    \begin{minipage}{0.49\linewidth}
    \includegraphics[width=0.99\columnwidth]{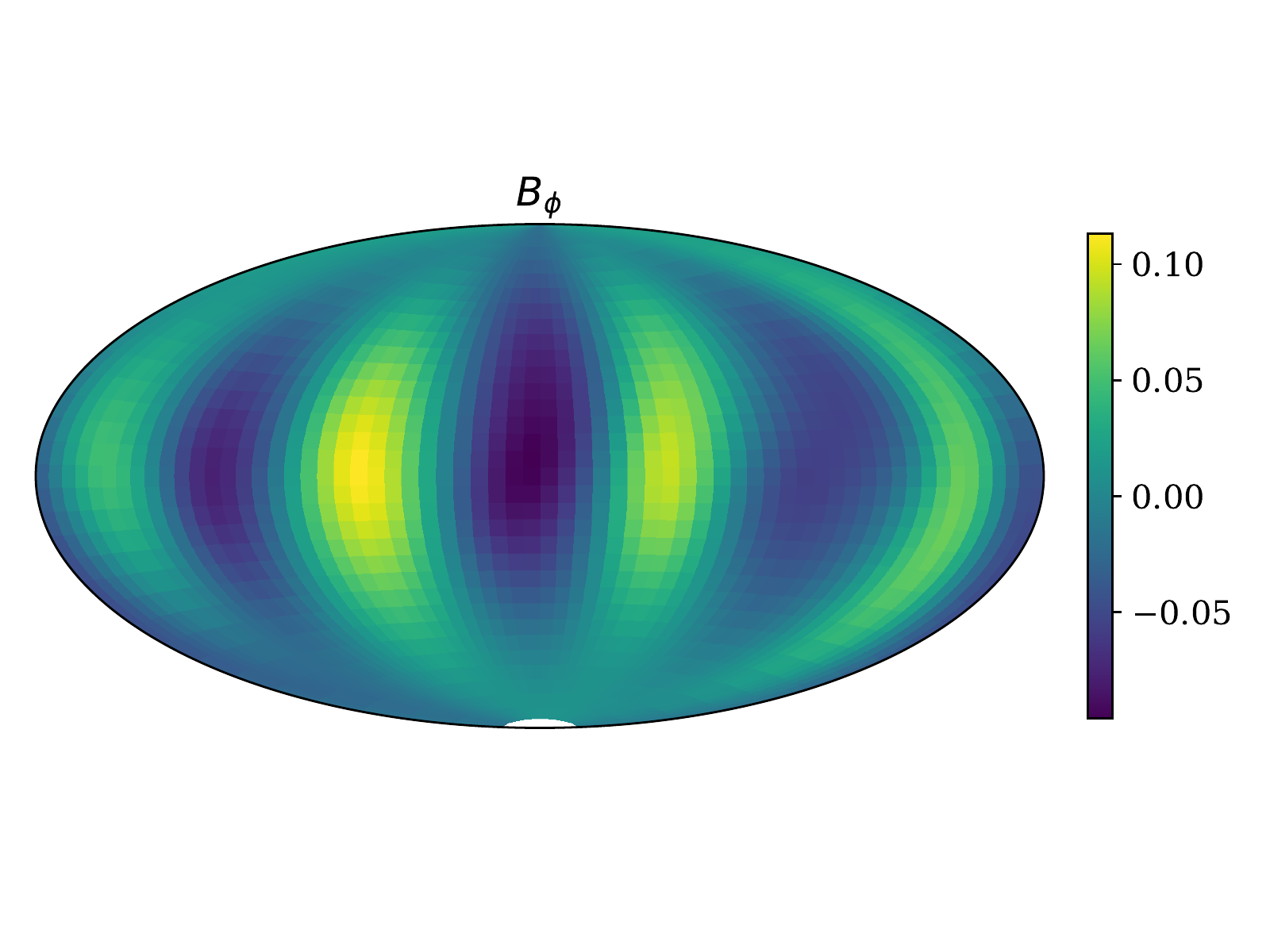}
    \end{minipage}
    \caption{Surface magnetic field for $r_\mathrm{cut} = 2$ at dimensionless time 21~Myr (top row) and at 161~Myr (bottom row).  }
    \label{fig:surface_field_map}
\end{figure*}

Small-scale magnetic field with $m=14$ emerges to the NS surface on a timescale of $2$~Myr. This structure is most noticeable in the $B_\phi$ component, which was absent in our initial conditions.  
In Figure~\ref{fig:surface_field_map} we show $B_r$ and $B_\phi$ components of magnetic field. This component has filaments stretching in the north-south direction. The $B_\phi$ component has a clearer $m=14$ structure. The surface pattern evolves and forms $m=4$ by 160~Myr. As is clear from the plots, the dipolar component stays dominant even at these long timescales, although the degree of dominance falls from 15 times to 4 times.

Unexpected small-scale magnetic fields were discovered in millisecond radio pulsars, see e.g. \cite{Bilous2019ApJ}. Our results indicate that ambipolar diffusion could give rise to higher order multipoles in old neutron stars.


\subsubsection{Crust failure}

\begin{figure}
    \centering
    \includegraphics[width=0.99\columnwidth]{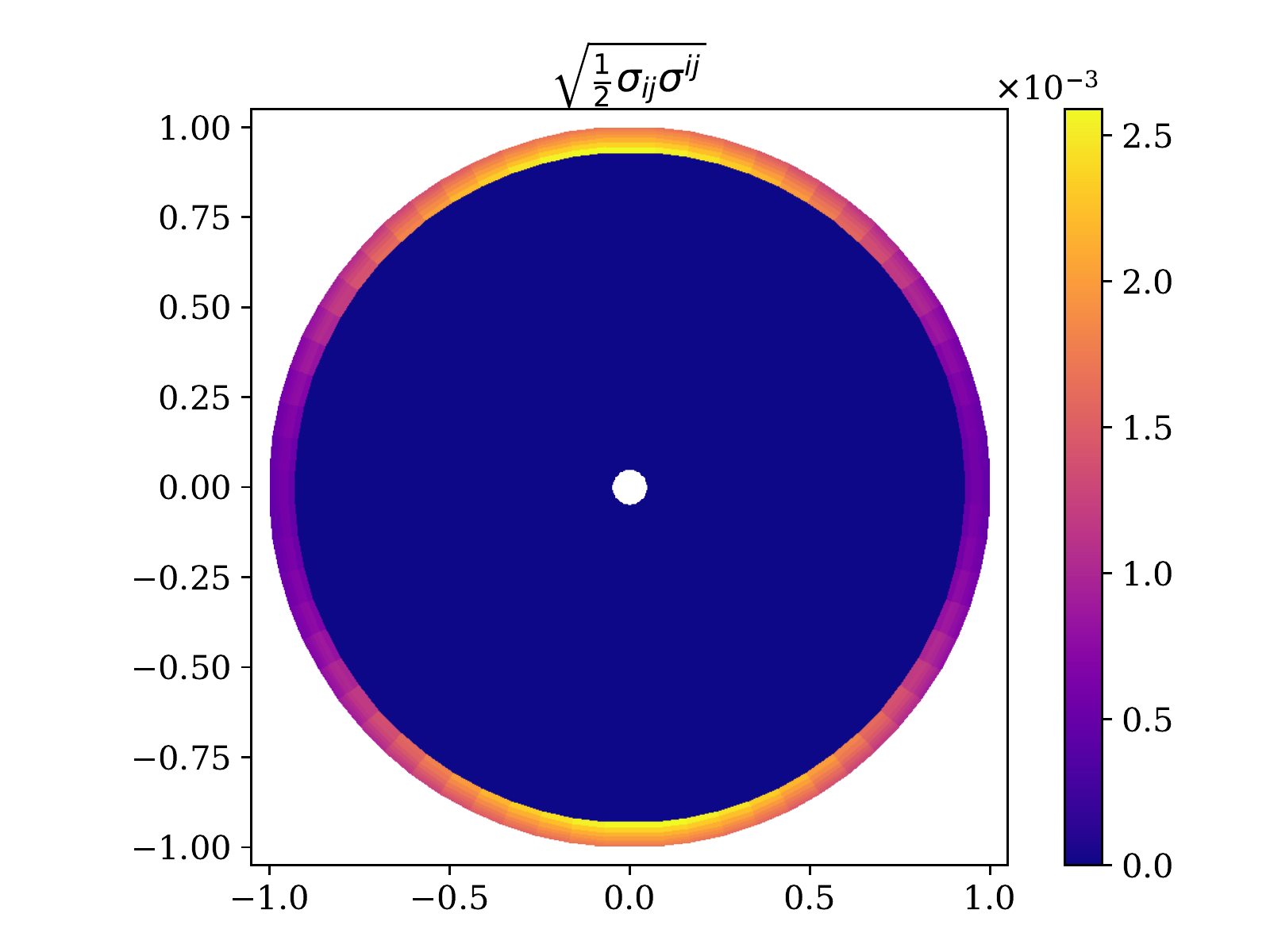}
    \caption{NS crust strain after 167~Myr, shown in a meridional cut. }
    \label{fig:sigma_ij}
\end{figure}

The electric current formed as a result of ambipolar diffusion could lead to crust failure. To check if it is the case we compute the elastic strain tensor $\hat \sigma$ following the prescription of \citep{Lander2015MNRAS,Gourgouliatos2022Symm}:
\begin{equation}
\sqrt{\frac{1}{2} \sigma_{ij} \sigma^{ij}} = \frac{1}{8\pi\mu_\mathrm{shear}} \sqrt{B^2B_0^2 + \frac{3}{2}B^4 + \frac{3}{2}B_0^4 - 4 (\vec B\cdot \vec B_0)^2},
\label{eq:sigma_ij}
\end{equation}
where we used the Einstein summation rule, and $\mu_\mathrm{shear}$ is the shear modulus of the NS crust, assumed to be $10^{30}$~dyn~cm$^{-2}$ \citep{Ruderman1969Natur}. More modern estimates for the shear modulus close to the core-crust boundary are $1.8\times 10^{30}$~dyn~cm$^{-2}$ \citep{Hoffman2012MNRAS}. In eq. (\ref{eq:sigma_ij}), $\vec B_0$ stands for the initial magnetic field configuration when the crust froze. The magnetic field $\vec B$ is the instantaneous magnetic field. 
The NS crust fails according to the von Mises criterion \citep{vonMises1913} when:
\begin{equation}
\sqrt{\frac{1}{2} \sigma_{ij} \sigma^{ij}} > 0.1,
\end{equation}
where $0.1$ is the maximum breaking strain. We plot the value of NS crust strain in Figure~\ref{fig:sigma_ij}. The maximum value after 15~Myr of evolution is $2.5\times 10^{-3}$ which is not enough to break the crust.

\section{Conclusions}

In this work we used the \texttt{Dedalus} code to study ambipolar diffusion in neutron star cores. Unlike the previous analysis in one and two dimensions, we integrate the equations in three dimensions using the spherical coordinate system. We also include the neutron star crust with finite conductivity. 

Our work has following caveats:

(1) A neutron star core is expected to be in a superconducting and superfluid state, which will probably significantly affect the magnetic field evolution described here. Despite the significant progress recently reached in investigations of how neutron vortices and magnetic flux tubes interact with each other, the detailed equations describing the evolution and its importance is a matter of active scientific debate. Therefore, we refrain from implementing the superfluidity and superconductivity at the moment.

(2) We assume negligible baryon velocity, which is not the case at the beginning of the simulations. The main expected impact of this assumption is that we underestimated the speed of ambipolar diffusion, which might be a factor of a few times faster than in our simulations. Because of this assumption we are able to write equations in the one-fluid approximation. In future work we plan to implement the two-fluid approximation.

(3) Given limitations in computing power we had to restrict radial profiles for coefficients $\xi_3$ and $\xi_4$ by introducing the parameter $r_\mathrm{cut} = 2$. Under realistic conditions we see the formation of a compact current sheet at the crust-core boundary in the equatorial plane with radial extent smaller than $50$~m, which corresponds to our finest resolution.

Given these caveats we discovered the instability of pure poloidal axisymmetric magnetic field under influence of ambipolar diffusion in weak coupling mode. This instability leads to development of wave-like $B_\phi$ (kind of toroidal component) which is composed of harmonics with $l=21,\ 25,\ 28$ with $m=10,\ 12,\ 14$. This well-resolved cluster of harmonics grows from initial perturbations by four orders of magnitude over the first 15~Myr (1.5 dimensionless times for $B_0 = 10^{14}$~G). The growth of instability is exponential with a typical timescale of 2~Myr (0.2 dimensionless time). The azimuthal magnetic field reaches saturation around 20~Myr. The instability induces strong electric current in the NS crust and leads to exponential decay of magnetic energy on a timescale of 60~Myr in our setup with initial $B_r = 10^{14}$~G at the pole.

Our work has the following potential astrophysical implications:
\begin{itemize}
\item We found that ambipolar diffusion creates electric currents in the deep crust and allows energy release at the level of $10^{29}$~erg~s$^{-1}$ on a 10~Myr timescale. Thus a NS could stay relatively hot with temperatures of $\approx 10^5$~K for millions of years if it had a strong initial magnetic field $\approx 10^{14}$~G.
NSs with these temperatures were discovered in the past using optical, UV and X-ray telescopes. Future missions such as the Large UV/Optical/IR Surveyor (LUVOIR; \citealt{Luvoir2019arXiv}) as well as the next generation of X-ray telescopes such as Strobe-X \citep{StrobeX2019arXiv} could be used to measure surface temperatures for large number of old neutron stars and confirm or reject our numerical results. More work is required to produce reliable surface maps which will be possible to compare with UV and soft X-ray lightcurves.  
\item In our simulations, the dipolar component of magnetic field decays on a timescale of 120~Myr, which is expected to be sensitive to the conductivity of the deep crust. Further numerical simulations are required to establish a firm relationship between decay timescale, initial magnetic field strength and configuration, and conductivity of the deep crust. Ultimately, these decay timescales will be used in pulsar population synthesis to decode evolutionary relations between different classes of neutron stars (such as magnetars, central compact objects, radio pulsars and dim isolated X-ray sources). 
\item The instability leads to development of azimuthal magnetic field with initial wavenumber $m=14$ which merges with time and simplifies its structure reaching $m=4$ by 160~Myr. Many old radio pulsars continue to operate below the classical death line for dipolar magnetic field \citep{Medin2007MNRAS}. If ambipolar diffusion operates in these stars, it could be an important mechanism to increase the curvature of open field lines near the crust and facilitate the pair production allowing a NS to shine as a radio pulsar. 
\item Ambipolar diffusion does not seem to cause any crust failure for magnetic field $10^{14}$~G.
\end{itemize}

\section*{Acknowledgements}
A.I.P. thanks Dr.~Girish Nivarti, Dr.~Anna Guseva and Dr.~Calum Skene for multiple fruitful discussions. A.I.P. is very grateful to the \texttt{Dedalus} developers for fast and helpful support. This work was supported by STFC grant no.\ ST/W000873/1, and was undertaken on ARC4, part of the High Performance Computing facilities at the University of Leeds, UK.

\section*{Data Availability}

The data underlying this article will be shared on reasonable request to the corresponding author.



\bibliographystyle{mnras}
\bibliography{example} 




\appendix

\section{Derivation of the equation for chemical equilibrium deviation}
\label{s:derivation_delta_mu}
We start with the same system of equations as \cite{Passamonti2017MNRAS}:

\begin{align*}
-\vec \nabla \mu_p - m_p^* \vec \nabla \Phi + e \left(\vec E +\frac{\vec v_p}{c}\times \vec B\right) &= \frac{m_p^* \vec w_{pn}}{\tau_{pn}} + \frac{m_p^*\vec w_{pe}}{\tau_{pe}}  ,\\
-\vec \nabla \mu_e - m_e^* \vec \nabla \Phi - e \left(\vec E + \frac{\vec v_e}{c}\times \vec B\right) &= \frac{m_e^* \vec w_{en}}{\tau_{en}} + \frac{m_e^*\vec w_{ep}}{\tau_{ep}}  ,\\
-\vec \nabla \mu_n - m_n^* \vec \nabla \Phi &=  \frac{m_n^* \vec w_{np}}{\tau_{np}} + \frac{m_n^*\vec w_{ne}}{\tau_{ne}},
\end{align*}    
where $\mu_p, \mu_e$ and $\mu_n$ are chemical potentials for protons, electrons and neutrons, $\Phi$ is the gravitational potential, $m_p^*$, $m_n^*$ and $m_e^*$ are effective masses of proton, neutron and electron respectively. Absolute velocities for different species are $\vec v_p$ and $\vec v_e$, while relative velocities between species are $\vec w_{pe} = \vec v_p - \vec v_e$. Here $\tau_{pn}$ are relaxation times for collisions between protons and neutrons.
This system contains one more equation in comparison to \cite{Goldreich1992} for motion of neutrons which are not fixed.

We add the first two equations and subtract the third equation:
\begin{equation}
-\nabla (\Delta \mu) - \vec \nabla \Phi (m_p^* + m_e^* - m_n^*) + \frac{\vec j \times \vec B}{c n_c } = \frac{m_p^* \vec w_{pn} }{\tau_{pn}} - \frac{m_n^* \vec w_{np} }{\tau_{np}}.
\end{equation}
In this equation we combine $\Delta \mu = \mu_p + \mu_e - \mu_n$.
The right-hand side does not contain any terms with $\vec w_{pe}$ because of conservation of momentum, so $n_p m_p^* / \tau_\mathrm{pe} = n_e m_e^* / \tau_\mathrm{ep}$ and $\vec w_{pe} = - \vec w_{ep}$, and electroneutrality $n_e \approx n_p = \nnc$. Following \cite{Passamonti2017MNRAS} we assume that electron-neutron interactions are much weaker in comparison to proton-neutron interactions which are mediated by the strong force. That is why we neglected terms with $\tau_\mathrm{en}$ and $\tau_\mathrm{ne}$. We also assume that contribution of electrons to NS mass is negligible, i.e. $m_p^* + m_e^* - m_n^* \approx 0$.
We combine the terms on the right as follows:
\begin{equation}
\frac{m_p^* \vec w_{pn} }{\tau_{pn}} - \frac{m_n^* \vec w_{np} }{\tau_{np}}  =  \frac{m_p^* \vec w_{pn} }{\tau_{pn}} + \frac{n_p}{n_n}\frac{m_p^* \vec w_{pn} }{\tau_{pn}} = \frac{m_p^* \vec w_{pn} }{x_n\tau_{pn}},
\end{equation}
where $x_n = n_n / (n_p + n_n)$. Overall, at this stage we have the following equation:
\begin{equation}
-\nabla (\Delta \mu)  + \frac{\vec f_B}{n_c} = \frac{m_p^* \vec w_{pn} }{x_n\tau_{pn}}, 
\label{e:nabla_delta_mu}
\end{equation}
where we define:
\begin{equation}
\vec f_B = \frac{\vec j \times \vec B}{c} = \frac{1}{4\pi} (\vec \nabla \times \vec B) \times \vec B.
\end{equation}
We take the divergence of eq. (\ref{e:nabla_delta_mu}) and multiply by $(-1)$:
\begin{equation}
\vec \nabla^2 (\Delta \mu) = \vec \nabla \cdot \left(\frac{\vec f_B}{n_c}\right) - \vec \nabla \cdot \left( \frac{m_p^* \vec w_{pn} }{x_n\tau_{pn}}  \right).
\label{e:subst}
\end{equation}
We expand the last term on the right, multiplying numerator and denominator by $\nnc$:
\begin{equation}
\vec \nabla \cdot \left( \frac{m_p^* \vec w_{pn} }{x_n\tau_{pn}}  \right) =  \frac{m_p^*  }{x_n n_c \tau_{pn}} \vec \nabla \cdot (n_c \vec w_{pn}) + n_c \vec w_{pn} \cdot \vec\nabla \left( \frac{m_p^*  }{x_n n_c \tau_{pn}} \right).
\label{e:b6}
\end{equation}
Following the assumption by \cite{Passamonti2017MNRAS} we similarly assume:
\begin{equation}
\vec \nabla \cdot (n_c \vec w_{pn}) = - \frac{\lambda \Delta \mu}{x_n}.
\end{equation}
Further we substitute $n_c \vec w_{pn}$ from eq. (\ref{e:nabla_delta_mu}) into eq. (\ref{e:b6}):
\begin{equation}
n_c \vec w_{pn} = \frac{x_n \tau_{pn} n_c}{m_p^*} \left(-\vec \nabla (\Delta \mu) + \frac{\vec f_B}{n_c}\right).
\end{equation}
Thus, the final equation is written as:
\begin{equation}
\vec \nabla^2 (\Delta \mu) - \frac{m_\mathrm{p}^* \lambda}{x_n^2 \nnc \taupn} \Delta \mu = \vec \nabla \cdot \left( \frac{\vec f_B}{\nnc} \right) - \frac{x_n \taupn \nnc}{m_\mathrm{p}^*} \left(-\vec \nabla (\Delta \mu) + \frac{\vec f_B}{\nnc}\right) \vec \nabla \cdot \left(\frac{m_\mathrm{p}^*}{x_n \nnc \taupn}\right).
\label{eq:delta_mu_final}
\end{equation}

\section{Verification of the code and choice of numerical resolution}
\label{appendix:code_verification}

Comparing our short simulations for a range of temperatures (see Figure~\ref{fig:v_amb_temp}) we notice that while the solution for $\Delta \mu$ looks very similar to figure 3 in \cite{Passamonti2017MNRAS}, our amplitude is approximately three times larger. The exact reason for this difference is unknown. Our guess is that the difference appears because we normalise the equations differently. In the absence of open-source code used by \cite{Passamonti2017MNRAS} the difference is nearly impossible to track.  Despite this difference in amplitude we successfully reproduce the ambipolar velocity speeds and its patterns. In all our simulations $\Delta \mu / \mu_0 \ll 1$ which justifies application of linear approximation for reaction rates. With the cooling of the NS when temperature drops from $T_9 = 1$ to $T_9 = 0.1$ the velocity pattern transforms from irrotational-dominated flow to solenoidal-dominated flow in agreement with \cite{Passamonti2017MNRAS}. We successfully reproduce the location of zeros in this flow pattern.

We notice that radial profiles for $\xi_3$ and $\xi_4$ span many orders of magnitude from NS centre to the core-crust boundary. \cite{Passamonti2017MNRAS} remarked that parameter $b$, see eq. (\ref{e:b}), is measured in km and decays towards the core-crust interface reaching values around 200~m. Although our resolution is sufficient to resolve structures with size $\approx 50$~m in radial direction at the core-crust boundary, we do not seem to resolve the process completely. This is the motivation for introducing the parameter $r_\mathrm{cut}$ which restricts the maximum value reached by $\xi_3$ and $\xi_4$. 

\begin{figure*}
    \centering
    \begin{minipage}{0.42\linewidth}
    \includegraphics[width=0.99\columnwidth]{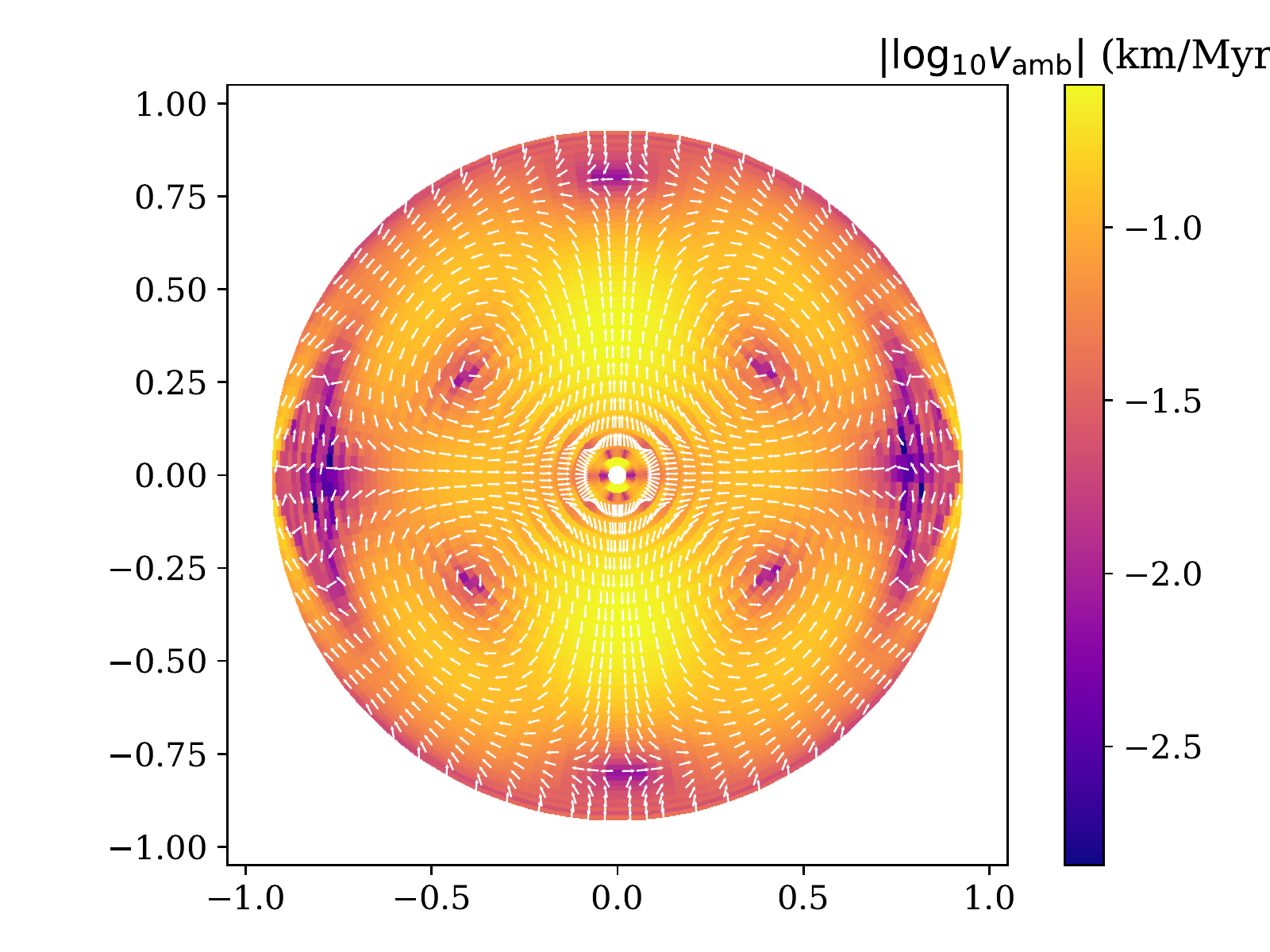}
    \end{minipage}
    \begin{minipage}{0.42\linewidth}
    \includegraphics[width=0.99\columnwidth]{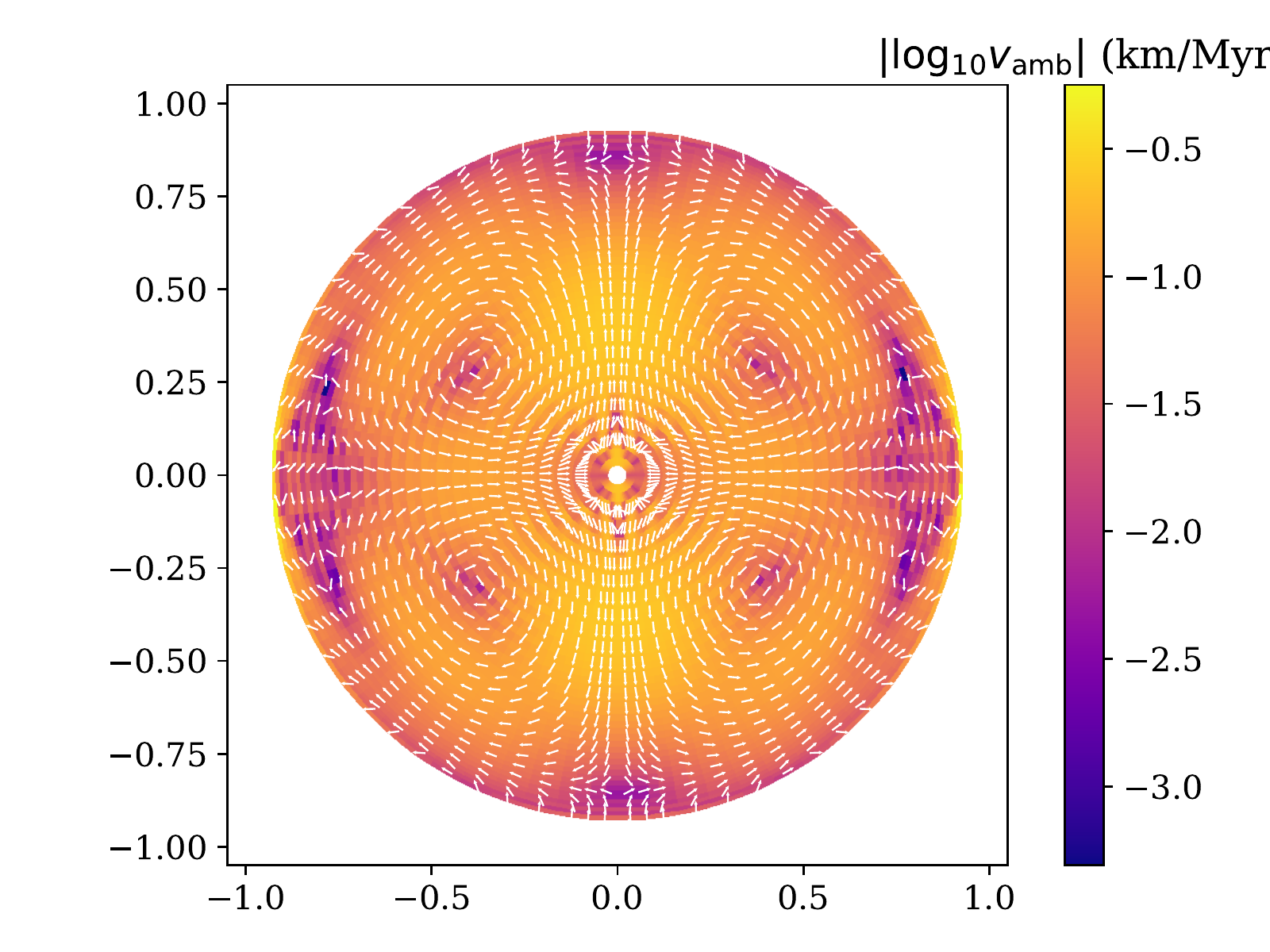}
    \end{minipage}
    \begin{minipage}{0.42\linewidth}
    \includegraphics[width=0.99\columnwidth]{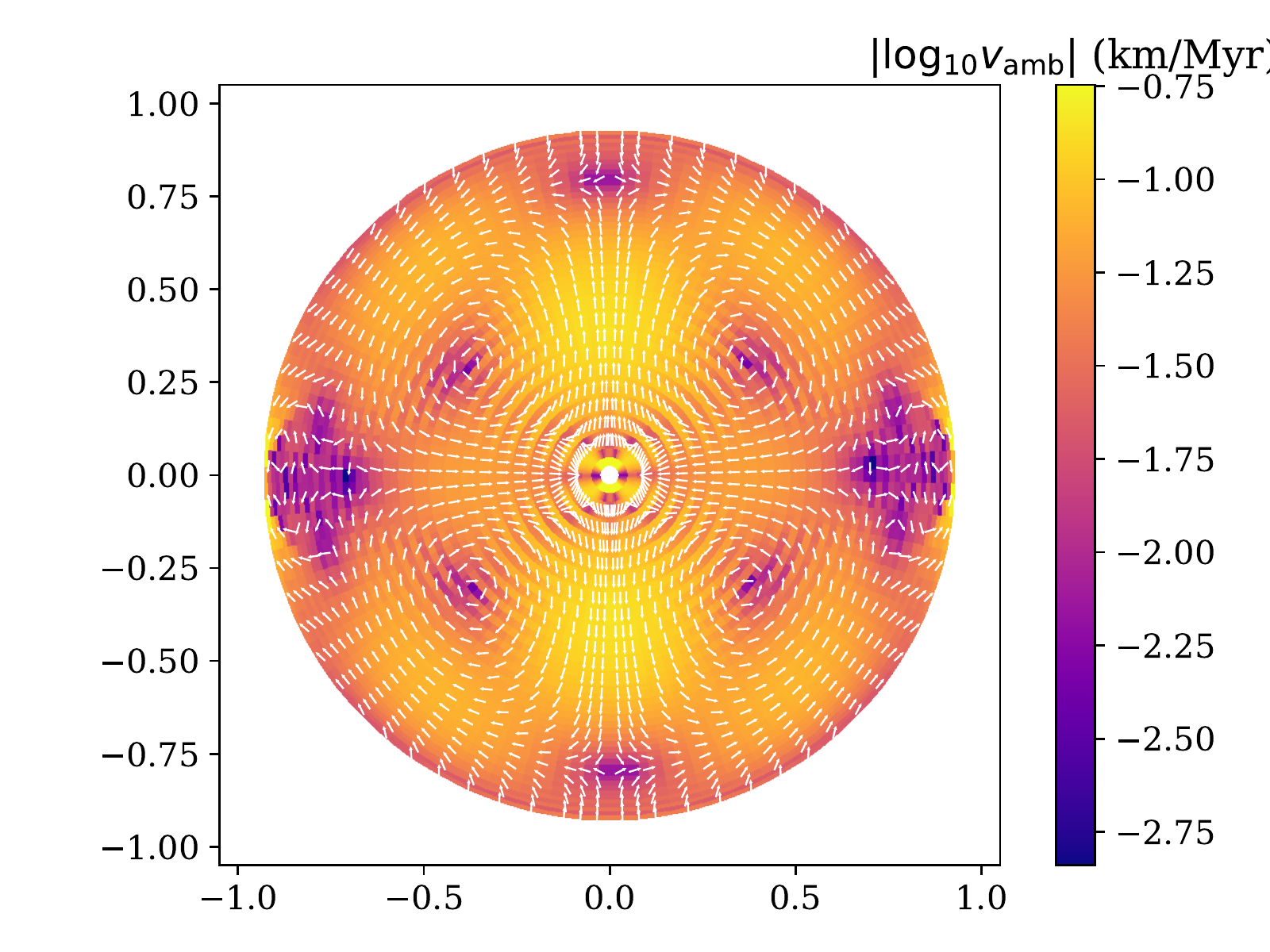}
    \end{minipage}
    \begin{minipage}{0.42\linewidth}
    \includegraphics[width=0.99\columnwidth]{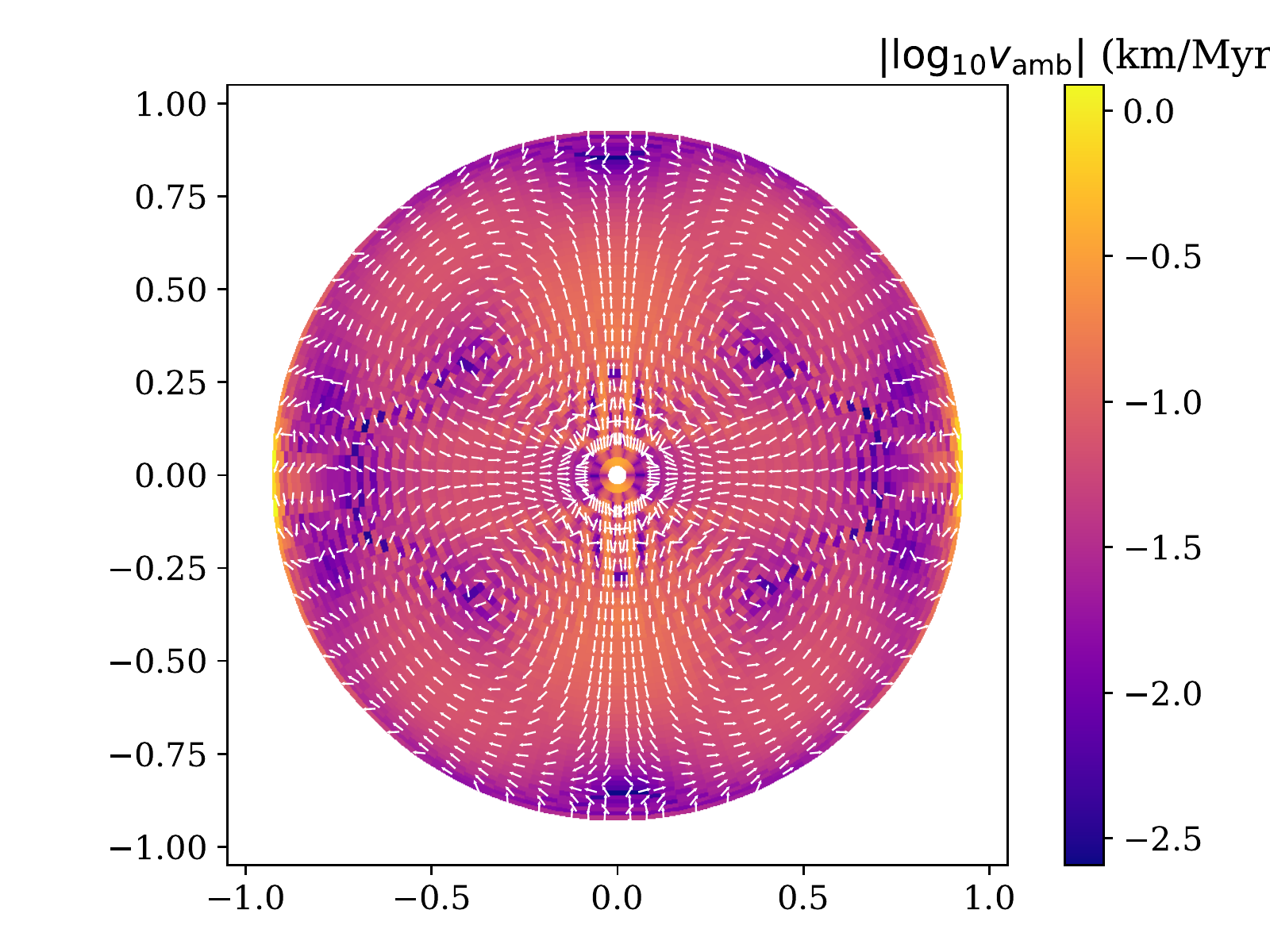}
    \end{minipage}
    \begin{minipage}{0.42\linewidth}
    \includegraphics[width=0.99\columnwidth]{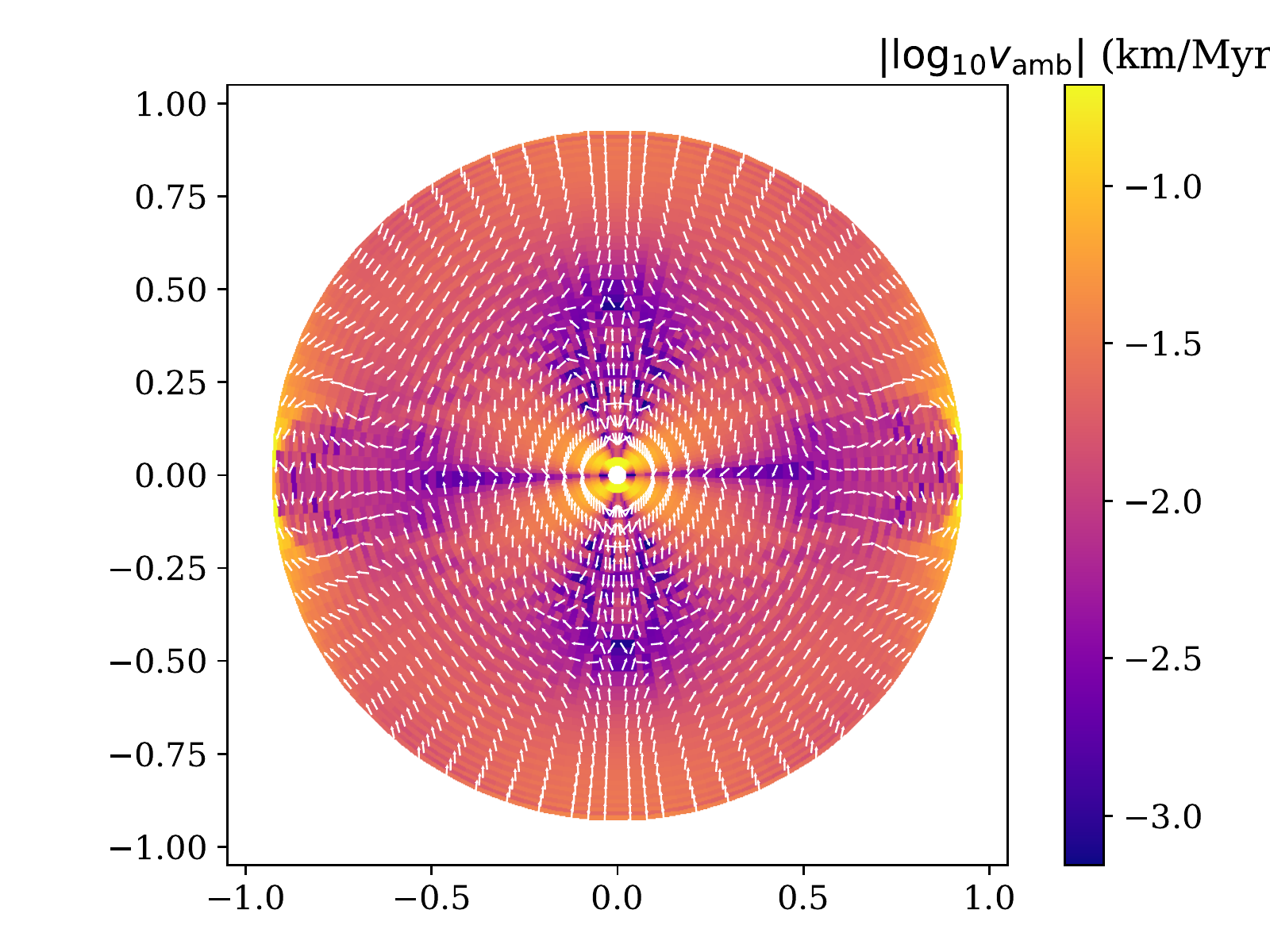}
    \end{minipage}
    \begin{minipage}{0.42\linewidth}
    \includegraphics[width=0.99\columnwidth]{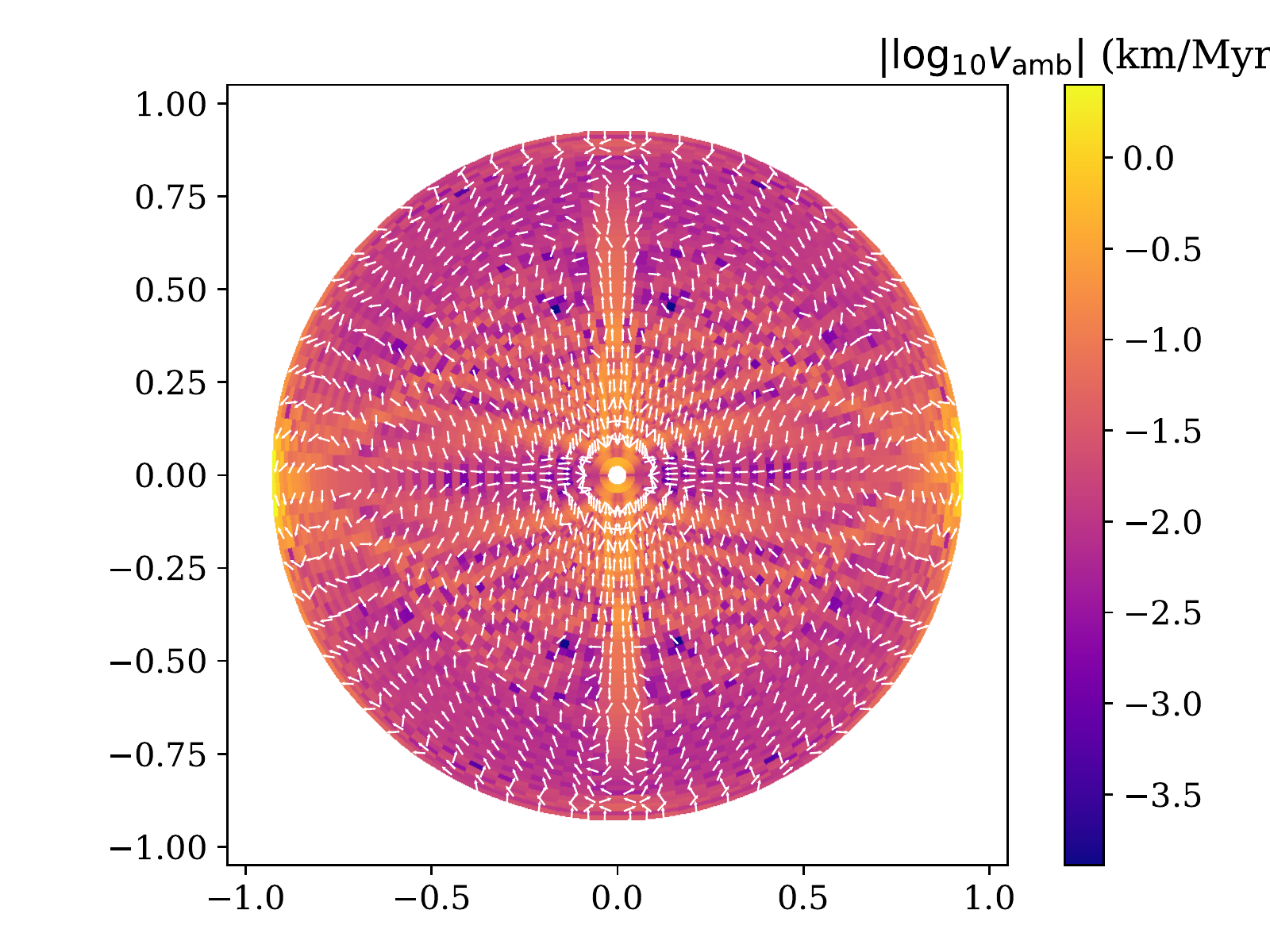}
    \end{minipage}
    \begin{minipage}{0.42\linewidth}
    \includegraphics[width=0.99\columnwidth]{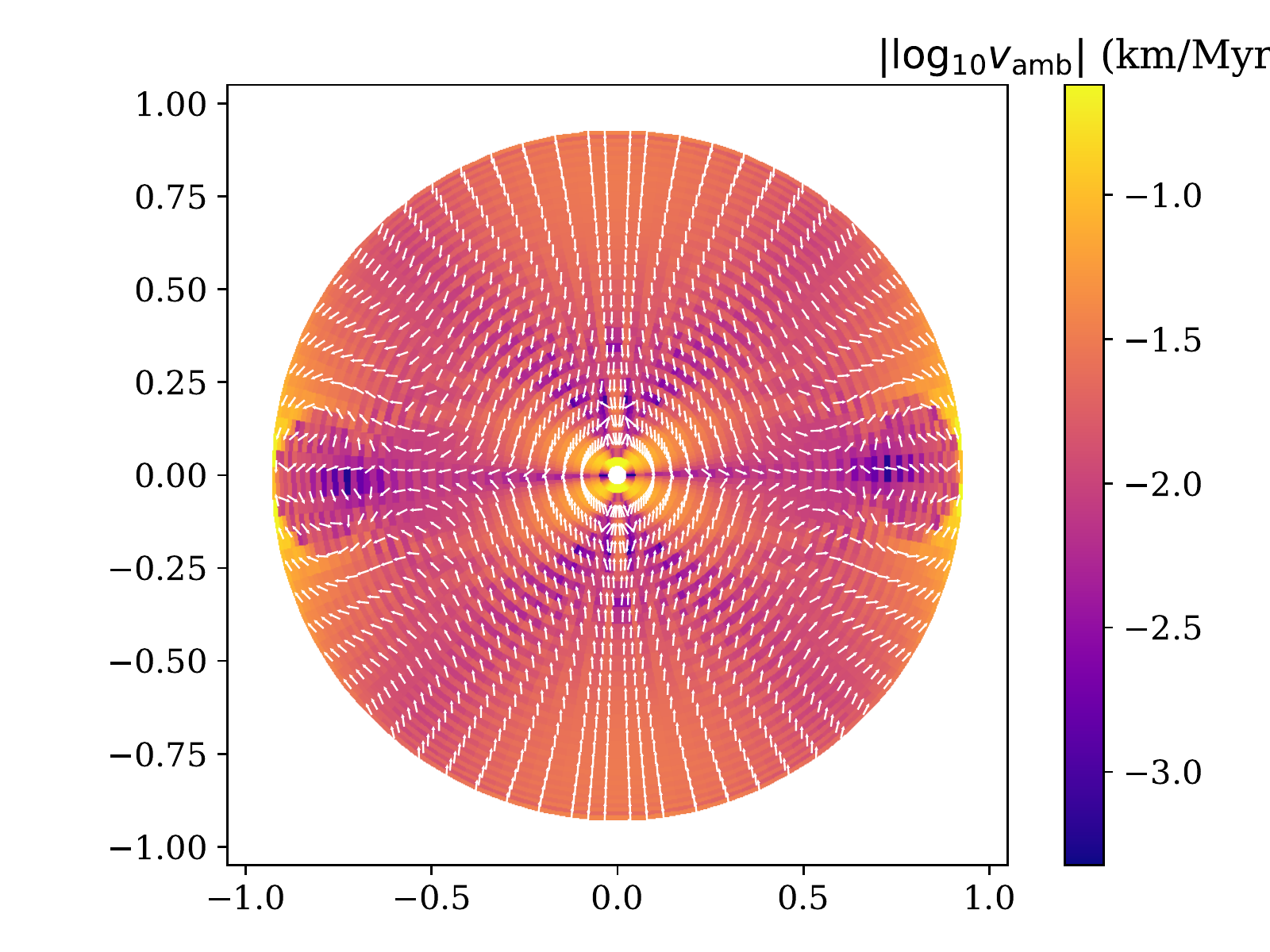}
    \end{minipage}
    \begin{minipage}{0.42\linewidth}
    \includegraphics[width=0.99\columnwidth]{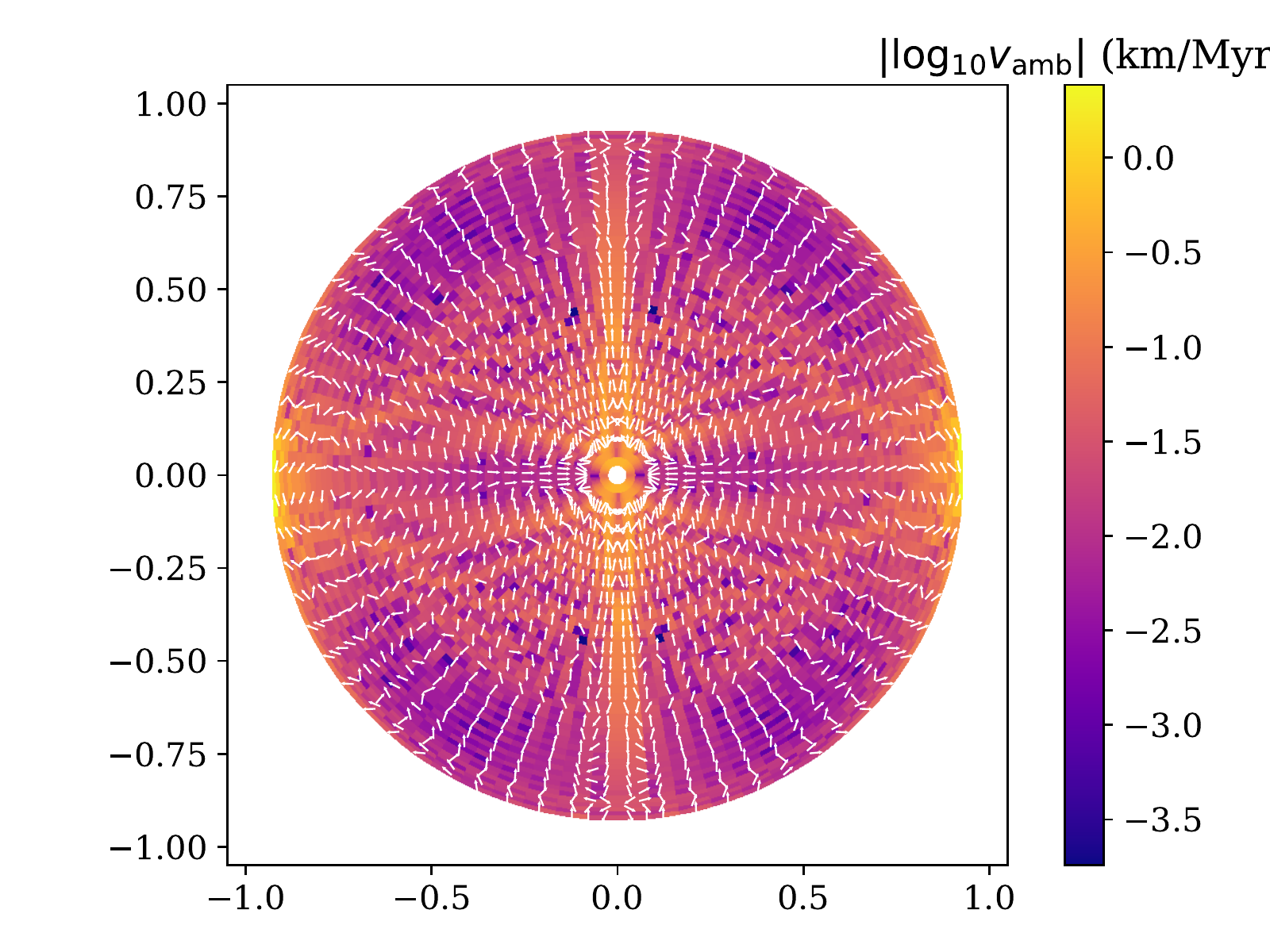}
    \end{minipage}
    \caption{Evolution of speed of ambipolar diffusion with fixed temperature $T_8 = 1$ and $r_\mathrm{cut}=2$ for $\xi_3$, $\xi_4$ (left column) and no $r_\mathrm{cut}$ (right column).
    From top to bottom the four rows are at ages 4 Kyr, 10 Kyr, 400 Kyr, and 1.5 Myr.}
    \label{fig:v_amb_temp_01_rdiff}
\end{figure*}

\begin{figure*}
    \centering
    \begin{minipage}{0.32\linewidth}
    \includegraphics[width=0.99\columnwidth]{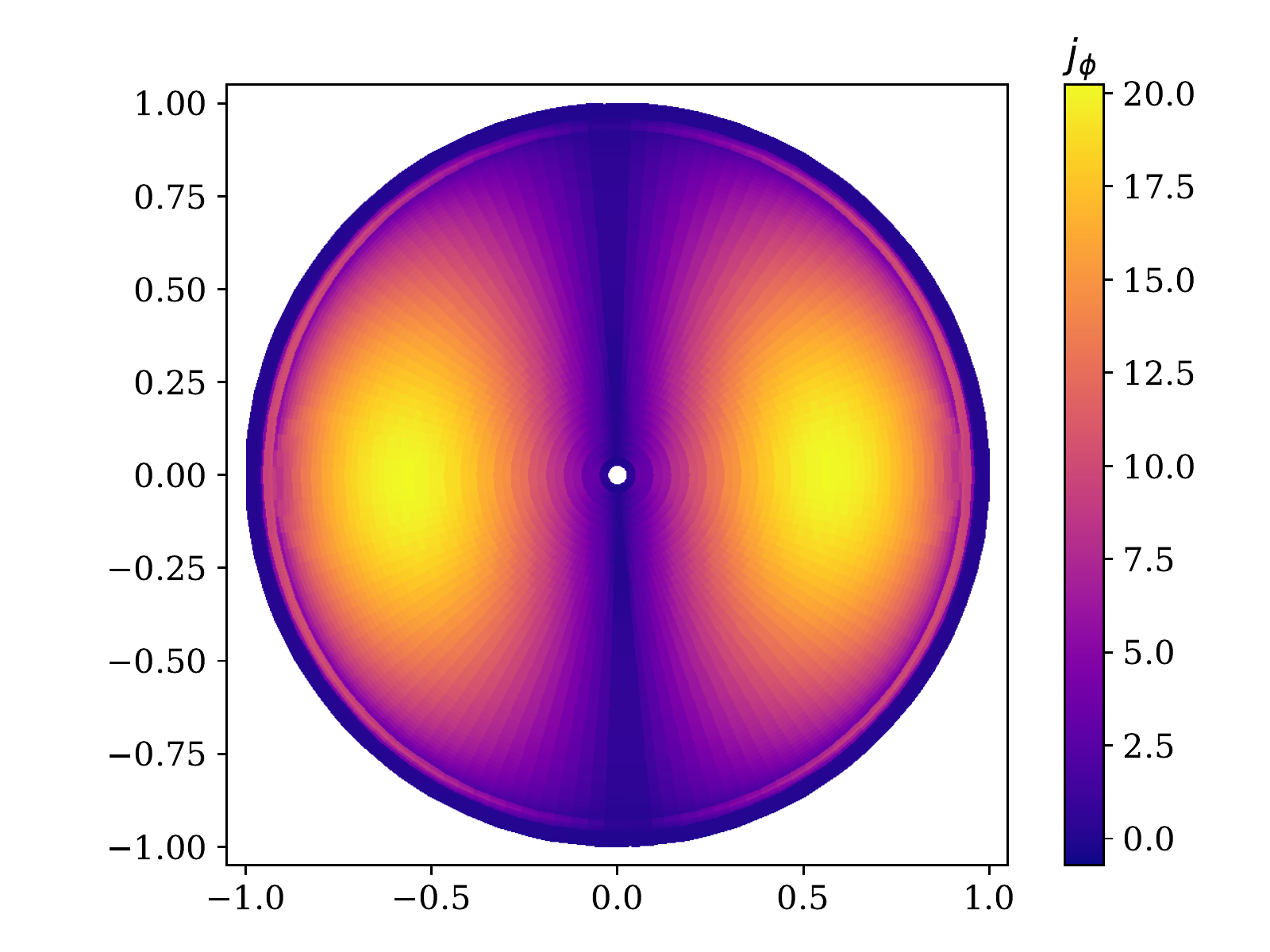}
    \end{minipage}
    \begin{minipage}{0.32\linewidth}
    \includegraphics[width=0.99\columnwidth]{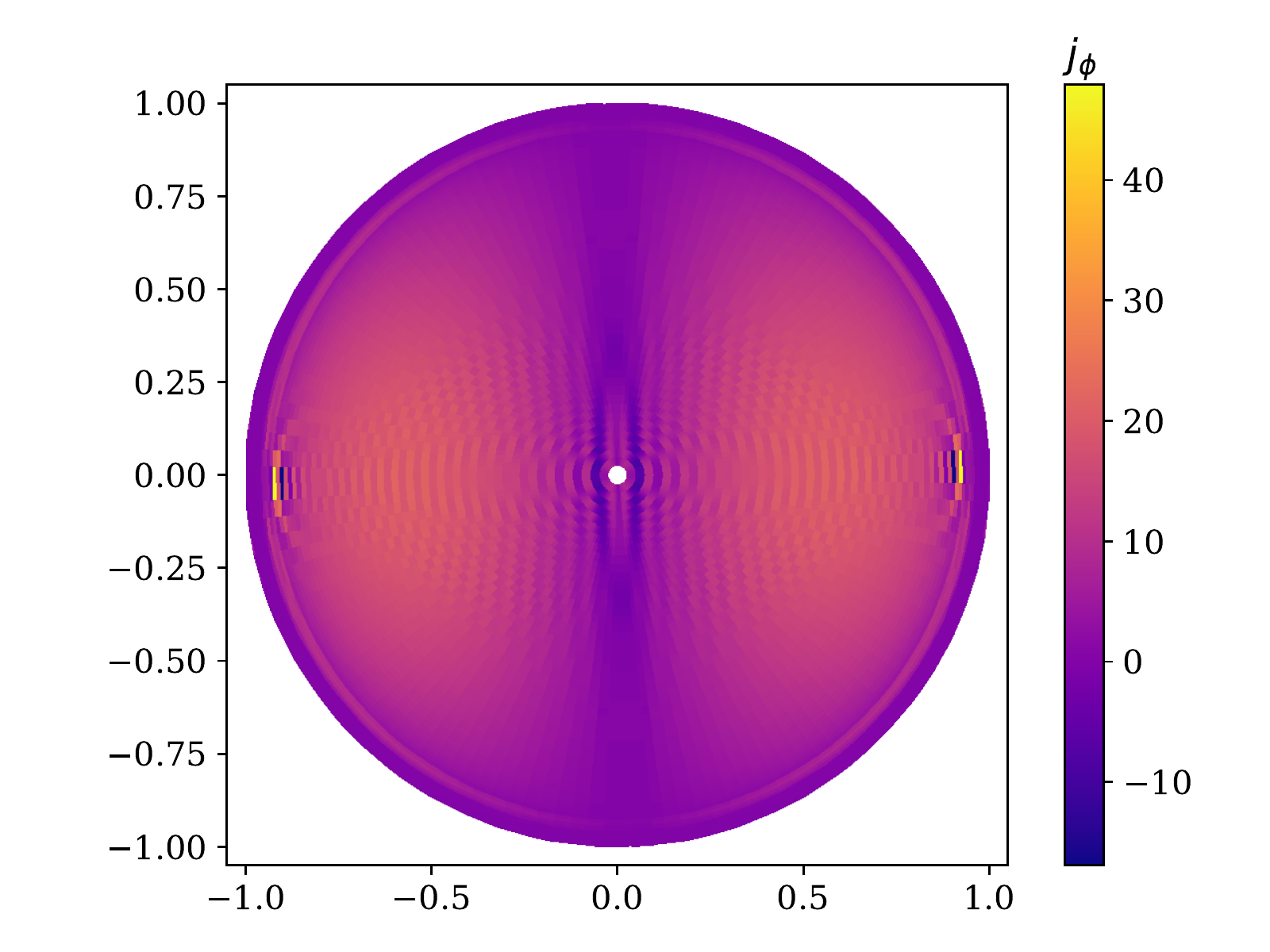}
    \end{minipage}
    \begin{minipage}{0.32\linewidth}
    \includegraphics[width=0.99\columnwidth]{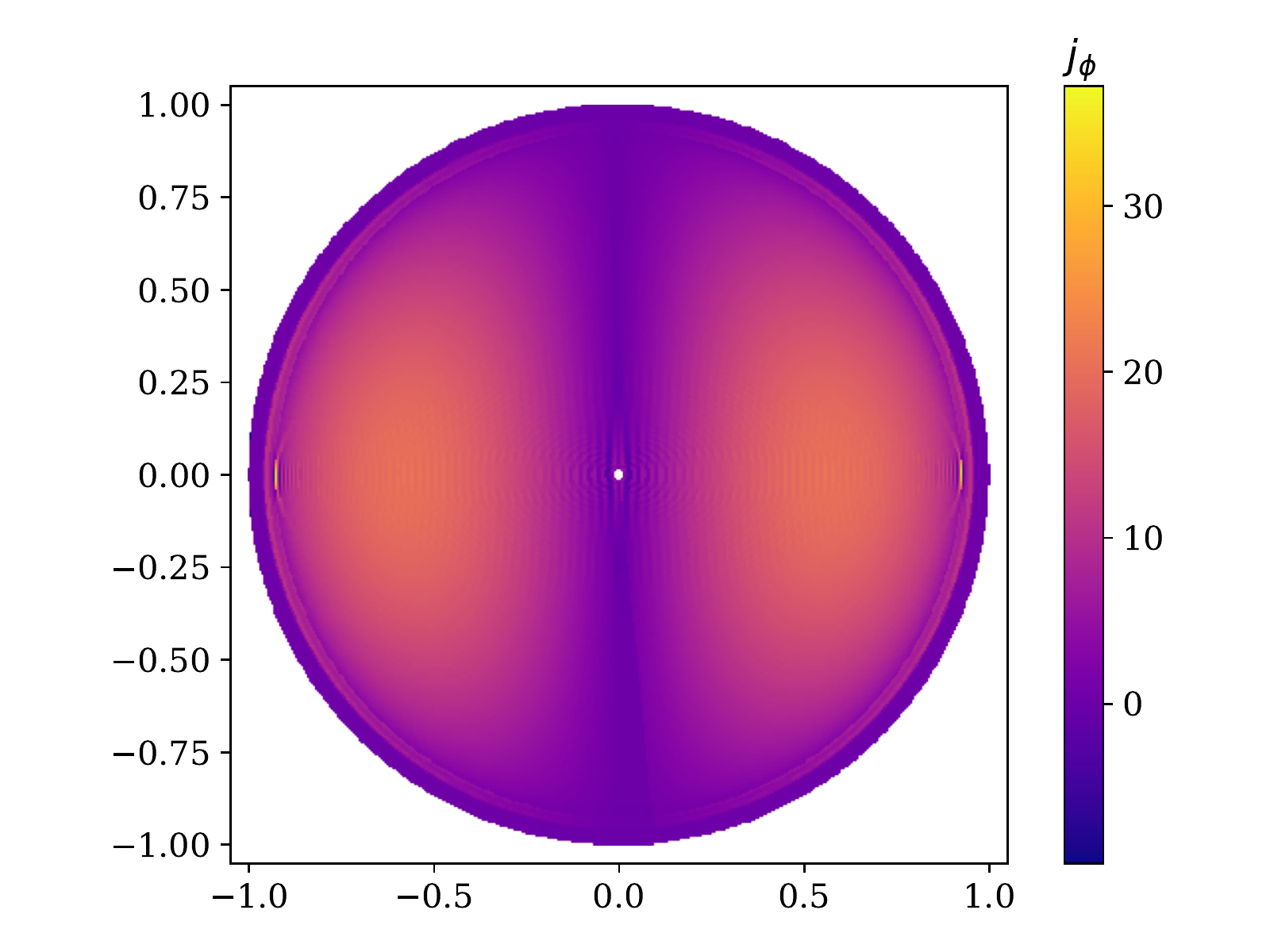}
    \end{minipage}
    \caption{Left panel: electric current $j_\phi$ at 600 Kyr in simulations with $r_\mathrm{cut} = 2$. Middle and right panels: no restriction on $\xi_3$, $\xi_4$ for resolution B (middle panel) and for resolution D (right panel).}
    \label{fig:j_phi_rcut}
\end{figure*}

We show the result of our simulations in Figure~\ref{fig:v_amb_temp_01_rdiff}. While the velocity field stays mostly smooth in the case of $r_\mathrm{cut} = 2$ (left panel), small-scale noise-like structures emerge if no cut is imposed (right panel). The rise of these structures coincides with appearance of a strong current at the core-crust boundary near $\theta = 90^\circ$, see Figure~\ref{fig:j_phi_rcut} for details. It is clear from this figure that large values of $\xi_3$ and $\xi_4$ cause appearance of compact current (see middle and right panels of Figure~\ref{fig:v_amb_temp_01_rdiff}). When we increase the numerical resolution the size of this current decays, but is still not fully resolved even with resolution D. It is to avoid these probable numerical artefacts that we introduced $r_\mathrm{cut} = 2$ in our basic simulations. When we introduce this restriction currents are well resolved and the velocity field looks much smoother.

\begin{figure*}
    \centering
    \begin{minipage}{0.49\linewidth}
    \includegraphics[width=0.99\columnwidth]{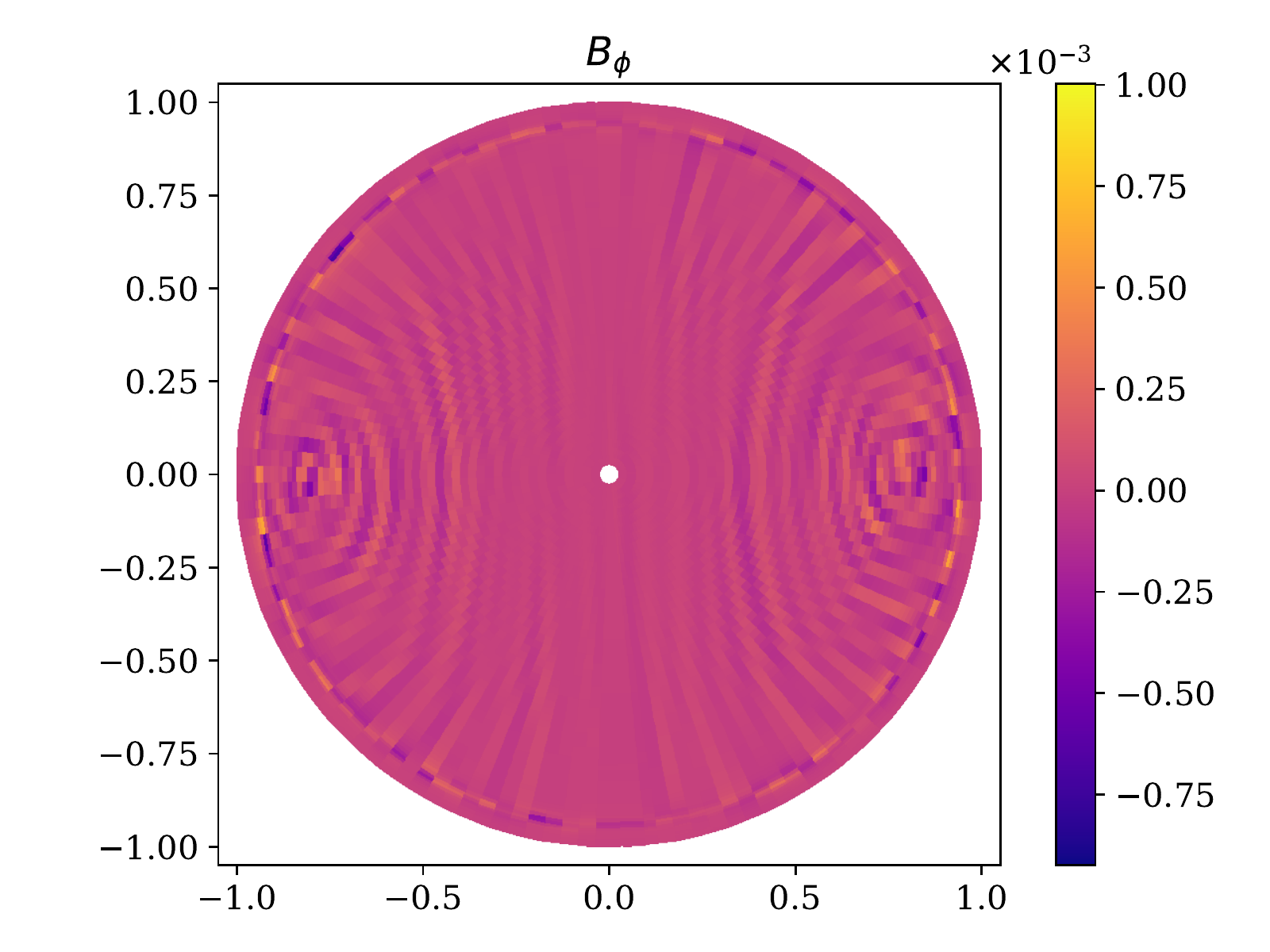}
    \end{minipage}
    \begin{minipage}{0.49\linewidth}
    \includegraphics[width=0.99\columnwidth]{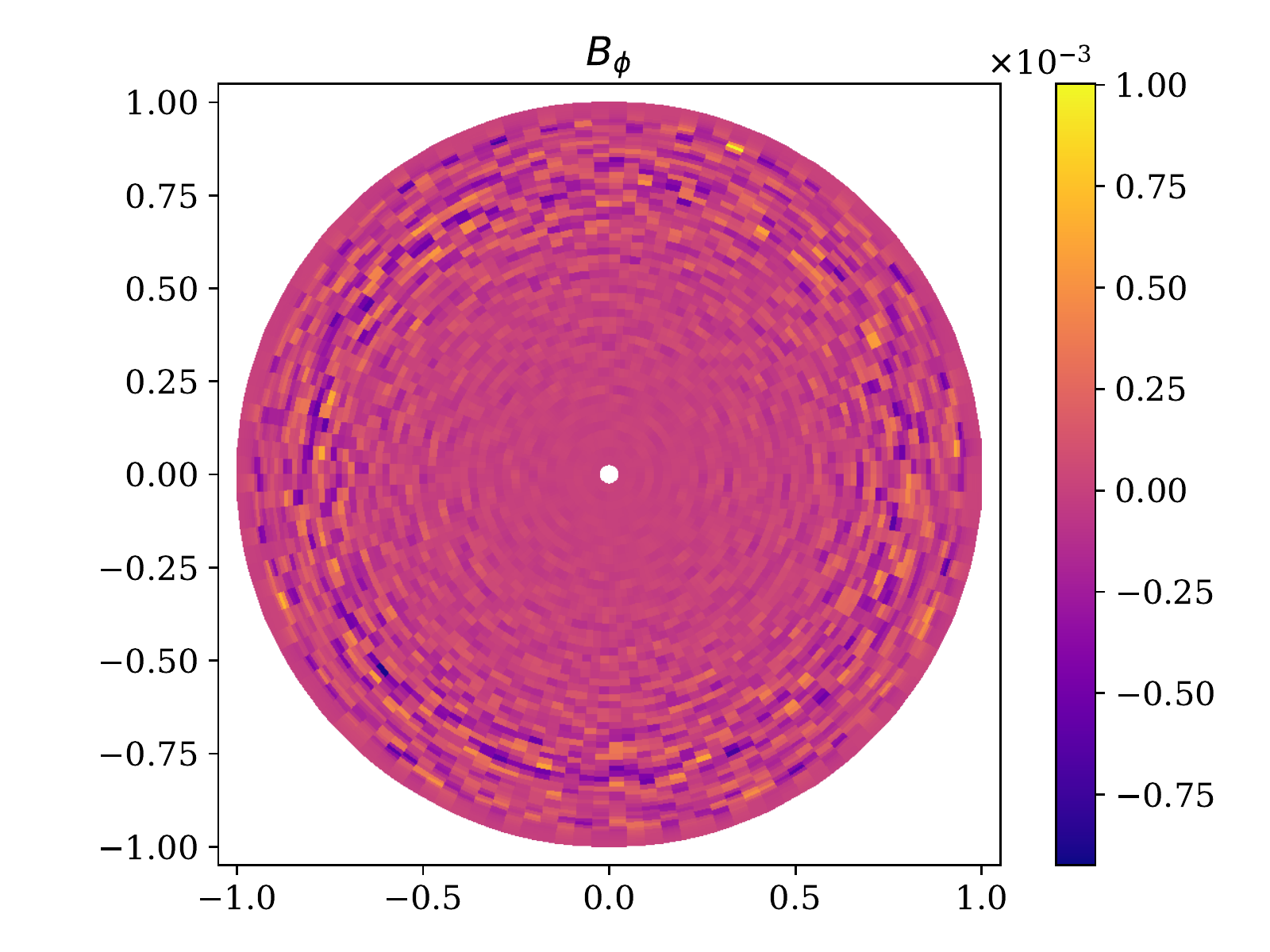}
    \end{minipage}
    \begin{minipage}{0.49\linewidth}
    \includegraphics[width=0.99\columnwidth]{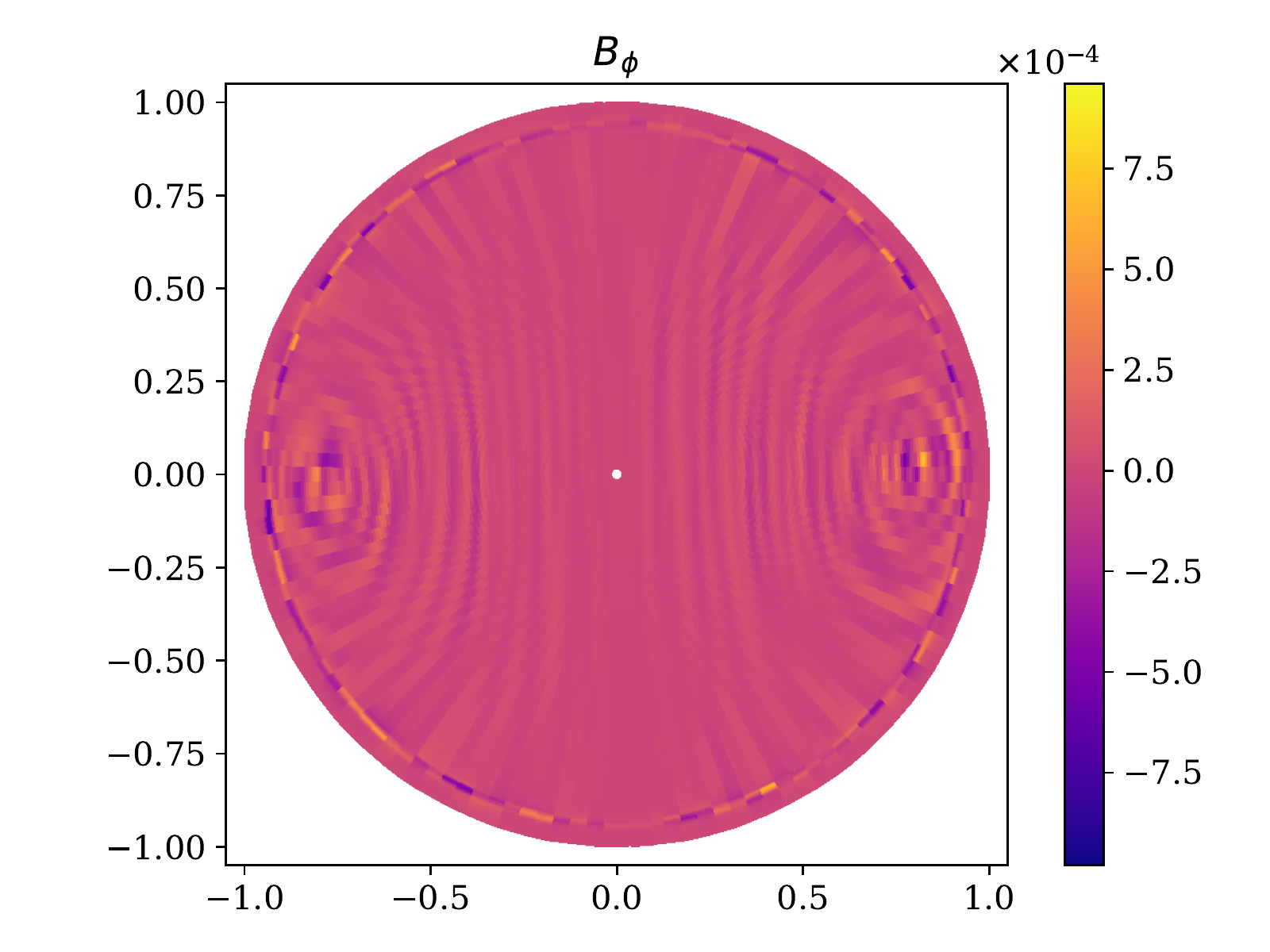}
    \end{minipage}
    \begin{minipage}{0.49\linewidth}
    \includegraphics[width=0.99\columnwidth]{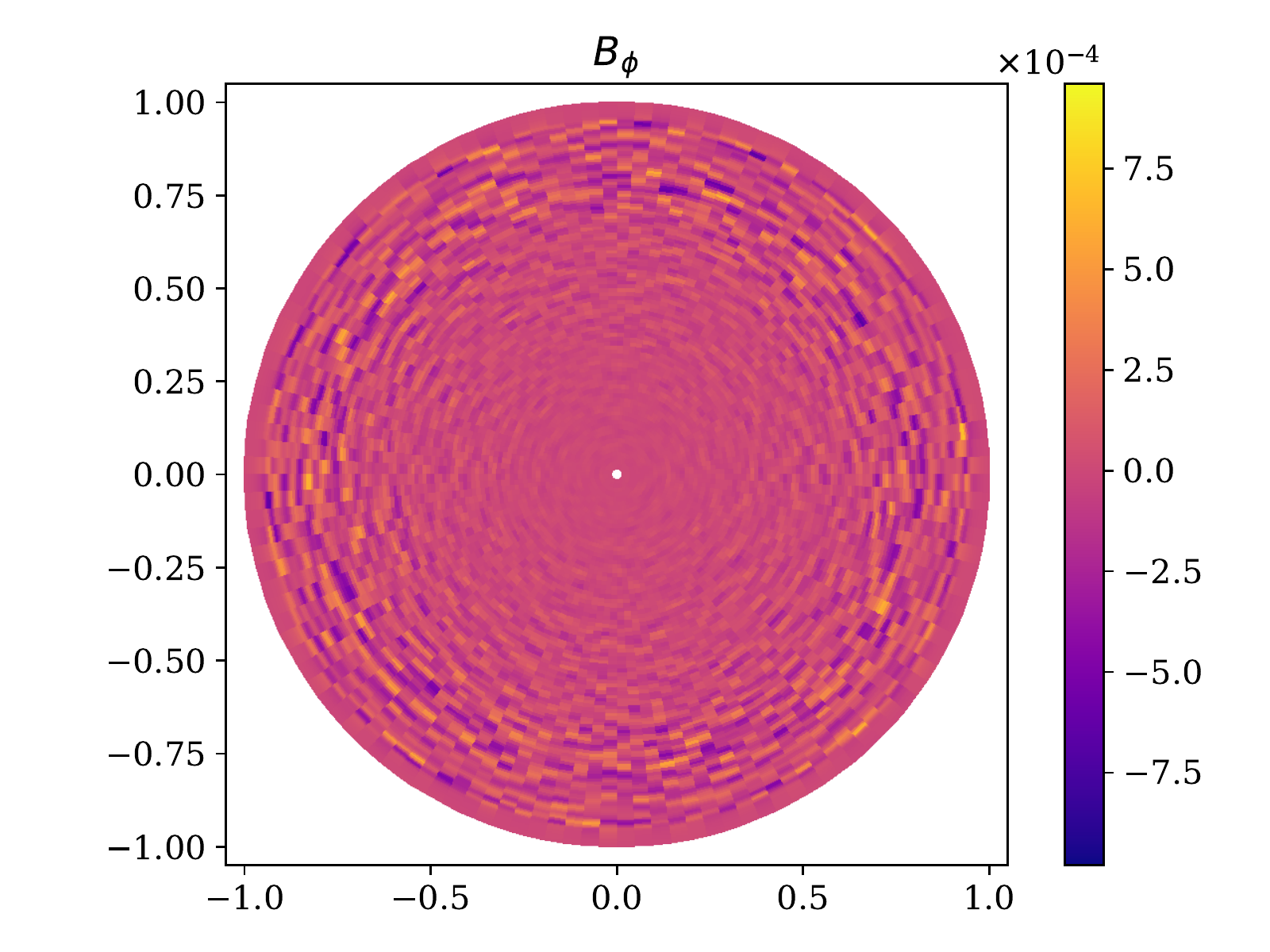}
    \end{minipage}
    \begin{minipage}{0.49\linewidth}
    \includegraphics[width=0.99\columnwidth]{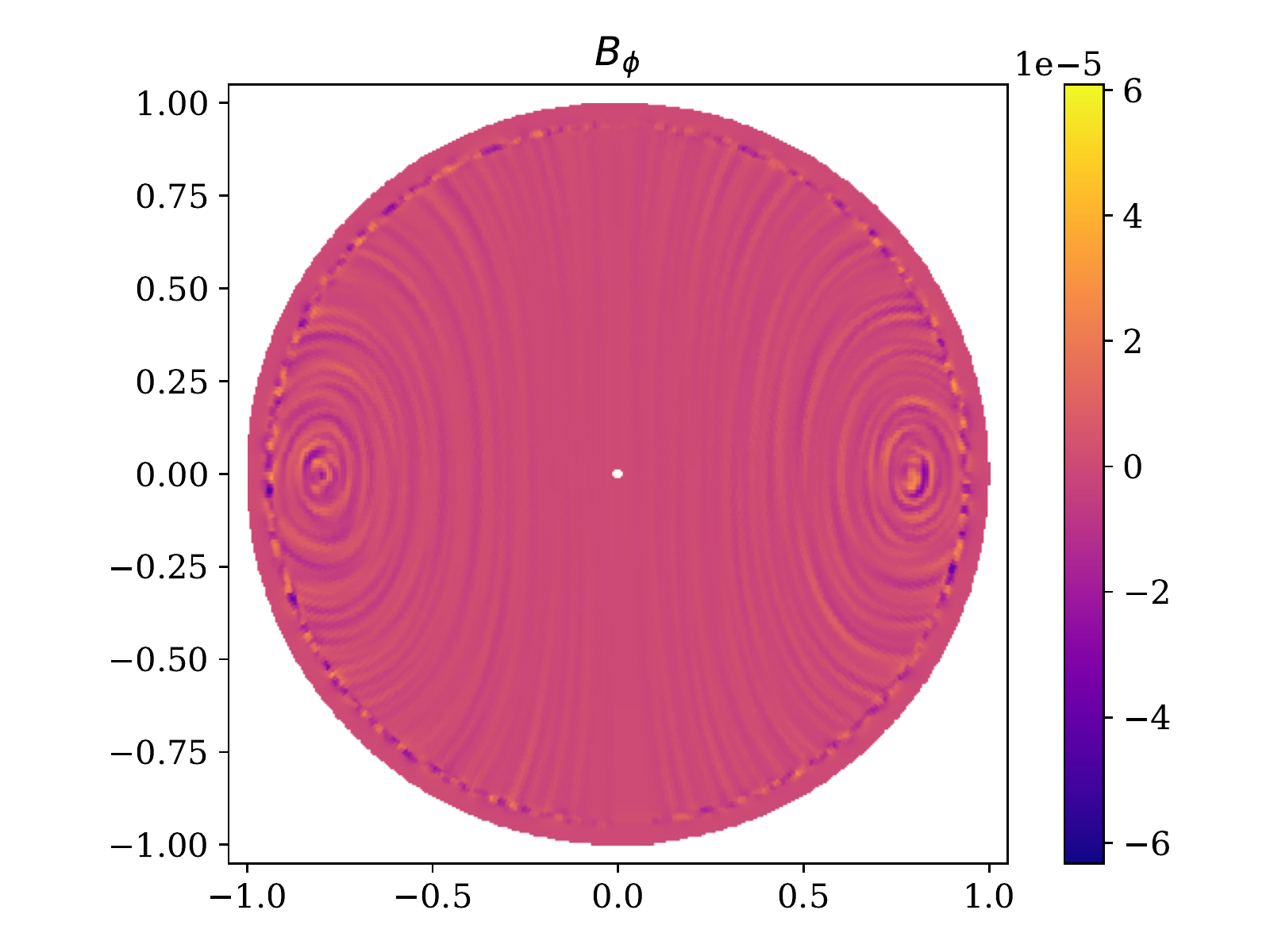}
    \end{minipage}
    \begin{minipage}{0.49\linewidth}
    \includegraphics[width=0.99\columnwidth]{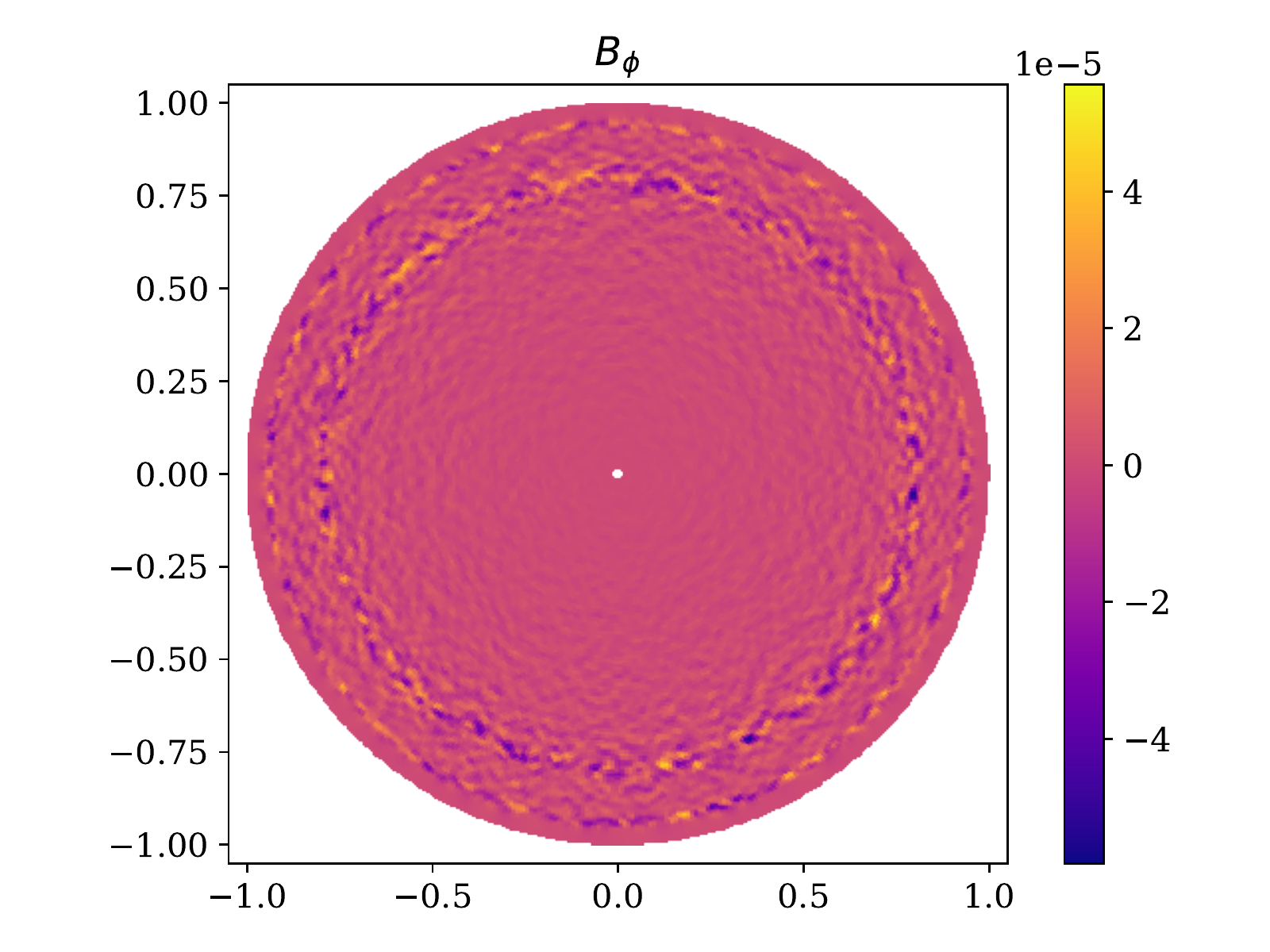}
    \end{minipage}
    \caption{Formation of regular non-axisymmetric toroidal magnetic field. The left column shows meridional cuts, the right column shows equatorial cuts,
    with times at 100 Kyrs.
    The top row shows the results of simulations with resolution B and $r_\mathrm{cut}=2$. The middle row shows the results of simulations with resolution C and $r_\mathrm{cut} = 20$. The last row shows the results of simulations with resolution D and no $r_\mathrm{cut}$ imposed.}
    \label{fig:non_axis_B_phi}
\end{figure*}

In order to check that behaviour which we identified in our simulation is actual physical behaviour and not a problem of the code, we run shorter simulations with different numerical resolution and varying parameter $r_\mathrm{cut}$. We demonstrate the results of these simulations in Figures~\ref{fig:non_axis_B_phi}.
A field with similar structure emerges in simulations with $r_\mathrm{cut}=20$ and better radial resolution. This toroidal magnetic field reaches maxima at radial distance of $r\approx 0.8$~R$_\mathrm{NS}$, i.e. well below the crust. When we increase the numerical resolution even further and removed restriction on radial profiles $\xi_3$ and $\xi_4$ we notice that a very similar toroidal magnetic field is formed. It requires significant computational resources to evolve simulation with resolution D on timescales of 10-20 Myr.

\section{Derivation of the initial condition for vector potential}
\label{a:vect_pot}
The initial condition by \cite{Akgun2013MNRAS} is written for poloidal and toroidal magnetic fields and not for their potentials. Namely, \cite{Akgun2013MNRAS} writes the magnetic field as:
\begin{equation}
\vec b = \hat \nabla \alpha \times \hat \nabla \phi + \beta \hat \nabla \phi,
\label{e:akgun}
\end{equation}
while the standard poloidal-toroidal decomposition is:
\begin{equation}
\vec b = \vec \nabla \times \vec \nabla \times (b_p \hat r) + \vec \nabla \times (b_t \hat r).
\end{equation}
Here $\hat \nabla \phi = \hat \phi / (r\sin \theta)$ and :
\begin{equation}
\alpha = f(x) \sin^2\theta,
\end{equation}
where $f(x)$ could be written as:
\begin{equation}
f(x) = \frac{35}{8} x^2 - \frac{21}{4} x^4 + \frac{15}{8} x^6.
\end{equation}
Here $x = r\in(0, 1]$. The respective scalar field for toroidal component is written as:
\begin{equation}
\beta = \left\{
\begin{array}{lcc}
(\alpha - 1)^2 & \mathrm{for} & \alpha \geq 1,\\
0              & \mathrm{for} & \alpha < 1. 
\end{array}
\right.
\end{equation}

First, we consider only the poloidal part of the magnetic field. We transform the first term of eq. (\ref{e:akgun}) and write it as:
\begin{equation}
\vec b_p = \vec \nabla \times (\alpha \vec \nabla \phi).
\end{equation}
In this case we just need to find $b_p$ such that $\vec \nabla \times (b_p \hat r) = \alpha \vec \nabla \phi$. Expanding the curl and assuming that the initial condition is axisymmetric we obtain:
\begin{equation}
-\frac{1}{r} \frac{\partial b}{\partial \theta} = \alpha \nabla \phi = \frac{f(x)}{r\sin \theta} \sin^2 \theta,
\label{e:eq_bp}
\end{equation}
using the same radial function $f(x)$ as by \cite{Akgun2013MNRAS}. The equation (\ref{e:eq_bp}) can be solved if we assume:
\begin{equation}
b_p (r, \theta) = f(r) \cos \theta.
\end{equation}

The same initial condition can also be written in terms of vector potential $\vec A$:
\begin{equation}
\vec\nabla \times \vec A = \vec B = \vec \nabla \times \left[(b_t \vec r) + \vec\nabla\times (b_p \vec r) \right],
\end{equation}
which we can write in components of the vector potential:
\begin{equation}
\begin{array}{ccc}
A_r      &=& b_t ,\\
A_\theta &=& \left[\vec \nabla \times (b_p \vec r)\right]_\theta ,\\
A_\phi   &=& \left[\vec \nabla \times (b_p \vec r)\right]_\phi.
\end{array}
\end{equation}
In our case:
\begin{equation}
\begin{array}{ccc}
A_r     &=& 0,\\
A_\theta &=& 0,\\
A_\phi   &=& \frac{f(r)}{r} \sin \theta.
\end{array}
\end{equation}

If we next add the toroidal magnetic field:
\begin{equation}
\beta \frac{\hat \phi}{r\sin\theta} = \nabla \times (b_t \vec r),
\end{equation}
then expanding the curl we obtain:
\begin{equation}
\frac{\partial b_t}{\partial \theta} = \frac{\beta}{\sin\theta}.
\end{equation}
We can solve this differential equation for cases when $\alpha \geq 1$:
\begin{equation}
b_t = \frac{1}{12} \left(\cos(3 \theta) - 9 \cos(\theta)\right) f^2(x) - 2 f(x) \cos\theta + \log\left(\cot \left[\frac{\theta}{2}\right]\right).
\end{equation}
For the cases $\alpha < 1$ we have to use a correct constant.


\bsp	
\label{lastpage}
\end{document}